\documentclass{ws-rv975x65}
\usepackage{subfigure}     % required only when side-by-side / subfigures are used
\usepackage{ws-rv-thm}     % comment this line when `amsthm / theorem / ntheorem` package is used
\usepackage{ws-rv-van}     % numbered citation/references (default)
\makeindex
\def\maprt#1#2{\smash{
    \mathop{\longrightarrow}\limits^{#1}_{#2} } }

\begin{document}

% The original title was
%\chapter[Energy and the Hamiltonian]{Energy and the Hamiltonian}

%\chapter{Gravitational energy: a covariant Hamiltonian--gauge theory perspective}
\chapter{Gravitational energy for GR and Poincar\'e gauge theories:\\ a covariant Hamiltonian approach}

%Would you like to consider to change the title to "Gravitational energy for general relativity and for Poincar e gauge theories: a covariant Hamiltonian approach" by adding "general relativity and for" for obvious reasons? The original one is also OK for me too.

\author[C.-M. Chen, J. M. Nester, R.-S. Tung]{Chiang-Mei Chen$^{1,2}$, James M. Nester$^{1,2,3,4}$, Roh-Suan Tung$^5$}
%\index[aindx]{Author, F.} % or \aindx{Author, F.}
%\index[aindx]{Author, S.} % or \aindx{Author, S.}

\address{$^1$Department of Physics, National Central University, Chungli 32054, Taiwan \\
$^2$Center for Mathematics and Theoretical Physics, National Central University, Chungli 32054, Taiwan \\
$^3$Graduate Institute of Astronomy, National Central University, Chungli 32054, Taiwan \\
$^4$Leung Center for Cosmology and Particle Astrophysics, National Taiwan University, Taipei 10617, Taiwan \\
$^5$Institute of Advanced Studies, Nanyang Technological University, Singapore \\
cmchen@phy.ncu.edu.tw, nester@phy.ncu.edu.tw, roh.suan.tung@gmail.com}

\begin{abstract}
Our topic concerns a long standing puzzle: the energy of gravitating systems.
More precisely we want to consider, for gravitating systems, how to best describe energy-momentum and angular momentum/center-of-mass momentum (CoMM).  It is known that these quantities cannot be given by a local density. The modern understanding is that (i) they are quasi-local (associated with a closed 2-surface), (ii) they have no unique formula, (iii) they have no reference frame independent description.
In the first part of this work we review some early history, much of it not so well known, on the subject of gravitational energy in Einstein's general relativity (GR), noting especially Noether's contribution.  In the second part we review (including some new results) much of our covariant Hamiltonian formalism and apply it to Poincar\'e gauge theories of gravity (PG), with GR as a special case.  The key point is that the Hamiltonian boundary term has two roles, it determines the quasi-local quantities, and, furthermore it determines the boundary conditions for the dynamical variables.   Energy-momentum and angular momentum/CoMM are associated with the geometric symmetries under Poincar\'e transformations.  They are best described in a local Poincar\'e gauge theory.  The type of spacetime that naturally has this symmetry is Riemann-Cartan spacetime, with a metric compatible connection having, in general, both curvature and torsion.
% Hence for a good description of energy etc.\ we developed a covariant Hamiltonian formulation applied to Poincar\'e gauge theories.
Thus our expression for the energy-momentum of physical systems is obtained via our covariant Hamiltonian formulation applied to the PG.
\end{abstract}

%\markright{Customized Running Head for Odd Page} % default is Chapter Title.
\body

%\section{outline}

%\begin{enumerate}
%\item Introduction
%\item historical background (esp Einstein, Hilbert, Klein, Noether)
%\item pseudotensor to Hamiltonian
%\item quasi-local
%\item gravity, geometry, connection \& gauge
%\item differential forms
%\item variational principles for form fields
%\item examples of Noether's theorems
%\item first order formalism
%\item Hamiltonian and the 3+1 split
%\item Hamiltonian and its boundary term
%\item asymptotics
%\item geometry
%\item variational principle for dynamic geometry
%\item first order form for PG and boundary term
%\item concluding discussion
%\item reference
%\end{enumerate}

%%%%%%%%%%%%%%%%%%%%%%%%%%%%%%%%%%%%%%%%%%%%%%%%%%%%%%%%%%%%%%%%%%%%%%
\section{Introduction}
%%%%%%%%%%%%%%%%%%%%%%%%%%%%%%%%%%%%%%%%%%%%%%%%%%%%%%%%%%%%%%%%%%%%%%
How to give a meaningful description of energy-momentum for gravitating systems (hence for all physical systems) has been an outstanding fundamental issue since Einstein began his search for his gravity theory, general relativity (GR).  It is deeply connected to the essential  nature of not only geometric gravity but of all the fundamental interactions---their inherent gauge nature.  Noether's paper that includes her two famous theorems relating global symmetries to conserved quantities and local gauge symmetries to differential identities was originally motivated by this very issue.
She showed that gravitational energy has no proper local description.  So investigators only found various expressions which were inherently non-tensorial (reference frame dependent \emph{pseudotensors}).
They have two inherent ambiguities: (i) there are many possible expressions, (ii) they are non-covariant---reference frame dependent. The modern view is that energy-momentum is \emph{quasi-local} (associated with a closed 2-surface).  Quasi-local proposals have analogous ambiguities.  These ambiguities  can be clarified by the Hamiltonian approach.  From a first order Lagrangian for quite general differential form fields, we have constructed a spacetime covariant Hamiltonian formalism, which incorporates the Noether conserved currents and differential identities.  The Hamiltonian that dynamically evolves a spatial region includes a boundary term. The explicit form of the boundary term depends on the boundary conditions and also on an appropriate reference choice. With a suitable vector field it gives expressions for the quasi-local quantities (energy-momentum, angular momentum/center-of-mass momentum) and also quasi-local energy flux.  A geometric gauge theory perspective provides the most appropriate dynamical variables. The geometry is Riemann-Cartan, with, in general, both curvature and torsion. For the PG (GR is a special case) with general source and gauge fields, we  identified a preferred Hamiltonian boundary expression along with a procedure for finding a ``best matched'' reference.  With this one can obtain  values for the quasi-local energy-momentum and angular momentum/center-of-mass momentum.\footnote{For an alternative to our Hamiltonian approach to energy-momentum and angular momentum for the PG see Kawai (1986).\cite{Kawai:1986wm}}

%%%%%%%%%%%%%%%%%%%%%%%%%%%%%%%%%%%%%%%%%%%%%%%%%%%%%%%%%%%%%%%%%%%%%%
Our topic here concerns the localization of energy-momentum.
%%%jmn418
The main aim of our research program has actually been to better understand the Hamiltonian
for dynamic spacetime geometry,  especially the role of the Hamiltonian boundary term.
It turns out that this sheds much light on the issue of the localization of energy.\cite{Energy10}
%%%jmn321 improve
A number of different ideas will be fit together to give a good picture of this long standing puzzle.
In addition to being mindful of Noether's results, we will use a Hamiltonian approach combined with a local gauge theory view of dynamic spacetime geometry.

This present work is largely just an application of Noether's result.
We will begin with some early history (much of it not so well known) regarding energy in the context of GR, especially Noether's contribution.  Next, we will show how in GR pseudotensors are connected with the Hamiltonian and introduce the quasi-local idea.
Following this we make some brief remarks about gravity, geometry, connection and gauge.  We then introduce and give a short review of our main tool: differential forms.  We develop in some detail variational principles with differential form fields.  With this we can give simple examples of applications of the two Noether theorems.  We then introduce the first order formalism followed by our Hamiltonian formulation and the 3+1 split. The Hamiltonian boundary term and its important roles are discussed next.  Asymptotic fall offs for the fields are noted.  We explain why Riemann-Cartan geometry is appropriate for our purposes.  Variational principles for dynamic spacetime geometry with quite general sources are developed, including the Noether conserved currents and differential identities.  We present a first order and Hamiltonian formulation for the PG along with the Hamiltonian boundary term and identify our preferred Hamiltonian boundary term for these dynamic spacetime geometry theories.  We also include a prescription for choosing the necessary reference values that are needed for the quasi-local energy-momentum and angular momentum expressions.

%%%%%%%%%%%%%%%%%%%%%%%%%%%%%%%%%%%%%%%%%%%%%%%%%%%%%%%%%%%%%%%%%%%%%%
\section{Background}
%%%%%%%%%%%%%%%%%%%%%%%%%%%%%%%%%%%%%%%%%%%%%%%%%%%%%%%%%%%%%%%%%%%%%%
As this present work approached its final form we received some very good news: all of the volumes of the Einstein papers published so far\footnote{At present up to 1923.}---both the originals and the English translations\cite{cpae}---are now freely available online.\cite{einsteinpapers}  An examination of Einstein's dozens of papers on gravity during the period 1913---1918, as well as his extensive correspondence with his contemporaries on the topic of gravity, shows that most of them include a significant consideration of the topic of gravitational energy.

\subsection{Some brief early history}
%%%%%%%%%%%%%%%%%%%%%%%%%%%%%%%%%%%%%%%%%%%%%%%%%%%%%%%%%%%%%%%%%%%%%%
We have only begun to look into the historical development of the modern ideas regarding gravitational energy; the topic merits much deeper study.  Here we can only give a brief report, relying on the Einstein papers as well as some of the many good historical investigations available, in particular regarding energy-momentum conservation.\cite{CattaniDeMaria,Trautman62,Brading2005}

We will rely on the Hamiltonian formalism applied to dynamical variables that are related to a local gauge theory of spacetime symmetry approach. (Earman\cite{Earman2003} has given a very interesting discussion on how the Hamiltonian approach connects with the gauge theory perspective.)

%%%jmn321 3a2

It seems not so well known that gravitational energy, or more precisely the proper description of the energy of gravitating systems (i.e., all real physical systems), has played a large role in the development of 20th century physics.

In the years 1912--1915 Einstein, when he was searching for satisfactory field equations, used a form of the equations that explicitly included an energy-momentum density for the gravitational field and were designed to satisfy the principle of energy-momentum conservation.\footnote{See Refs.~[\refcite{Janssen2005,Zurich,JanssenRenn07,Pais,Straumann:2011ev,Norton89}] for  discussions of
how Einstein found his field equations.\cite{AE1915}}
Thus an expression for the Einstein energy-momentum pseudotensor already existed even before he found the correct field equations.  It should be appreciated that general covariance brought with it features that had never before been encountered in any theory. (Indeed there is still controversy up to the present day\cite{Stachel89,BrownBrading2002,Norton93, Norton03}.) For a couple of years Einstein very much doubted that a generally covariant theory could be found;\footnote{One reason was his famous ``hole'' argument.\cite{Janssen2005,Pais}} he proposed that  energy conservation would select the preferred physical coordinate frame.
Initially Hilbert followed Einstein in this belief (see the proofs of his first note in Ref.~[\refcite{genesisHil}]).

Although Einstein began using variational principles in 1914\cite{AE1914}
%\footnote{29 May 1914, \textit{Zeitschrift f\"ur Mathematik und Physik} \textbf{63} (1914) 215.}
this was not his path to the field equations.  Hilbert was the first to identify a generally covariant Lagrangian\footnote{Einstein and Hilbert had quite different agendas;\cite{Vizgin,Rowe2001,Sauer2005,todorov} Hilbert in his Foundation of Physics papers, based on the work of Einstein and Mie, was using his axiomatic method with the objective of finding a unified field theory of gravity and electromagnetism.\cite{genesisHil,Hilbert,Sauer1999,CorryHilbert,BradingRyckman08,RS99}} (proportional to the Riemannian scalar curvature). He also constructed (in a complicated way that was not easy to understand) his ``conserved energy vector'', a vector with vanishing divergence associated with the general coordinate invariance (i.e., diffeomorphism invariance) of his Lagrangian.

Einstein's energy-momentum pseudotensor was criticized\cite{AE1918a} for giving ``unphysical'' values (Schr\"odinger\cite{schroedinger1918} noted that one could choose the coordinates to give a vanishing value outside a fluid sphere, and Bauer\cite{bauer1918} noted that one could choose the coordinates to give a vanishing energy value for empty Minkowski space).

Lorentz, Levi-Civita and Klein argued that the Einstein curvature tensor $G_{\mu\nu}$ was the only proper gravitational energy-momentum density; hence one should regard the Einstein equation in the form
\begin{equation}
-\frac{1}{\kappa} G_{\mu\nu} + T_{\mu\nu} = 0
\end{equation}
as describing the vanishing sum of gravitational and material energy-momentum.
(This idea has been advanced more recently by Cooperstock\cite{cooperstock}.) In our modern perspective for GR their idea is quite correct---but a \emph{density} is not the whole story.  There is more to energy-momentum than just a density.

\subsection{From Einstein's correspondence}
%%%%%%%%%%%%%%%%%%%%%%%%%%%%%%%%%%%%%%%%%%%%%%%%%%%%%%%%%%%%%%%%%%%%%%
Here are some excerpts from Einstein's correspondence concerning the Einstein pseudotensor, Hilbert's energy vector and Noether's contribution.\cite{Noether,Brading2005,BradingBrown2003,BrownBrading2002,Kosmann-Schwarzbach,Rowe1999}  They reveal the difficulties and the extent of understanding these people had at that time.
All these are quoted from the Einstein papers\cite{cpae,einsteinpapers} Vol.~8.

\begin{quotation}
``Highly esteemed Colleague, \dots
I am sitting over your relativity paper, \dots, and am honestly toiling over it.  I do admire your method, as far as I have understood it.  But at certain points I cannot progress and therefore ask that you assist me with brief instructions.
\dots
I still do not grasp the energy principle at all, not even as a statement.'' (Doc.~221 to Hilbert 25 May 1916)
\end{quotation}

\begin{quotation}
``Your explanation of equation (6) of your paper was charming.  Why do you make it so hard for poor mortals by withholding the technique behind your ideas? \dots
In your paper everything is understandable to me now except for the energy theorem. Please do not be angry with me that I ask you about this again. \dots
How is this cleared up? It would suffice, of course, if you would charge Miss Noether with explaining this to me.''\\
$\hbox{ }$\qquad (Doc.~223 to Hilbert 30 May 1916)
\end{quotation}

\begin{quotation}
\noindent ``My $\mathfrak{t}^\mu_\sigma$'s are being rejected by everyone as unkosher.''\\ $\hbox{ }$\qquad (Doc.~503 to Hilbert 12 April 1918)
\end{quotation}

\begin{quotation}
\noindent``\dots Only (24) is an identity \dots The relations here are exactly analogous to those for nonrelativistic theories.''\\
$\hbox{ }$\qquad (Doc.~480 to Klein 13 March 1918)
\end{quotation}

\begin{quotation}
\noindent ``I have succeeded in discovering the organic formation law for Hilbert's energy vector'' (Doc.~588 from Klein 15 July 1918)
\end{quotation}

\begin{quotation}
\noindent ``The only thing I was unable to grasp in your paper is the conclusion at the top of page 8 that $\varepsilon^\sigma$ was a vector.''\\
$\hbox{ }$\qquad (Doc.~638 to Klein 22 Oct 1918)
\end{quotation}

\begin{quotation}
\noindent ``Thank you very much for the transparent proof, which I understood completely.'' (Doc.~646 to Klein 8 Nov 1918).
\end{quotation}

\begin{quotation}
\noindent ``\dots Meanwhile, with Miss Noether's help, I understand that the proof for the vector character of $\varepsilon^\sigma$ from ``higher principles'' as I had sought was already given by Hilbert on pp.\ 6, 7 of his first note, although in a version that does not draw attention to the essential point.'' (Doc.~650 from Klein 10 Nov 1918)
\end{quotation}

%Regarding Noether

%Doc 548 To Hilbert 24 May 1918

%``Yesterday I received a very interesting paper by Ms. Noether about the generation of invariants.
%It impresses me that these things can be surveyed from such a general point of view.
%It would not have harmed the G\"ottingen old guard to have been sent to Miss Noether for schooling.
%She seems to know her trade well!''

%Doc 650 From Klein 10 Nov 1918

Briefly, after a couple of years Klein clarified Hilbert's energy-momentum ``vector''; he related it to Einstein's pseudotensor, but (as we will discuss in more detail shortly) disagreed with Einstein's physical interpretation of divergenceless expressions.\footnote{For these investigators these things were not as easy as they are for us today; in particular the Bianchi identity and its contracted version were not known to these people,\cite{Pais,RoweBianchi} so they had, in effect, to rediscover that identity from effectively the diffeomorphism invariance of the Lagrangian.}
Enlisted by Hilbert and Klein, it was Emmy Noether who solved the primary puzzle regarding gravitational energy.

\subsection{Noether's contribution}
%%%%%%%%%%%%%%%%%%%%%%%%%%%%%%%%%%%%%%%%%%%%%%%%%%%%%%%%%%%%%%%%%%%%%%
If one had to describe 20th century physics in one word a good choice would be \emph{symmetry}.
Most of the new theoretical physics ideas involved symmetry.
Essentially they can be seen as applications of Noether's theorems.
Briefly, Noether's first theorem associates conserved quantities with global symmetries, and
Noether's second theorem concerns local symmetries: it is the mathematical foundation of the modern gauge theories.
Unfortunately her work\footnote{For discussions see Refs.~[\refcite{Noether,Brading2005,BradingBrown2003,BrownBrading2002,Kosmann-Schwarzbach,Rowe1999}].} was largely overlooked for about 50 years\cite{Kosmann-Schwarzbach}.
Why did Noether make her investigation?  To clarify the issue of gravitational energy.

Klein was looking into Einstein's theory and the relationship between Einstein's pseudotensor and Hilbert's energy vector.  Some of the correspondence between Hilbert and Klein was published in a paper.\cite{Klein}  We quote some excerpts:\footnote{We do not know of any English translation of Klein's papers; the translations of the following excerpts are quoted from Ref.~[\refcite{Kosmann-Schwarzbach}] p 66.}
\medskip

Klein wrote
\begin{quotation}
You know that Miss Noether advises me continually regarding my work, and that in fact it
is only thanks to her that I have understood these questions. When I was speaking recently
to Miss Noether about my result concerning your energy vector, she was able to inform me
that she had derived the same result on the basis of developments of your note (and thus not
from the simplified calculations of my section 4) more than a year ago, and that she had then
put all of that in a manuscript (which I was subsequently able to read). She simply did not
set it out as forcefully as I recently did at the Mathematical Society (22 January [1918]).
%%%jmn321 cite translation source
\end{quotation}

Hilbert responded
\begin{quotation}
I fully agree in fact with your statements on the energy theorems: Emmy Noether, on whom
I have called for assistance more than a year ago to clarify this type of analytical questions
concerning my energy theorem, found at that time that the energy components that I
had proposed---as well as those of Einstein---could be formally transformed, using the Lagrange
differential equations (4) and (5) of my first note, into expressions whose divergence
vanishes identically, that is to say, without using the Lagrange equations (4) and (5).
%One would say in a more modern language that the conservation laws are valid.
%%%jmn31 cite translation source
\end{quotation}

Also
\begin{quotation}
Indeed I believe that in the case of general relativity, i.e., in the case of the general invariance
of the Hamiltonian function, the energy equations which in your opinion correspond
to the energy equations of the theory of orthogonal invariance do not exist at all; I can even
call this fact a characteristic of the general theory of relativity.
%%%jmn321 cite translation source
\end{quotation}

What Hilbert here calls the Hamiltonian function we now refer to as the Lagrangian.  Noether wrote her 1918 paper to clarify the situation.

\subsection{Noether's result}
%%%%%%%%%%%%%%%%%%%%%%%%%%%%%%%%%%%%%%%%%%%%%%%%%%%%%%%%%%%%%%%%%%%%%%
Many have heard of Noether's theorems, but the full scope of what she actually did is not so generally well known.  So we take this opportunity to quote (from the highly recommended book of Kosmann-Schwartzbach\cite{Kosmann-Schwarzbach}) in full her key results. (It should be mentioned that her Lagrangians were quite general, they could depend on any finite number of derivatives.)

\begin{quotation}
\noindent\textbf{Theorem I}. If the integral $I$ is invariant under a [finite continuous group with $\rho$ parameters] $G_{\rho}$, then there are $\rho$ linearly independent combinations among the Lagrangian expressions which become divergences---and conversely, that implies the invariance of $I$ under a [group] $G_{\rho}$. The theorem
remains valid in the limiting case of an infinite number of parameters.
\end{quotation}

\begin{quotation}
\noindent\textbf{Theorem II}. If the integral $I$ is invariant under a [an infinite continuous group] $G_{\infty \rho}$ depending on arbitrary functions and their derivatives up to order $\sigma$, then there are $\rho$ identities among the Lagrangian expressions and their derivatives up to order $\sigma$. Here as well the converse
is valid.
\end{quotation}

Furthermore she has another important result, although it follows easily from Theorem II, in our opinion, both because of its importance and the fact that it was the key issue motivating her investigation,  it could have been set off and called
\textbf{Theorem III}:

\begin{quotation}
\noindent Given $I$ invariant under the group of translations, then the energy relations are improper if and only if $I$ is invariant
under an infinite group which contains the group of translations as a subgroup.
\end{quotation}

Regarding this latter result, she ends her paper with the remarks

\begin{quotation}
As Hilbert expresses his assertion, the lack of a proper law of energy constitutes
a characteristic of the ``general theory of relativity.'' For that assertion to be literally
valid, it is necessary to understand the term ``general relativity'' in a wider sense than
is usual, and to extend it to the aforementioned groups that depend on $n$ arbitrary
functions.$^{27}$
\end{quotation}

The footnote that ends her paper is also of interest:

\begin{quotation}
\noindent $^{27}$ This confirms once more the accuracy of Klein's remark that the term ``relativity'' as it is used in physics should be replaced by ``invariance with respect to a group.''
\end{quotation}

Her result regarding the lack of a proper law of energy applies not just to Einstein's general relativity theory, but in fact to all geometric theories of gravity: for all such theories there is no proper conserved energy-momentum density.

As a well known textbook expresses it:
\begin{quotation}
\noindent Anyone who looks for a magic formula for ``local gravitational energy-momentum'' is looking for the right answer to the wrong question.  Unhappily, enormous time and effort were devoted in the past to trying to ``answer this question'' before investigators realized the futility of the enterprise.\\
$\hbox{ }\qquad $ Misner, Thorne \& Wheeler \textit{Gravitation}\cite{MTW73} p 467.
\end{quotation}

%%%%%%%%%%%%%%%%%%%%%%%%%%%%%%%%%%%%%%%%%%%%%%%%%%%%%%%%%%%%%%%%%%%%%%
\section{The Noether energy-momentum current ambiguity}
%%%%%%%%%%%%%%%%%%%%%%%%%%%%%%%%%%%%%%%%%%%%%%%%%%%%%%%%%%%%%%%%%%%%%%
Let us begin our technical discussion by first reviewing some background.

As we will soon show, a well known result is that
from a classical field Lagrangian density, ${\cal L}(\varphi_A, \partial_\mu \varphi_A)$, via Noether's first theorem, the translational symmetry of Minkowski spacetime leads to a simple formula for the ``conserved''
\emph{canonical energy-momentum density} %%%jmn419 here should cite some classical fied theory books.
\begin{equation}
T^\mu {}_\nu := \delta^\mu_\nu {\cal L}-\frac{\partial {\cal L}}{\partial \partial_\mu \varphi_A} \partial_\nu \varphi_A .\label{canEMT}
\end{equation}
The divergence of this expression satisfies the identity\footnote{A consequence of assuming that  the Lagrangian depends on position only through the field $\varphi_A$.}
\begin{equation}
\partial_\mu T^\mu{}_\nu \equiv \frac{\delta {\cal L}}{\delta \varphi_A} \partial_\nu \varphi_A,\label{concanEM}
\end{equation}
which can easily be directly verified using the definition of the \emph{Euler-Lagrange variational derivative} 
\begin{equation}
\frac{\delta {\cal L}}{\delta \varphi_A} := \frac{\partial {\cal L}}{\partial \varphi_A} - \partial_\mu \left( \frac{\partial {\cal L}}{\partial\partial_\mu \varphi_A} \right).
\end{equation}
The canonical energy-momentum density is a conserved current in the sense that ``on shell'' (i.e., when the Euler-Lagrange field equations are satisfied: ${\delta {\cal L}}/{\delta \varphi_A}=0)$ its divergence vanishes.

The above energy-momentum density has the usual conserved current ambiguity:
\begin{equation}
T'^\mu{}_\nu:=T^\mu{}_\nu + \partial_\lambda U^{[\mu\lambda]}{}_\nu    \label{T+dU}
\end{equation}
is likewise conserved but defines different energy-momentum values.  Essentially, one can always adjust by a ``curl'' a divergence free current.

At first thought one might be inclined to follow the rule of sticking with the results obtained directly from the Lagrangian and the above formula. But sometimes the results so obtained are not so suitable physically.

%%%jmn419 the following has been revised to address the point raised in 3a9
A simple example in Minkowski space is the Lagrangian density
\begin{equation}
{\cal L} = - \frac14 F^{\mu\nu} F_{\mu\nu}, \qquad F_{\mu\nu} = \partial_\mu A_\nu - \partial_\nu A_\mu.
\end{equation}
If one regards this Lagrangian density according to the above paradigm as a function of $A_\mu$ and $\partial_\mu A_\nu$ then
the above formula leads directly to the conserved expression
\begin{equation}
T^\mu{}_\nu = F^{\mu\alpha} \partial_\nu A_\alpha - \frac14 \delta^\mu_\nu F^{\alpha\beta} F_{\alpha\beta}. \label{canMax}
\end{equation}
Now the above Lagrangian density (up to a suitable overall scaling coefficient that can be set aside for the purposes of this section) can be used to describe Maxwell electrodynamics.
As is physically appropriate, this Lagrangian density and the field equations obtained from it are gauge invariant under the local ``gauge'' transformation $A_\mu\to A_\mu +\partial_\mu \chi$.
However the above canonical energy-momentum density \emph{is not gauge invariant} (nevertheless, as we will see later, it can still be useful physically). Naturally, one would generally prefer to have a gauge invariant energy-momentum density for electrodynamics.  In this particular case there are several ways that one can find an alternative to~(\ref{canMax}): (i) one can exploit the abovementioned freedom (\ref{T+dU}) and thereby find ``by hand'' an ``improved'' gauge invariant expression, (ii) one could regard the Lagrangian as being a function of a one-form $A$ and its differential (this is really the proper way to treat electromagnetism---but then one needs an extension of the above classical field theory formalism that can accommodate form fields; we will discuss such a formalism below), or (iii) one can consider that physically any time one has material energy-momentum one must also have gravity: the gravitational equations will include an unambiguous gauge invariant energy-momentum density.\footnote{The existence of a gravitational field will reveal the location of a source with energy-momentum even if it has not otherwise been detected.  An important example of this is---assuming gravity is well described by GR---from astronomical observations it seems that there is a large amount ``dark matter'' in the universe.   Clearly, the issue of the proper description of energy for gravitating systems can have  major consequences for our conception of the physical world.}  From a specific gravity theory one gets then a specific formula for the energy-momentum density. In this way the ambiguity in the canonical energy-momentum density for any classical field can be entirely removed when we consider their gravitational effects.
%
%%%jmn321 comment
%
%%%jmn418 3b14
Specifically, in the case of GR, knowing the curvature gives, via the Einstein tensor, the \emph{symmetric Hilbert energy-momentum density}.  In particular for the electromagnetism example this is
\begin{equation}
T^\mu{}_\nu = F^{\mu\alpha} F_{\nu\alpha} - \frac14 \delta^\mu_\nu F^{\alpha\beta} F_{\alpha\beta},
\label{HilbertMax}
\end{equation}
which is no doubt a good choice for the energy-momentum density for Maxwell electrodynamics; in fact it is,  as we shall see, the same as the energy-momentum density that one obtains by regarding the vector potential as a one-form.

While for electromagnetism one has another criteria (gauge invariance) that can be used to arrive at a physically suitable energy-momentum density, for most other sources one can only turn to gravity to identify a unique energy-momentum density.
Hence gravity \emph{uniquely} detects the energy-momentum density of its sources.  It may thus seem somewhat ironic that \emph{for gravity itself} there is no proper energy-momentum density.

%%%%%%%%%%%%%%%%%%%%%%%%%%%%%%%%%%%%%%%%%%%%%%%%%%%%%%%%%%%%%%%%%%%%%%
\section{Pseudotensors}\label{S:pseudotensors}
%%%%%%%%%%%%%%%%%%%%%%%%%%%%%%%%%%%%%%%%%%%%%%%%%%%%%%%%%%%%%%%%%%%%%%
%\subsection{Klein-Einstein}

The Einstein Lagrangian for GR differs from the Hilbert scalar curvature Lagrangian by a certain total divergence which removes the 2nd derivatives of the metric:%
\footnote{Here $\Gamma^\alpha{}_{\beta\gamma}:=\frac12g^{\alpha\lambda}(\partial_\beta g_{\lambda\gamma}+\partial_\gamma g_{\lambda\beta}- \partial_\lambda g_{\beta\gamma})$ is the well-known Christoffel/Levi-Civita connection,
$\delta^{\mu\nu}_{\beta\sigma}:=2\delta^\mu_{[\beta}\delta^\nu_{\sigma]}$, and
$\kappa:=8\pi G/c^4$.}
%\begin{eqnarray}%%%jmn322 3b10
%
\begin{eqnarray}%%%jmn513 3b10
2\kappa{\cal L}_{\rm E}(g_{\alpha\beta}, \partial_\mu g_{\alpha\beta})&:=&
-\sqrt{-g}g^{\beta\sigma} \Gamma^\alpha{}_{\gamma\mu} \Gamma^\gamma{}_{\beta\nu}\delta^{\mu\nu}_{\alpha\sigma}\nonumber \label{EinsteinL}\\
&\equiv&
\sqrt{-g}R-\partial_\mu(\sqrt{-g}g^{\beta\sigma}\Gamma^\alpha{}_{\beta\gamma}\delta^{\mu\gamma}_{\alpha\sigma})\nonumber\\
& =:&
2\kappa{\cal L}_{\rm H}(g_{\alpha\beta}, \partial_\mu g_{\alpha\beta}, \partial_{\mu\nu} g_{\alpha\beta}) - \partial_\mu{\cal K^\mu}.
\end{eqnarray}
They give the same field equations.
The Einstein pseudotensor can be obtained from ${\cal L}_{\rm E}$ using the aforementioned formula for the canonical energy-momentum tensor~(\ref{canEMT}):
\begin{equation}\label{EpseudoT}
\mathfrak{t}_{\rm E}^\mu{}_\nu := \delta^\mu_\nu{\cal L}_{\rm E} - \frac{\partial{\cal L}_{\rm E}}{\partial \partial_\mu g_{\alpha\beta}} \partial_\nu g_{\alpha\beta}.
\end{equation}
(Here, following tradition, gothic letters indicate densities.)
We have from (\ref{concanEM}) using the Einstein equation
\begin{eqnarray}
\partial_\mu (\mathfrak{t}_{\rm E}^\mu{}_\nu)&\equiv&
\frac{\delta {\cal L}_{\rm E}}{\delta g_{\alpha\beta}}\partial_\nu g_{\alpha\beta}  %\nonumber\\&\equiv&
\equiv-(2\kappa)^{-1}\sqrt{-g} G^{\alpha\beta}\partial_\nu g_{\alpha\beta}%\nonumber %\\
%&=&
=-\frac12\mathfrak{T}^{\alpha\beta}\partial_\nu g_{\alpha\beta}.
%&=&
\end{eqnarray}
Hence, using the vanishing covariant divergence of the material energy-momentum,
\begin{equation}
0=\nabla_\mu(\mathfrak{T}^\mu{}_\nu)= \partial_\mu(\mathfrak{T}^\mu{}_\nu)-\Gamma^\gamma{}_{\nu\mu} \mathfrak{T}^\mu{}_\gamma=\partial_\mu(\mathfrak{T}^\mu{}_\nu)-\frac12\mathfrak{T}^{\alpha\beta}\partial_\nu g_{\alpha\beta}, \label{divT}
\end{equation}
we obtain
\begin{equation}
\partial_\mu (\mathfrak{T}^\mu{}_\nu+\mathfrak{t}_{\rm E}^\mu{}_\nu)=0,
\end{equation}
a vanishing ordinary divergence, i.e., a conserved total energy-momentum ``current''.
Here we assumed the vanishing of the covariant divergence of the material energy-momentum tensor and used Einstein's equations to obtain an ordinary divergence conserved current.  But one can argue the other way around, as Einstein did in 1916.

\subsection{Einstein, Klein, and superpotentials}
%%%%%%%%%%%%%%%%%%%%%%%%%%%%%%%%%%%%%%%%%%%%%%%%%%%%%%%%%%%%%%%%%%%%%%
Einstein\cite{AE1916} obtained results of the form
\begin{eqnarray}
\partial_\mu (\mathfrak{T}^\mu{}_\nu+\mathfrak{t}_{\rm E}^\mu{}_\nu)&=&0,\label{divTotalT}\\
\mathfrak{T}^\mu{}_\nu+\mathfrak{t}_{\rm E}^\mu{}_\nu&=&\partial_\lambda \mathfrak{s}^{\mu\lambda}{}_\nu,\label{TotalTpot}\\
\partial_{\mu\lambda} \mathfrak{s}^{\mu\lambda}{}_\nu&\equiv&0, \label{diffeoident}\\
\mathfrak{s}^{\mu\lambda}{}_\nu&:=&g^{\mu\alpha}\frac{\partial{\cal L}_{\rm E}}{\partial \partial_\lambda g^{\nu\alpha}}.\label{S}
\end{eqnarray}
Klein regarded the first three relations as mathematical identities and argued that energy-momentum conservation in GR was fundamentally different from that in classical mechanics.\cite{Klein,Brading2005}
Einstein did not agree with either of these statements; he regarded only~(\ref{diffeoident}) as an identity,
which he obtained using a general coordinate invariance argument.  Now taking the divergence of~(\ref{TotalTpot}) using~(\ref{diffeoident}) gives~(\ref{divTotalT}), which---reversing the computation in the previous subsection---leads to~(\ref{divT}).  In this way Einstein showed that the local coordinate invariance identity plus his field equations---which are equivalent to~(\ref{TotalTpot})---gives the conservation of material energy momentum, without ever having to use any matter field equations.
This type of argument is referred to as \emph{automatic conservation of the source} (see MTW\cite{MTW73}, section 17.1); effectively it uses  a Noether second theorem type of argument to obtain current conservation.  Weyl used the same type of argument for the conservation of the electromagnetic current in his seminal gauge theory papers\cite{Weyl1918,Weyl1929}, whereas modern field theory books generally use Noether's first theorem in connection with current conservation.\cite{Brading2002}

The identity~(\ref{diffeoident}) is equivalent to the contracted Bianchi identity, $\nabla_\mu G^\mu{}_\nu\equiv0$, where $G^\mu{}_\nu$ is the Einstein curvature tensor.  In those days the \emph{Bianchi identity}, $\nabla_{[\mu}R^{\alpha\beta}{}_{\nu\gamma]}\equiv0$, was not generally well known (it was first used in GR by Levi-Civita in 1917\cite{RoweBianchi}).  For any Lagrangian constructed out of the metric and its derivatives, it is now well known that local diffeomorphism invariance (with $\delta g_{\mu\nu}=\pounds_\xi g_{\mu\nu}=\nabla_\mu \xi_\nu+\nabla_\nu \xi_\mu$) of the associated action leads to a divergence identity:
\begin{equation}
\int \frac{\delta{\cal L}}{\delta g_{\mu\nu}}\delta g_{\mu\nu}d^4x=0 \quad \Longrightarrow \quad \nabla_\mu\frac{\delta{\cal L}}{\delta g_{\mu\nu}}\equiv0.
\end{equation}
In general such identities can involve higher derivatives of the curvature, however for the Hilbert scalar curvature Lagrangian of GR this Noether second theorem type argument yields a divergence which coincides with the contraction of the Bianchi identity.

By the way, Einstein had been using essentially the set of energy-momentum conservation relations~(\ref{divTotalT})--(\ref{S}) for some years in connection with the (non-covariant) ``Entwurf'' equations that he worked out with Marcel Grossmann.\cite{Entwurf}  However, his Lagrangian for that scheme (see CPAE\cite{cpae,einsteinpapers}, Vol.~6, Doc.~2)  was not---up to an exact differential---diffeomorphically invariant, so that~(\ref{diffeoident}) was in that case a relation that selected a preferred set of coordinates.

Since for the Einstein Lagrangian~(\ref{diffeoident}) is an identity, one might wonder why did Einstein (and others) favor an invariance argument rather than just directly calculating it?  It turns out that the invariance argument is considerably easier. Let us look into this a little further. The detailed form of~(\ref{S}) can be written out with the help of some formulas given by Tolman\cite{Tolman}.  From his eqs (87.5) and (89.9) we have
\begin{eqnarray}
\mathfrak{s}^{\mu\lambda}{}_\nu&=&-\mathfrak{g}^{\alpha\mu}W^\lambda{}_{\nu\alpha}
+\frac12\delta^\mu_\nu\mathfrak{g}^{\alpha\beta}W^\lambda{}_{\alpha\beta}, \\
W^\lambda{}_{\alpha\beta}&:=& \frac{\partial{\cal L}_{\rm E}}{\partial \partial_\lambda \mathfrak{g}^{\alpha\beta}} =-\Gamma^\lambda{}_{\alpha\beta}+\delta^\lambda_{(\alpha}\Gamma^\sigma{}_{\beta)\sigma}.
\end{eqnarray}
To show directly that this expression satisfies~(\ref{diffeoident}) is not short or simple.  The only published calculation that we know of is in M{\o}ller,\cite{Mol58} some 40 years later.

If $\mathfrak{s}^{\mu\lambda}{}_\nu$ were antisymmetric in its upper indices the identity would be trivially satisfied.  Stated another way, if~(\ref{divTotalT}) is to be satisfied then there should exist a \emph{superpotential} $\mathfrak{U}^{\mu\lambda}{}_\nu\equiv \mathfrak{U}^{[\mu\lambda]}{}_\nu$ such that
\begin{equation}
\mathfrak{T}^\mu{}_\nu+\mathfrak{t}^\mu{}_\nu=\partial_\lambda \mathfrak{U}^{\mu\lambda}{}_\nu.
\end{equation}
Einstein's $\mathfrak{s}^{\mu\lambda}{}_\nu$ is not antisymmetric; it is not the right kind of potential.
A suitable superpotential for the Einstein pseudotensor was found over 20 years later by Freud:\cite{Freud}
\begin{equation}
\mathfrak{U}_{\rm F}^{\mu\lambda}{}_\nu:=-\mathfrak{g}^{\beta\sigma}
\Gamma^\alpha{}_{\beta\gamma}\delta^{\mu\lambda\gamma}_{\alpha\sigma\nu}.\label{UF}
\end{equation}
To obtain this  he did not follow Einstein's path.  He started with the basics, the Einstein equations, and rearranged them using some formulas from Weyl's book\cite{Weyl} and some complicated identities he found in Pauli's 1921 encyclopedia article.\cite{Pauli}  Later we will give a simple derivation of Freud's superpotential using a better technique.

\subsection{Other GR pseudotensors}
%%%%%%%%%%%%%%%%%%%%%%%%%%%%%%%%%%%%%%%%%%%%%%%%%%%%%%%%%%%%%%%%%%%%%%
The presence of a non-vanishing energy-momentum density necessarily produces gravity (i.e., the curvature of spacetime). In
curved spacetime the total source energy-momentum tensor satisfies~(\ref{divT}).
Without the second term we would have an expression suitable for integrating to obtain a conservation law. The second term represents a local interaction exchanging energy-momentum between the source and the gravitational field. To have a good conservation law we would like to rewrite~(\ref{divT}) in the form of~(\ref{divTotalT})
%\begin{equation}
%\partial_\mu[ \sqrt{-g} (T^\mu{}_\nu + t^\mu{}_\nu)] = 0
%\end{equation}
for some suitable {\em gravitational energy-momentum density} $\mathfrak{t}^\mu{}_\nu$.  In fact this can be done in an infinite number of ways, {\em and}, in all cases, the quantity $\mathfrak{t}^\mu{}_\nu$ is not a tensor.  (For some good overviews of such pseudotensors and their properties see Refs.~[\refcite{Goldberg:1958zz,JuliaSilva,Trautman62,Szabados:1991dd}].)

%\cite{Bergmann:1953jz, Gold53,  Jackiw, JuliaSilva}.
%%%jmn322 3b15

Here is a construction.  Select an object (referred to as a {\em superpotential}) with suitable symmetries: ${\mathfrak{U}^{\nu\lambda}{}_\mu}\equiv \mathfrak{U}^{[\nu\lambda]}{}_\mu$.
Now use it to split the Einstein tensor, defining a {\em gravitational energy-momentum pseudotensor} according to
\begin{equation}
2\kappa {\mathfrak{t}^\mu{}_\nu} := -2\sqrt{-g} G^\mu{}_\nu +  \partial_\lambda {\mathfrak{U}^{\mu\lambda}{}_\nu}.
\end{equation}
Then Einstein's equation, $G^\mu{}_\nu = \kappa T^\mu{}_\nu$, takes a form (analogous to Maxwell's equation) with a {\em total}
effective energy-momentum pseudotensor as its source:
\begin{equation}
\partial_\lambda {\mathfrak{U}^{\mu\lambda}{}_\nu} = %2 \kappa \sqrt{-g}{\mathfrak{T}}^\mu{}_\nu :=
2 \kappa  ({\mathfrak{t}^\mu{}_\nu} + \mathfrak{T}^\mu{}_\nu).
\end{equation}
The essential feature of such source expressions is that they are equated to a derivative of a {\it superpotential} in such a way that their divergence {\em automatically} vanishes. Thus, as a consequence of the symmetry of $\mathfrak{U}^{\mu\lambda}{}_\nu$ we have
(similar to Maxwell's theory) ``automatic conservation of the source'': $\partial_\mu ( {{\mathfrak{T}}^\mu{}_\nu}+\mathfrak{t}^\mu{}_\nu) \equiv 0$.
This expression can be integrated to define the total conserved energy-momentum within any volume $V$:
\begin{equation}
P_\nu(V) := \int_V ({{\mathfrak{T}}^\mu{}_\nu} +\mathfrak{t}^\mu{}_\nu) d^3\Sigma_\mu = \frac1{2 \kappa} \int_V \partial_\lambda
{\mathfrak{U}^{\mu\lambda}{}_\nu} d^3\Sigma_\mu \equiv \frac1{2\kappa} \oint_{\partial V} \mathfrak{U}^{\mu\lambda}{}_\nu \frac12 d^2S_{\mu\lambda}. \label{totP}
\end{equation}%%%jmn322 3b17
From the volume integral on the left one would expect that the results would be highly ambiguous---depending on the choice of
reference frame throughout the volume of interest.
However, from the last surface form, one can see that the situation is not quite so bad. The result does not depend on the choice of reference frame within the volume, it is
\emph{quasi-local}, i.e., it depends on the fields and choice of reference frame only on the boundary.  It should be noted that (for any given a reference frame on the boundary) the value of $P_\nu(V)$ is well defined by the above integral.  Its value, however, comes from a mixture of physics and a quasi-local reference frame; still it can be useful if one is mindful of its nature.

%
 %%%jmn322 3b20a
%
%%%jmn322 3b19

%
%\cite{Goldberg:1958zz,Trautman62,Szabados:1991dd,Chang:1998wj,Nester04,SoNChen}
%
There are some variations on the above formulation. The classical pseudotensorial total energy-momentum density
complexes, ${\cal T}^\mu{}_\nu:={{\mathfrak{T}}^\mu{}_\nu} +\mathfrak{t}^\mu{}_\nu$, all follow from suitable superpotentials according to
one of the patterns
\begin{equation}
2\kappa {\cal T}^\mu{}_\nu=\partial_\lambda
\mathfrak{U}^{\mu\lambda}{}_\nu,\quad
2\kappa {\cal T}^{\mu\nu}=\partial_\lambda \mathfrak{U}^{\mu\lambda\nu},%
\quad 2\kappa {\cal
T}^{\mu\nu}=\partial_{\alpha\beta}\mathfrak{H}^{\alpha\mu\beta\nu},
\end{equation}
where the superpotentials have certain symmetries which
automatically guarantee conservation: specifically
$\mathfrak{U}^{\mu\lambda}{}_\nu\equiv \mathfrak{U}^{[\mu\lambda]}{}_\nu$,
$\mathfrak{U}^{\mu\lambda\nu}\equiv
\mathfrak{U}^{[\mu\lambda]\nu}$, while $\mathfrak{H}^{\alpha\mu\beta\nu}$ has the
algebraic symmetries of the Riemann tensor (this latter form yields a \emph{symmetric} pseudotensor and, hence, a simpler conservation of angular momentum description, see MTW\cite{MTW73}, \S20.3). We have already considered the Einstein total
energy-momentum density which follows from the Freud superpotential~(\ref{UF}).
For completeness we list the other well-known ones.
The Bergmann-Thompson\cite{Bergmann:1953jz}, Landau-Lifshitz\cite{LL62}, Papapetrou\cite{Papapetrou:1948jw}, Weinberg\cite{Wein72} (also used in MTW\cite{MTW73}) and
M{\o}ller\cite{Mol58} total energy-momentum complex
expressions can be obtained respectively from
\begin{eqnarray}
\mathfrak{U}_{\rm BT}^{\mu\lambda\nu}&:=&g^{\nu\delta}U_{\rm
F}^{\mu\lambda}{}_\delta,\label{UBT}\\
\mathfrak{U}_{\rm LL}^{\mu\lambda\nu}&:=&|g|^{\frac12}U_{\rm
BT}^{\mu\lambda\nu},\quad {\mathrm{equivalently}} \quad \mathfrak{H}_{\rm
LL}^{\alpha\mu\beta\nu}:=|g|\delta^{\mu\alpha}_{ma}
g{}^{a\beta}g^{m\nu},\label{ULL}\label{HLL}\\
\mathfrak{H}_{\rm
P}^{\alpha\mu\beta\nu}&:=&\delta^{\mu\alpha}_{ma}\delta^{\nu\beta}_{nb}\bar
g{}^{ab}(|g|^{\frac12}g^{mn}),\label{HP}\\
\mathfrak{H}_{\rm
W}^{\alpha\mu\beta\nu}&:=&\delta^{\mu\alpha}_{ma}\delta^{\nu\beta}_{nb}|\bar
g|^{\frac12}\bar g{}^{ab}(-{\bar g}^{mc}{\bar g}^{nd}+{\frac12}{\bar
g}^{mn}{\bar g}^{cd}) g_{cd},\label{HW}\\
\mathfrak{U}_{\rm
M}^{\mu\lambda}{}_\nu&:=&-|g|^{\frac12}g^{\beta\sigma}\Gamma^\alpha{}_{\beta\nu}\delta^{\mu\lambda}_{\alpha\sigma}
\equiv |g|^{\frac12}g^{\beta\mu}g^{\lambda\delta}(\partial_\beta
g_{\delta\nu}-\partial_\delta g_{\beta\nu}) \label{UM}.
\end{eqnarray}
Here $\bar g^{ab}$ is the Minkowski metric, all indices in these expressions refer to spacetime and range from 0 to 3,
otherwise our conventions follow MTW.\cite{MTW73}

People have often looked askance at such pseudotensors, e.g.,  the above quote from MTW and Schr\"odinger refers to them as ``sham''.\cite{Scroedinger1950}
As we noted, there are no doubt two unsatisfactory aspects: (i) which of the many possible expressions should one use? (ii) and which \emph{quasi-local} (in view of~(\ref{totP})) reference frame should be used.  On the other hand one should also be mindful that (a) they do provide a description of energy-momentum conservation, (b) they (like connection coefficients) really are geometric objects, with well defined values in each reference frame (this issue has been rigourously addressed using fiber bundle formulations\cite{Frau89,Szabados:1991dd,Pitts:2009km}).

All of these pseudotensors (except for M{\o}ller's) give the expected total energy-momentum values at spatial infinity.  On the other hand, \emph{none} of them give the desired positivity of energy for small vacuum regions to lowest non-vanishing order\cite{SoNChen}, however a set of new pseudotensors depending on several parameters with this desirable property has been constructed.\cite{SoN09a} How can one understand the physical significance of these various pseudotensors?  We have found a way using the Hamiltonian approach.

\subsection{Pseudotensors and the Hamiltonian}
%%%%%%%%%%%%%%%%%%%%%%%%%%%%%%%%%%%%%%%%%%%%%%%%%%%%%%%%%%%%%%%%%%%%%%
To see how one can be led to the Hamiltonian one need merely redo the calculation of~(\ref{totP}) as an identity (``off shell'').  For some fixed reference frame, with a (constant in the present reference frame) vector field $Z^\mu$
inserted we find\footnote{The sign in this expression is dictated by the condition for positive energy determined by the Hamiltonian using our local Minkowski signature convention: $P_\mu=(-E/c,\vec p)$.}
\begin{eqnarray}%%%jmn322 2b20
-Z^\mu P_\mu(V) &:=& - \int_V Z^\mu {{\mathfrak{T}}^\nu{}_\mu} \sqrt{-g}d^3\Sigma_\nu
\nonumber\\
&\equiv& \int_V \left[ Z^\mu \sqrt{-g} \left( \frac1{\kappa} G^\nu{}_\mu - T^\nu{}_\mu \right) - \frac1{2\kappa} \partial_\lambda \left( Z^\mu {\mathfrak{U}^{\nu\lambda}{}_\mu} \right) \right] d^3\Sigma_\nu
\nonumber\\
&\equiv& \int_V Z^\mu {\cal H}^{\rm GR}_\mu + \oint_{S=\partial V} {\cal B}^{\rm GR}(Z) \equiv H(Z, V). \label{basicHam}
\end{eqnarray}%%%jmn322 3b21
Here ${\cal H}^{\rm GR}_\mu$ can be recognized as the covariant expression which, when expressed in terms of the appropriate canonical variables, is just the ADM Hamiltonian density (i.e.,~the superhamiltonian and
supermomentum), see, e.g., Refs~[\refcite{adm, IN80}] and MTW\cite{MTW73} Chapter 21. The expression includes a {\em Hamiltonian boundary term}, a 2-surface integral of ${\cal B}^{\rm GR}(Z) = - Z^\mu (1/2\kappa) {\mathfrak{U}^{\nu\lambda}{}_\mu} (1/2) d^2S_{\nu\lambda}$, i.e., it is entirely determined by the superpotential.
%%%jmn322 3b21
The value of the
Hamiltonian on a solution is entirely determined by this boundary term; the initial value constraints ensure that the Hamiltonian density in the spatial volume integral vanishes ``on shell''
(i.e., when the field equations are satisfied).
In a similar way the value given by any pseudotensor can be regarded as the value of the Hamiltonian with a certain boundary term.\cite{Chang:1998wj}
From the Hamiltonian variation, as we will discuss below, one gets important information that tames the ambiguity in the boundary term---namely boundary conditions---and thereby determines the physical significance of the various quasi-local values.  The energy-momentum values obtained for the various pseudotensors can all be regarded as values of the Hamiltonian with different boundary conditions.

%%%%%%%%%%%%%%%%%%%%%%%%%%%%%%%%%%%%%%%%%%%%%%%%%%%%%%%%%%%%%%%%%%%%%%
\section{The quasi-local view}
%%%%%%%%%%%%%%%%%%%%%%%%%%%%%%%%%%%%%%%%%%%%%%%%%%%%%%%%%%%%%%%%%%%%%%
The modern idea, due to Penrose\cite{Penrose82} in 1982, is that energy-momentum  is \emph{quasi-local}: i.e., it is associated with a closed 2-surface (while the pseudotensor energy-momentum complexes always had this property, its essential importance became much more appreciated after this work of Penrose which introduced this convenient term).
There is a comprehensive review
% (with 582 references!)
of this topic:
%%%jmn322 3b23
Szabados (2009).\cite{Sza09}
%
%L.B. Szabados, Quasi-Local Energy-Momentum and Angular Momentum in General Relativity,\textit{http://www.livingreviews.org/lrr-2009-4}.
The many recent works cited in this review show that this is still a topic of considerable interest.
%%%jmn419 3b23

In  a brief summary one can find the statement:

\begin{quotation}
``... contrary to the high expectations of the 1980s, finding an appropriate quasi-local
notion of energy-momentum has proven to be surprisingly difficult. Nowadays, the state of the art is
typically postmodern: although there are several promising and useful suggestions, we not only have
no ultimate, generally accepted expression for the energy-momentum and especially for the angular
momentum, but there is not even a consensus in the relativity community on general questions ...
or on the list of the criteria of reasonableness of such expressions.''
\end{quotation}

However if one takes a more specific approach, one can come to a more satisfactory conclusion.  In particular the Hamiltonian view quite changes the prospects, especially when used along with a gauge perspective.

\section{Currents as generators}%%%jmn322 3b25
%%%%%%%%%%%%%%%%%%%%%%%%%%%%%%%%%%%%%%%%%%%%%%%%%%%%%%%%%%%%%%%%%%%%%%
Noether's work was entirely Lagrangian based.
Her results can be taken a further step when they are combined with the Hamiltonian formulation.  As we will see, the Hamiltonian formulation offers a handle on the Noether current ambiguity.

One key feature can be seen already in Hamiltonian mechanics.   A quantity $Q$ conserved under the time evolution generated by a Hamiltonian $H=H(q,p)$ is more than a just a conserved quantity, it is also \emph{the canonical generator} of a one parameter transformation on phase space $(q(\lambda),p(\lambda))$ which is a symmetry of the Hamiltonian.
\begin{equation}
0 = \frac{dQ}{dt} = [Q, H] \quad \Longrightarrow \quad \frac{dH}{d\lambda} = [H, Q] = 0.
\end{equation}

In Hamiltonian field theory, the conserved currents are the generators of the associated symmetry.
In particular, the generator of a local spacetime ``translation'' (an infinitesimal diffeomorphism) is the Hamiltonian; energy-momentum is the associated conserved quantity.  Conversely, for spacetime translations, the associated Noether conserved current expression (i.e., the energy-momentum density) \emph{is} the Hamiltonian density---the canonical generator of spacetime displacements.  As we will see, because it can be varied this translation generator gives a handle on the associate conserved current ambiguity.  The Lagrangian formulation affords no such handle, because in terms of Lagrangian variables the translation current is not a generator that can be varied.

%%%%%%%%%%%%%%%%%%%%%%%%%%%%%%%%%%%%%%%%%%%%%%%%%%%%%%%%%%%%%%%%%%%%%%
\section{Gauge and geometry}
%%%%%%%%%%%%%%%%%%%%%%%%%%%%%%%%%%%%%%%%%%%%%%%%%%%%%%%%%%%%%%%%%%%%%%
For the early history of gauge theory see O'Raifeartaigh\cite{DawningGauge}.
Briefly, the milestone works are Hermann Weyl's treatments of electromagnetism: Weyl (1918)\cite{Weyl1918}, Weyl (1929)\cite{Weyl1929}, then the generalization to non-Abelian groups by Yang \& Mills (1954)\cite{YangMills} and Utiyama (1956, 1959)\cite{Utiyama, Utiyama59}.
Explicitly treating gravity as a gauge theory was pioneered by Utiyama\cite{Utiyama, Utiyama59}, using the Lorentz group and Riemannian geometry. Sciama\cite{Sciama61} also used  the Lorentz group but with Riemann-Cartan geometry (i.e., non-vanishing torsion). Kibble\cite{Kibble} put things in their proper place, he gauged the Poincar\'e group (i.e., the inhomogeneous Lorentz group, including translations).

For accounts of gravity as a spacetime symmetry gauge theory, see Hehl and coworkers\cite{HHKN,Hehl1980,MAG,Gronwald:1995em,Hehl:2012pi}, Mielke\cite{Mielke} and Blagojevi\'c\cite{Blagojevic2002}.  A comprehensive reader with  summaries, discussions, and many reprints has recently appeared: Blagojevi\'c \& Hehl\cite{BlagojevicHehl}.  For the observational constraints on torsion see Ni (2010).\cite{Ni10}

To us it is rather surprising that the idea of regarding gravity as a gauge theory is not better known.   Examined more closely, one finds that gravity played an important role in the argument used in both of the above mentioned seminal works of Weyl, and  thus in all of the above---except for the Yang-Mills paper. Furthermore, later in 1974 Yang himself published a paper\cite{Yang:1974kj} where he proposed a certain treatment of gravity as a gauge
theory.\footnote{The aforementioned reader includes a chapter with a critical discussion of Yang's gauge theory of gravity. Recently Yang was asked about his 1974 paper, he said: ``I do not believe that paper is correct.''\cite{YangNTU}}

According to our understanding, properly speaking, GR can be understood as the original gauge theory.  After all, it was the first physical theory where local gauge freedom (in the guise of general coordinate invariance) played a key role.\footnote{It is true that the electrodynamics potentials along with their gauge freedom were known long before GR (in fact a Lagrangian which is locally gauge invariant had already been presented\cite{Schw1903}), but this gauge invariance was not seen as having any important role in connection with the nature of the interaction, the conservation of current,  or a differential identity---until the seminal work of Weyl, which post-dated (and was inspired by) GR.}

The conserved quantities, energy-momentum and angular momentum/center-of-mass momentum are associated with with the geometric symmetry of Minkowski spacetime, the spacetime translations and Lorentz rotations, i.e., the Poincar\'e group.  Furthermore this group is used to classify physical particles according to mass and spin.  So a local Poincar\'e gauge theory is quite appropriate both geometrically and physically.

To give a good account, one should also be mindful of the parallel development of the closely related concept of a  connection in differential geometry.
Here we just briefly mention that the main ideas were due to Hessenberg, Levi-Civita, Schouten, Weyl, Cartan, Ehresmann, and Koszul; for discussions of connections see Nomizu\cite{Nomizu}, Kobayashi \& Nomizu\cite{KobNom} and Spivak\cite{Spivak}.

As we will see in more detail, Riemann-Cartan (with a metric and a metric compatible connection, having both curvature and torsion) is the most appropriate geometry for a dynamic spacetime geometry theory: its local symmetries are just those of the local Poincar\'e group.  So in this presentation we will be considering the Pincar\'e gauge theories of gravity (PG); GR is included in this class as a special case.

%%%%%%%%%%%%%%%%%%%%%%%%%%%%%%%%%%%%%%%%%%%%%%%%%%%%%%%%%%%%%%%%%%%%%%
\section{Dynamical spacetime geometry and the Hamiltonian}
%%%%%%%%%%%%%%%%%%%%%%%%%%%%%%%%%%%%%%%%%%%%%%%%%%%%%%%%%%%%%%%%%%%%%%
We will consider geometric gravity theories with both a metric and an a priori metric compatible connection.  Both curvature and torsion are allowed. The variational principles are developed. The Noether symmetries and the associated conserved quantities and differential identities are discussed. From a first order Lagrangian formalism using differential forms, we construct a spacetime covariant Hamiltonian formalism. The Hamiltonian boundary term gives appropriate expressions for the quasi-local quantities, energy-momentum, angular momentum and center-of-mass momentum, as well as quasi-local energy flux. The formalism easily specializes to teleparallel theory and Einstein's GR.

The Hamiltonian approach reveals certain aspects of a theory, including the constraints, gauges, and degrees-of-freedom, as well as expressions for energy-momentum and angular momentum. However the usual ADM approach achieves this at a heavy cost: the loss of manifest 4D-covariance. Our alternative approach is complementary: a major benefit is manifestly 4D-covariant expressions for the quasi-local quantities: energy-momentum and angular momentum/center-of-mass momentum.

\subsection{The main ideas}
%%%%%%%%%%%%%%%%%%%%%%%%%%%%%%%%%%%%%%%%%%%%%%%%%%%%%%%%%%%%%%%%%%%%%%
The Hamiltonian for physical systems and dynamic spacetime geometry generates the evolution of a spatial region along a vector field. It includes a boundary term which determines the boundary conditions and supplies the value of the Hamiltonian. The Hamiltonian value gives the quasi-local quantities: energy-momentum and angular-momentum/center-of-mass momentum.
A spacetime gauge theory perspective identifies suitable geometric variables. We found a certain preferred Hamiltonian boundary term. The Hamiltonian boundary term depends not only on the dynamical variables but also on their reference values; they determine the ground state---the state with vanishing quasi-local quantities.
To determine the ``best matched'' reference metric and connection values for our preferred boundary term
%for Einstein's GR,
 we propose on the boundary 2-surface: (i) 4D isometric matching, and (ii) extremizing the energy.

\subsection{Some comments}
%%%%%%%%%%%%%%%%%%%%%%%%%%%%%%%%%%%%%%%%%%%%%%%%%%%%%%%%%%%%%%%%%%%%%%
Before we begin our technical discussion of our work, let us make a few general comments. We work in 4D spacetime, but most of this can be extended to other dimensions in a straightforward fashion (except for the reference construction). The class of dynamical Lagrangians we will consider does not allow for any derivatives of curvature or torsion.
%\footnote{There is no obstruction to extending the general scheme to include such terms.}
Our concerns are entirely classical.

We focus here on Riemann-Cartan spacetime geometry (i.e., spacetimes with a metric and a metric compatible connection, having both curvature and torsion) and the PG; our general analysis can be specialized both to Riemannian geometry (vanishing torsion) and teleparallel geometry (vanishing curvature);
it includes GR and the teleparallel equivalent of GR as two special cases.
Here we assume a metric compatible connection; elsewhere we will present the generalization which includes non-metricity.
The extension to non-metricity and the special case of teleparallel geometry each offer further insight into gravitational energy; we believe those insights are best appreciated when compared to the results presented here for the Riemann-Cartan geometry with the PG.

%
%

%%%%%%%%%%%%%%%%%%%%%%%%%%%%%%%%%%%%%%%%%%%%%%%%%%%%%%%%%%%%%%%%%%%%%%
\section{Differential forms}%%%jmn Done except to maybe add some references.
%%%%%%%%%%%%%%%%%%%%%%%%%%%%%%%%%%%%%%%%%%%%%%%%%%%%%%%%%%%%%%%%%%%%%%
In this work we mainly use differential forms\cite{West,Frankel,HO}. The reader may wonder why we use this less widely familiar idiom. The simple brief explanation is that they have some qualities that are technically very convenient for our needs.
%Why a \emph{form\/}alism?
%Why do we choose to use forms?
Differential forms are multiplied using the (graded Grassmann) wedge product. They can be differentiated using $d$, the \emph{exterior differential}, a  graded derivation which enjoys the property $d^2 \equiv 0$, so a differential equation $d\alpha = \beta$ has the integrability condition $0 = d\beta$, furthermore $d\beta = 0 \Longrightarrow \beta = d\alpha$, at least locally. The integrals of forms satisfy the general boundary theorem\footnote{This generalization of the fundamental theorem of calculus is often referred to as the generalized Stokes theorem. Special cases include the Ostrogradsky-Gauss and Stokes theorem of vector analysis.}
\begin{equation}
\int_U d\beta \equiv \oint_{\partial U} \beta. \label{boundarythm}
\end{equation}
Also, they are well suited to representing interacting physical fields, especially gauge fields, and, as we shall see, they give a succinct representation of the main geometric objects: connection, curvature, coframe, torsion, etc. Moreover, as will be explained below, they are quite convenient for the essential 3+1 spacetime decomposition of derivatives that is needed for a dynamical Hamiltonian formulation.\cite{Nes84}

Regarding notation, here the contraction (or \emph{interior} product) with a vector field is denoted by $i_X \alpha(., ., \dots) := \alpha(X, ., \dots) $ (some authors use the notation of left contraction: $X \rfloor \alpha$).
The \emph{Lie derivative} on forms is given by $\pounds_X \equiv i_X d + d i_X$, it has the nice property $d \pounds_X \equiv \pounds_X d$.
We are concerned here with the case of 4-dimensional spacetime, which has a local Minkowski structure, having a metric with Lorentz signature. The metric determines the unit volume 4-form $\eta$ with components $\eta_{\mu \nu \alpha \beta} = \eta_{[\mu \nu \alpha \beta]}$, $\eta_{0123} = \sqrt{|g|}$ which is used to construct the Hodge dual that maps $k$-forms to $(4-k)$-forms.

From the coframe $\vartheta^\alpha$ one can construct a basis for $k$-forms $\vartheta^{\alpha \beta \dots} := \vartheta^\alpha \wedge \vartheta^\beta \wedge \dots$ and a useful dual basis $\eta^{\alpha \beta \dots} := *\vartheta^{\alpha \beta \dots}$. They are related by various identities, especially
\begin{eqnarray}
\vartheta^\rho \wedge \eta_{\mu\nu\lambda} &\equiv& \delta^\rho_\lambda \eta_{\mu\nu} + \delta^\rho_\mu \eta_{\nu\lambda} + \delta^\rho_\nu \eta_{\lambda\mu},
\\
\vartheta^{\alpha\beta} \wedge \eta_{\mu\nu\lambda} &\equiv& \delta^{\alpha\beta}_{\mu\nu} \eta_\lambda +
\delta^{\alpha\beta}_{\nu\lambda} \eta_\mu + \delta^{\alpha\beta}_{\lambda\mu} \eta_\nu,
\\
\vartheta^\rho \wedge \eta_{\mu\nu} &\equiv& \delta^\rho_\nu \eta_\mu - \delta^\rho_\mu \eta_\nu.
\end{eqnarray}

Maxwell's electrodynamics is a good example of the utility of differential forms.
Charge identifies the charge-current $3$-form (density):
\begin{equation}
Q(V) = \int_V J.
\end{equation}
Charge is conserved:
\begin{equation}
dJ = 0 \quad \Longrightarrow \quad J = dH. \label{maxJ}
\end{equation}

The electromagnetic field is represented by a 2-form $F$.  An elementary way to see why this is appropriate is to examine the motion of a point test charge.
One should begin with kinematics in Minkowski space.  Consider the motion of a point particle as a function of proper time: $x^\mu=x^\mu(\tau)$.  The 4-velocity $v^\mu:=dx^\mu/d\tau$ has constant magnitude: $v^\mu v_\mu=-c^2$ so the 4-acceleration is Lorentz orthogonal to the 4-velocity.  Hence the 4-force must be orthogonal to the 4-velocity. Consequently the 4-force must depend on the velocity. The simplest case is for a 4-force linear in the 4-velocity.  Thus the simplest dynamical law has the form
\begin{equation}%%%jmn419 3b36
\frac{dp_\mu}{d\tau} = q F_{\mu\nu} v^\nu,
\end{equation}
where $p_\mu:=mv_\mu$ is the 4-momentum, $q$ is a coupling constant and $F_{\mu\nu}$ is some tensor field which is \emph{antisymmetric}, i.e, it is a 2-form.
The Lorentz force law of electrodynamics has this form.
The force law identifies a certain field strength $2$-form $F$ which includes the electric and magnetic fields.
Conservation of magnetic flux through a closed 2 surface $S = \partial V$ gives
\begin{equation}
0 = \oint_{S} F = \int_V dF, \quad \Longrightarrow \quad dF = 0 \quad \Longleftrightarrow \quad F = dA, \label{maxF}
\end{equation}
and there are local gauge transformations: $A \to A + d\chi$. The vacuum constitutive relation is $H = *F/Z_0$ ($Z_0$ is the vacuum impedance). This covariant formulation for Maxwell's electrodynamics is valid for all dynamic geometry gravity theories and does not depend upon using a particular set of units, for a detailed, comprehensive and instructive presentation see Hehl \& Obukhov\cite{HO}.

%%%%%%%%%%%%%%%%%%%%%%%%%%%%%%%%%%%%%%%%%%%%%%%%%%%%%%%%%%%%%%%%%%%%%%
\section{Variational principle for form fields}%%%jmn Done except perhaps for some references.
%%%%%%%%%%%%%%%%%%%%%%%%%%%%%%%%%%%%%%%%%%%%%%%%%%%%%%%%%%%%%%%%%%%%%%
Why do we use variational principles?\cite{Lanczos} The answer is pragmatics: \emph{because they work}. With appropriate symmetries they give consistent interacting field equations along with conserved Noether currents for all the desired quantities.
%
%%%jmn322 3c27
As far as we know, all the known good dynamical evolution equations for the fundamental interacting classical field theories have a variational formulation.

In the usual formulations most dynamical fields satisfy second order equations. We refer to such formulations as {\it second order}.

Let $\varphi^A$ be some kind of vector field. The label ``$A$'' stands for some collection of indices, e.g., spinor, spacetime, isospin. Allow $\varphi^A$ to also be a differential form of rank $f$ where $f = 0, 1, 2$, or $3$, e.g., $\varphi^A = \frac12 \varphi^A_{\mu\nu} \, \vartheta^\mu \wedge \vartheta^\nu = \frac12 \varphi^A_{ij} dx^i \wedge dx^j$ for $f = 2$.

The {\it Lagrangian density} is a $4$-form:
\begin{equation}
{\cal L} = {\cal L}(\varphi^A, d \varphi^A).
\end{equation}
Note that there is no explicit appearance of the coordinates $x^i$ or the coordinate partials $\partial_i$; $d \varphi^A$ is an $(f+1)$-form which geometrically includes partial derivatives of the components of $\varphi^A$, but only in an antisymmetric combination. (Here we explicitly consider just one $f$-form field. The generalization to include several fields of different grades, is straightforward.)

Our convention is to vary fields off to the left (other conventions would differ only by some signs).
The variation of ${\cal L}$ is thus
\begin{equation}
\delta {\cal L} = \delta d \varphi^A \wedge \frac{\partial {\cal L}}{\partial d \varphi^A} + \delta \varphi^A \wedge \frac{\partial {\cal L}}{\partial \varphi^A}.
\end{equation}
This implicitly defines ${\partial {\cal L}/\partial d\varphi^A}$ as a $(3-f)$-form and $\partial{\cal L}/\partial\varphi^A$ as a $(4-f)$-form. Next, interchange the order (i.e., $\delta \, d = d \, \delta$) to get
\begin{eqnarray}
\delta {\cal L} &=& d \delta \varphi^A \wedge \frac{\partial {\cal L}}{\partial d \varphi^A} + \delta \varphi^A \wedge \frac{\partial {\cal L}}{\partial \varphi^A}
\nonumber\\
&\equiv& d \left( \delta \varphi^A \wedge \frac{\partial {\cal L}}{\partial d \varphi^A} \right) - (-1)^f \delta \varphi^A \wedge d \left( \frac{\partial {\cal L}}{\partial d \varphi^A} \right) + \delta \varphi^A \wedge \frac{\partial {\cal L}}{\partial \varphi^A}.
\end{eqnarray}
(Upon integration over some spacetime region this last step is ``integration by parts'', with the total differential term becoming a boundary term.) From the above it follows that the basic variational relation
\begin{equation} \label{3:thebasicidentity}
\delta {\cal L} = d (\delta \varphi^A \wedge p_A) + \delta \varphi^A \wedge \frac{\delta {\cal L}}{\delta \varphi^A}
\end{equation}
can be regarded as implicitly defining the {\it conjugate field momentum} and the Euler-Lagrange {\it variational derivative}, which have the respective explicit definitions
\begin{equation}
p_A := \frac{\partial {\cal L}}{\partial d \varphi^A},
\end{equation}
\begin{equation}
\frac{\delta {\cal L}}{\delta \varphi^A} := \frac{\partial {\cal L}}{\partial \varphi^A} - \varsigma \, d \left( \frac{\partial {\cal L}}{\partial d \varphi^A} \right).
\end{equation}
A small price for using form fields is the appearance of occasional sign factors like $\varsigma := (-1)^f$.

\subsection{Hamilton's principle}%%%jmn ok
%%%%%%%%%%%%%%%%%%%%%%%%%%%%%%%%%%%%%%%%%%%%%%%%%%%%%%%%%%%%%%%%%%%%%%
Our first application of~(\ref{3:thebasicidentity}) is Hamilton's principle ({\em the principle of least action}). Let the action within a region $U$ be given by $S := \int_U {\cal L}$. Then
\begin{equation}
\delta S \equiv \int_U \delta {\cal L} \equiv \int_U d(\delta \varphi^A \wedge p_A) + \delta \varphi^A \wedge \frac{\delta {\cal L}} {\delta \varphi^A} \equiv \int_U \delta \varphi^A \wedge \frac{\delta {\cal L}} {\delta\varphi^A} + \oint_{\partial U} \delta\varphi^A \wedge p_A.
\end{equation}
Now require the action $S$ to be extreme (i.e., $\delta S = 0$) for $\delta \varphi^A$ vanishing on the boundary of $U$. This yields the field equation ${\delta {\cal L} / \delta \varphi^A} = 0$.

\subsection{Compact representation}%%%jmn ok
%%%%%%%%%%%%%%%%%%%%%%%%%%%%%%%%%%%%%%%%%%%%%%%%%%%%%%%%%%%%%%%%%%%%%%
For a compact general discussion it is convenient to suppress the field component index.\cite{Nester08} (This could be represented in matrix notation;
our basic fields $\varphi$ and their differential could be regarded as row vectors.) The Lagrangian then has the form ${\cal L} = {\cal L}(\varphi, d \varphi)$ and the variational scheme proceeds as
\begin{eqnarray}
\delta {\cal L} &=& d \delta \varphi \wedge \frac{\partial {\cal L}}{\partial d \varphi} + \delta \varphi \wedge \frac{\partial {\cal L}}{\partial d \varphi}
\nonumber\\
&=& d \left( \delta \varphi \wedge \frac{\partial {\cal L}}{\partial d  \varphi} \right) + \delta \varphi \wedge \left[ - \varsigma \, d \left( \frac{\partial {\cal L}}{\partial d \varphi} \right) + \frac{\partial {\cal L}}{\partial \varphi} \right],
\end{eqnarray}
hence the key Lagrangian variational identity takes the form
\begin{equation} \label{3:thebasicidentity0}
\delta {\cal L} \equiv d (\delta \varphi \wedge p) + \delta \varphi \wedge \frac{\delta {\cal L}}{\delta \varphi}.
\end{equation}
In this succinct alternative to~(\ref{3:thebasicidentity}) the conjugate momentum and the variational derivative can be regarded as form-valued column vector fields.

%%%%%%%%%%%%%%%%%%%%%%%%%%%%%%%%%%%%%%%%%%%%%%%%%%%%%%%%%%%%%%%%%%%%%%
\section{Some simple examples of the Noether theorems}%%%jmn ok
%%%%%%%%%%%%%%%%%%%%%%%%%%%%%%%%%%%%%%%%%%%%%%%%%%%%%%%%%%%%%%%%%%%%%%
Here we present simple examples of Noether's two theorems.\cite{Noether} Later we shall use the same types of arguments in more complicated situations.

\subsection{Noether's first theorem: energy-momentum}\label{S:N1 energy-momentum} %%%jmn ok
%%%%%%%%%%%%%%%%%%%%%%%%%%%%%%%%%%%%%%%%%%%%%%%%%%%%%%%%%%%%%%%%%%%%%%
Further applications of the basic variational identity~(\ref{3:thebasicidentity}) or~(\ref{3:thebasicidentity0})
yield the Noether theorems. Their applications to physical systems are discussed in many works, e.g., Konopleva and Popov\cite{Konopleva-Popov}.
Here we introduce our particular use of them using two specific important cases.

Noether's first theorem states that for a constant parameter symmetry there is a conserved current.

For our concerns the most important example is the conservation of energy-momentum. As our specific relevant simple case exemplifying the argument, we specialize in this subsection to Minkowski spacetime, which is homogeneous and thus naturally has a geometric symmetry under translations. Dynamically let us  assume symmetry also of the action under constant translations. The symmetry depends on a continuous parameter; it is sufficient to consider the infinitesimal case. Geometrically an infinitesimal translation corresponds to a constant vector field $Z$. Under such a transformation the change in the components of form fields is given by the Lie derivative. We have:
\begin{eqnarray}
\Delta \varphi &=& - \pounds_Z \varphi = - (d \, i_Z + i_Z \, d) \varphi,
\\
\Delta {\cal L} &=& - \pounds_Z {\cal L} = - d \, i_Z {\cal L}.
\label{3:dizl}
\end{eqnarray}
Equation~(\ref{3:thebasicidentity0}) under these specific variations (i.e., replacing the general $\delta$ by these specific changes) should be an identity (since the Lagrangian ${\cal L}$ depends on the position only through the fields $\varphi$). Rearranging leads to
\begin{equation} \label{3:noetherem}
\Delta {\cal L} - d (\Delta \varphi \wedge p) \equiv \Delta \varphi \wedge \frac{\delta {\cal L}}{\delta \varphi}.
\end{equation}
%.

Because of~(\ref{3:dizl}) the l.h.s. of eq.~(\ref{3:noetherem}) is a total differential---a total differential which moreover vanishes if the Euler-Lagrange field equations are imposed:
\begin{equation}
d ( - i_Z {\cal L} + \pounds_Z  \varphi \wedge p) \equiv - \pounds_Z \varphi \wedge \frac{\delta {\cal L}}{\delta  \varphi}.
\end{equation}
This identifies a \emph{conserved current density} (a 3-form with vanishing differential \emph{on shell}, i.e., when the field equations are satisfied) called the generalized {\it canonical stress energy-momentum density} ($3$-form):
\begin{equation} \label{3:canem}
%%%jmn322 3c40 I keep {\cal T} to distinguish this from the torsion, we do not use it much. We use {\mathfrak{T}} for a pseudotensor.
{\cal{T}}(Z) := i_Z{\cal L}-\pounds_Z \varphi \wedge p.
\end{equation}
For a zero-form field it takes the special shape ${{T}}^\alpha{}_\mu Z^\mu \, \eta_\alpha$ where (with $L=*{\cal L}$)
\begin{equation}%%%%jmn322 for a zero form we use the very standard energy-momentum index notation
{{T}}^\alpha{}_\mu = \delta^\alpha_\mu L- \partial_\mu \varphi \frac{\partial L}{\partial \partial_\alpha  \varphi},
\end{equation}
which is the well known expression mentioned earlier~(\ref{canEMT}).

\subsection{Noether's second theorem: gauge fields}%%%jmn ok
%%%%%%%%%%%%%%%%%%%%%%%%%%%%%%%%%%%%%%%%%%%%%%%%%%%%%%%%%%%%%%%%%%%%%%

%
\emph{Note: For the rest of the present section there is no need to restrict the spacetime geometry in any way.  Our considerations apply quite generally.}

Now we want to consider invariance under {\it local} gauge transformations:
\begin{equation} \label{3:fieldlocalgauge}
\Delta \varphi = \alpha^p \, \varphi \, T_p,
\end{equation}
where the $\alpha^p$ are position dependent parameters and the $T_p$ are the matrix generators (in our representation the fields are on the left and the matrix on the right) of the gauge group. Replace $d \varphi$ by the {\it gauge covariant differential}:
\begin{equation}
D \varphi := d \varphi + A^p \wedge \varphi \, T_p,
\end{equation}
containing a certain compensating field, the {\em gauge vector potential} (a.k.a. the {\em gauge connection one-form}): $A^p = A^p{}_j dx^j$, which has a special non-homogeneous gauge transformation:
\begin{equation} \label{3:conngauge}
\Delta A^p = - D \alpha^p := -( d \alpha^p + A^q C^p{}_{qr} \, \alpha^r),
\end{equation}
where $C^p{}_{qr}$ are the gauge group structure constants: $[T_q, T_r] = C^p{}_{qr} T_p$. Then
\begin{eqnarray}
\Delta (D \varphi) &:=& (\Delta{D}) \varphi + D \Delta\varphi
\nonumber\\
&=& - (D \alpha^p) \wedge \varphi T_p + D(\alpha^p \varphi T_p)
\nonumber\\
&=& \alpha^p (D \varphi) T_p.
\end{eqnarray}
Thus $D \varphi$ transforms just like $\varphi$.

Rather than starting with a Lagrangian $4$-form of the type ${\cal L} = {\cal L}( \varphi, d \varphi, A^p, d A^p)$ and then discovering that the variables $A^p, d \varphi, dA^p$ can only appear in nice covariant combinations, let us proceed more covariantly, beginning with the {\it Lagrangian $4$-form}
\begin{equation} \label{3:Lgauge}
{\cal L} = {\cal L}(\varphi, D \varphi, A^p, F^p),
\end{equation}
where $F^p$ is the {\it field strength} or {\it gauge curvature} $2$-form:
\begin{equation}
F^p := d A^p + \frac12 C^p{}_{qr} \, A^q \wedge A^r. \label{F2}
\end{equation}
Since $A^p$ is still allowed to appear independently in~(\ref{3:Lgauge}), there is no loss of generality.
The variation of ${\cal L}$~(\ref{3:Lgauge}) is
\begin{equation} \label{3:varLgauge}
\delta {\cal L} = d \left(\delta \varphi \wedge \frac{\partial {\cal L}}{\partial D \varphi} + \delta A^p \wedge \frac{\partial {\cal L}}{\partial F^p} \right) + \delta \varphi \wedge \frac{\delta {\cal L}}{\delta \varphi} + \delta A^p \wedge \frac{\delta {\cal L}}{\delta A^p}.
\end{equation}

Now we are set for an example of Noether's 2nd theorem---that for each local invariance there is a differential identity.

Assume that ${\cal L}$~(\ref{3:Lgauge}) is invariant under the special changes $\Delta \varphi, \Delta A$ of eqs~(\ref{3:fieldlocalgauge}, \ref{3:conngauge}). From~(\ref{3:varLgauge}) we then have the identity
\begin{equation} \label{3:N2ident}
0 \equiv d \left( \alpha^p \varphi T_p \wedge \frac{\partial {\cal L}}{\partial D \varphi} - D \alpha^p \wedge \frac{\partial {\cal L}}{\partial F^p} \right) + \alpha^p \varphi \wedge T_p \frac{\delta {\cal L}}{\delta \varphi} - D \alpha^p \wedge \frac{\delta {\cal L}}{\delta A^p}.
\end{equation}
The second term in the parenthesis may be rewritten as $ -d (\alpha^p \frac{\partial {\cal L}}{\partial F^p}) + \alpha^p D \frac{\partial {\cal L}}{\partial F^p}$, then using $d^2 \equiv 0$ gives
\begin{eqnarray}
0 &\equiv& d \left[ \alpha^p \left( \varphi T_p \wedge \frac{\partial {\cal L}}{\partial D \varphi} + D \frac{\partial {\cal L}}{\partial F^p} \right) \right] + \alpha^p \varphi T_p \wedge \frac{\delta {\cal L}}{\delta  \varphi} - D \alpha^p \wedge \frac{\delta {\cal L}}{\delta A^p}
\label{YangMillsNoether2a}\\
&\equiv& D \alpha^p \left(  \varphi T_p \wedge \frac{\partial {\cal L}}{\partial D \varphi} + D \frac{\partial {\cal L}}{\partial F^p} - \frac{\delta {\cal L}}{\delta A^p} \right)
\nonumber\\
&& + \alpha^p \left[ D \left( \varphi T_p \wedge \frac{\partial {\cal L}}{\partial D \varphi} + D \frac{\partial {\cal L}}{\partial F^p} \right) + \varphi T_p \wedge \frac{\delta {\cal L}}{\delta \varphi} \right]. \label{YangMillsNoether2b}
\end{eqnarray}

For a {\em local symmetry} the quantities $\alpha^p$ and $D \alpha^p$ are pointwise independent; their coefficients must vanish separately. The coefficient of $\alpha^p$ identifies a Noether I type conserved current:
\begin{equation}
J_p := \varphi T_p \wedge \frac{\partial {\cal L}}{\partial D \varphi} + D \frac{\partial {\cal L}}{\partial F^p}, \label{3:N1curr}
\end{equation}
which satisfies the ``conservation'' law
\begin{equation} \label{GaugeCurrCons}
D J_p \equiv - \varphi T_p \wedge \frac{\delta{\cal L}}{\delta \varphi}.
\end{equation}
The r.h.s. vanishes ``on shell'' (i.e., when the field equations are satisfied).

From the coefficient of $D \alpha^p$ we obtain an \emph{algebraic identity} relating the Noether I current to a variational derivative:
\begin{equation} \label{3:dyncurrentid}
J_p \equiv \frac{\delta {\cal L}}{\delta A^p},
\end{equation}
thereby the Noether I current conservation becomes a {\it differential identity}
\begin{equation} \label{3:diffeid}
D \frac{\delta {\cal L}}{\delta A^p} \equiv - \varphi T_p \wedge\frac{\delta {\cal L}}{\delta \varphi},
\end{equation}
between the variational derivatives. Note that to obtain these results there is no need for the explicit form of the field equations.

Another way to argue is to replace the last term in~(\ref{YangMillsNoether2a}) with a total differential minus a compensating term, bringing that relation into the form
\begin{equation}
0 \equiv d \left[ \alpha^p \left( \varphi T_p \wedge \frac{\partial {\cal L}}{\partial D \varphi} + D \frac{\partial {\cal L}}{\partial F^p} - \frac{\delta {\cal L}}{\delta A^p} \right) \right] + \alpha^p \left( \varphi T_p \wedge \frac{\delta {\cal L}}{\delta \varphi} + D \frac{\delta {\cal L}}{\delta A^p} \right).
\end{equation}
If one integrates this over any region the total differential term gives rise to an integral over the boundary.  To have a vanishing value for all possible gauge parameters with small support, the coefficient of the gauge parameter everywhere within the region and the coefficient of the gauge parameter everywhere on the boundary must both vanish identically. This again yields~(\ref{3:diffeid}) and
\begin{equation}
\varphi T_p \wedge \frac{\partial {\cal L}}{\partial D \varphi} + D \frac{\partial {\cal L}}{\partial F^p} - \frac{\delta {\cal L}}{\delta A^p} \equiv 0,
\end{equation}
which is equivalent to~(\ref{GaugeCurrCons}) with~(\ref{3:dyncurrentid}).

\subsection{Field equations with local gauge theory}%%%jmn322 3c43 suggested this; I am not so sure.
%%%%%%%%%%%%%%%%%%%%%%%%%%%%%%%%%%%%%%%%%%%%%%%%%%%%%%%%%%%%%%%%%%%%%%
It should be noted that the Noether invariance argument yields the differential identities just found involving the Euler-Lagrange expressions without any need to have the explicit form of the Euler-Lagrange expressions.  Of course if one explicitly computes the Euler-Lagrange expressions one could go on to verify these identities directly.  Furthermore, if one has the Euler-Lagrange expressions one could (probably not so easily)  directly discover such identities, even if one was not aware of the local symmetry.

To compute the field equations the explicit variations
\begin{eqnarray}
\delta D \varphi &=& D \delta \varphi + \delta A^p \wedge \varphi T_p,
\\
\delta F^p &=& d \delta A^p + C^p{}_{qr} A^q \wedge \delta A^r = D \delta A^p, \label{deltaF2}
\end{eqnarray}
are needed. The variation of the Lagrangian $4$-form $\cal L$~(\ref{3:varLgauge}) is
\begin{eqnarray}
\delta {\cal L} &=& \delta D \varphi \wedge \frac{\partial {\cal L}}{\partial D \varphi} + \delta \varphi \wedge \frac{\partial {\cal L}}{\partial \varphi} + \delta F^p \wedge \frac{\partial {\cal L}}{\partial F^p} + \delta A^p \wedge \frac{\partial {\cal L}}{\partial A^p}
\nonumber\\
&=& ( D \delta \varphi + \delta A^p \wedge \varphi T_p) \wedge \frac{\partial {\cal L}}{\partial D \varphi} + \delta \varphi \wedge \frac{\partial {\cal L}}{\partial \varphi} + D \delta A^p \wedge \frac{\partial {\cal L}}{\partial F^p} + \delta A^p \wedge \frac{\partial {\cal L}}{\partial A^p}
\nonumber\\
&=& D \left( \delta \varphi \wedge \frac{\partial {\cal L}}{\partial D \varphi} + \delta A^p \wedge \frac{\partial {\cal L}}{\partial F^p} \right) + \delta \varphi \wedge \left( - \varsigma D \frac{\partial {\cal L}}{\partial D \varphi} + \frac{\partial {\cal L}}{\partial \varphi} \right)
\nonumber\\
&& + \delta A^p \wedge \left( D \frac{\partial {\cal L}}{\partial F^p} + \frac{\partial {\cal L}}{\partial A^p} + \varphi T_p \wedge \frac{\partial {\cal L}}{\partial D  \varphi} \right).
\end{eqnarray}
Comparing the explicit form of $\delta {\cal L}/\delta A^p$ found here with~(\ref{3:dyncurrentid}, \ref{3:N1curr}) shows that~(\ref{3:dyncurrentid}) means
\begin{equation} \label{3:connindep}
\frac{\partial {\cal L}}{\partial A^p} \equiv 0.
\end{equation}
Thus local gauge invariance means: no explicit dependence on the gauge potential; all dependence on $A^p$ comes through $D\varphi$ and $F^p$. Furthermore if one makes the usual \emph{minimal coupling} assumption,
\begin{equation}
{\cal L} = {\cal L}_A(A^p, F^p) + {\cal L}_\varphi( \varphi, D \varphi, A^p),
\end{equation}
the identities~(\ref{3:connindep}, \ref{3:diffeid}) apply separately to each term. Hence, the Lagrangian $4$-form  must have the simpler form
\begin{equation}
{\cal L} = {\cal L}_A( F^p) + {\cal L}_\varphi( \varphi, D \varphi),
\end{equation}
and the differential identity~(\ref{3:diffeid}) becomes the two identities
\begin{eqnarray}
D \frac{\delta {\cal L}_A}{\delta A^p} \equiv 0, \qquad D \frac{\delta {\cal L}_\varphi}{\delta A^p} \equiv -\varphi T_p\wedge \frac{\delta {\cal L}_\varphi}{\delta \varphi}.\label{gaugeDI}
\end{eqnarray}
The first relation is explicitly
\begin{equation}
0\equiv D^2\frac{\partial {\cal L}_A}{\partial F^p}\equiv -F^q C^r{}_{qp}\wedge \frac{\partial {\cal L}_A}{\partial F^r}.
\end{equation}
The latter is a kind of gauge current ``conservation'', as the r.h.s. vanishes since
\begin{equation}
\frac{\delta {\cal L}_\varphi}{\delta \varphi} \equiv \frac{\delta {\cal L}}{\delta \varphi} = 0
\end{equation}
on shell.
In more detail, this gauge current ``conservation'' relation has the form
\begin{equation}
0 = D J_p = d J_p - A^q C^r{}_{qp}\wedge J_r.
\end{equation}
Thus it has some similarities to the vanishing covariant differential of the material energy-momentum. Just as in that case, one can rearrange the field equation to obtain a conserved gauge \emph{pseudocurrent}.

We have gone into considerable detail in this relatively simple example. We have done this to prepare the reader, because we are going to use a very similar argumentation in connection with the rather more complicated case involving gravity and the  dynamic spacetime geometric symmetries associated with energy-momentum and angular momentum. It will be seen that almost every step used in our later argument and every expression has an analogue with what we have done in this subsection.

%

%%%%%%%%%%%%%%%%%%%%%%%%%%%%%%%%%%%%%%%%%%%%%%%%%%%%%%%%%%%%%%%%%%%%%%
\section{First order formulation}%%%jmn ok
%%%%%%%%%%%%%%%%%%%%%%%%%%%%%%%%%%%%%%%%%%%%%%%%%%%%%%%%%%%%%%%%%%%%%%

\emph{In this section we discuss the general formulation of the first order formalism; the spacetime geometry has no restrictions.}

\medskip

We proceed from the action principle. Any action principle can be rewritten in an equivalent form, which (following, e.g., ADM\cite{adm} and Kucha\v{r}\cite{Kuchar}) we refer to as {\it first order}; this is the most convenient form for our purposes.
Here we present a simple argument (essentially the same Legendre transform idea as is used in classical mechanics to construct the Hamiltonian) which is applicable to a large class of 2nd order Lagrangians.

Given a 2nd-order Lagrangian $4$-form ${\cal L}(\varphi, d \varphi)$ we define its associated canonical momentum in the usual way:
\begin{equation}
p := \frac{\partial{\cal L}}{\partial d\varphi}(\varphi, d\varphi). \label{simpleLegendre}
\end{equation}
Next we define a $4$-form by
\begin{equation}
\Lambda(\varphi, d\varphi, p) := d\varphi \wedge p - {\cal L}.
\end{equation}
Now consider the variation of $\Lambda$:
\begin{eqnarray}
\delta\Lambda &=& \delta (d\varphi) \wedge p + d\varphi \wedge \delta p - \delta{\cal L}
\nonumber\\
&=& \delta (d\varphi) \wedge \left( p - \frac{\partial{\cal L}}{\partial d\varphi} \right) + d\varphi \wedge \delta p - \delta\varphi \wedge \frac{\partial{\cal L}}{\partial\varphi}
\nonumber\\
&=& d\varphi \wedge \delta p - \delta\varphi \wedge \frac{\partial{\cal L}}{\partial \varphi}.
\end{eqnarray}
Which shows that $\Lambda$ can be regarded as a function only of $\varphi, p$.\footnote{The procedure becomes technically somewhat more complicated if~(\ref{simpleLegendre}) cannot be inverted for $d\varphi$ in terms of $\varphi, p$. In that case one must introduce some additional variables that appear in $\Lambda$ only algebraically and thus function as Lagrange multipliers introducing some algebraic constraints. We will not go into such complications in our general development here. Later in our treatment of Einstein's GR we will see a concrete example. Examples of how this has been dealt with in field theory in practice can be found in.\cite{HRT,sundermeyer82,sundermeyer14}}\cite{}

%

%\subsubsection{first order form}

This construction takes one from the usual 2nd order Lagrangian to a \emph{first order} type of variational principle:
\begin{equation} \label{3:1ordL}
{\cal L}^{1\text{st}}(\varphi,d\varphi,p) = d  \varphi \wedge p - \Lambda(\varphi, p),
\end{equation}
where $\varphi$ and $p$ are now regarded as being independent variables and are varied independently. Varying~(\ref{3:1ordL}) gives
\begin{eqnarray}
\delta {\cal L}^{1\text{st}} &=& \delta d \varphi \wedge p + d \varphi \wedge \delta p - \delta \Lambda
\nonumber\\
&=& d (\delta \varphi \wedge p) - \varsigma \delta \varphi \wedge d p + d \varphi \wedge \delta p - \delta \varphi \wedge \frac{\partial \Lambda}{\partial \varphi} - \frac{\partial \Lambda}{\partial p} \wedge \delta p,
\nonumber\\
\therefore \quad \delta{\cal L}^{1\text{st}} &=& d (\delta \varphi \wedge p) + \delta \varphi \wedge \frac{\delta {\cal L}^{1\text{st}}}{\delta  \varphi} + \frac{\delta {\cal L}^{1\text{st}}}{\delta p} \wedge \delta p. \label{4:the basic identity1}
\end{eqnarray}
(Note: we find it more convenient to vary our momentum fields $p$ off to the right. This reduces a little the number of appearances of the sign factor $\varsigma$, and merely amounts to a sign convention on the definition of ${\partial \Lambda / \partial p}$.)
%%%jmn419 3c47 done

Using independent $p$ and $\varphi$ variations gives a pair of 1st order field equations for the differentials of the fields:
\begin{equation}
0 = \frac{\delta {\cal L}^{1\text{st}}}{\delta \varphi} = - \varsigma d p - \frac{\partial \Lambda}{\partial \varphi}, \qquad 0 = \frac{\delta {\cal L}^{1\text{st}}}{\delta p} = d \varphi - \frac{\partial \Lambda}{\partial p}. \label{4:1ordfe}
\end{equation}

%%%%%%%%%%%%%%%%%%%%%%%%%%%%%%%%%%%%%%%%%%%%%%%%%%%%%%%%%%%%%%%%%%%%%%
\section{The Hamiltonian and the 3+1 space-time split}%%%jmn ok
%%%%%%%%%%%%%%%%%%%%%%%%%%%%%%%%%%%%%%%%%%%%%%%%%%%%%%%%%%%%%%%%%%%%%%
\emph{Here we introduce the Hamiltonian and the space-time split.  In this introductory subsection we use for motivation some well-know elementary expressions in Minkowski spacetime.  In the subsequent subsections the spacetime geometry is quite general.}
\bigskip

A key feature of the canonical Hamiltonian formulation is that the field equations are decomposed into two sets: the initial value constraint equations and the dynamic equations. A familiar example which illustrates many of the ideas is Maxwell's vacuum electrodynamics.\footnote{$Z_0$ is the vacuum impedance, $\varepsilon_0 = (Z_0 c)^{-1}$ is the vacuum permittivity, and $\mu_0 = Z_0 c^{-1}$ is the vacuum permeability. Here we are taking for simplicity $\mu_0 \varepsilon_0 = c^{-2} = 1$ in relativistic space-time units.} The 4-covariant equations were given earlier (\ref{maxJ}, \ref{maxF}): $d*F = Z_0 J$, $dF = 0$, or in tensor index form
\begin{equation}
\partial_\mu (\sqrt{-g} F^{\nu\mu}) = Z_0 \sqrt{-g} J^\nu, \qquad \partial_{[\alpha} F_{\mu\nu]} = 0.
\end{equation}
They split (in Minkowski spacetime) into the familiar initial value constraints (spatial projections, with no time derivatives):
\begin{equation}
{\bf \nabla \cdot E} = \frac{\rho} {\epsilon_0}, \qquad {\bf \nabla \cdot B} = 0,
\end{equation}
and the time projections, a pair of dynamic equations:
\begin{equation}
{\bf \dot B + \nabla \times E} = 0, \qquad {\bf \nabla \times B} = \mu_0 {\bf J} + \dot{\bf E},
\end{equation}
which contain the first time derivatives of the dynamical fields linearly.

The canonical Hamiltonian form of these equations is in terms of the 4-vector potential (which satisfies $F = dA$ and splits into the scalar and vector potential). The familiar vector form is
\begin{eqnarray}
\hbox{constraint} &\qquad& {\bf \nabla \cdot E} = {\frac{\rho}{\epsilon_0}}, \label{4:emconstraint}
\\
\hbox{dynamic} &\qquad& {\bf \dot A = - E - \nabla} \Phi, \quad {\bf - \dot E = - \nabla \times (\nabla \times A)} + \mu_0 {\bf J}. \label{4:dynamicpotential}
\end{eqnarray}
The scalar potential field appears here, but it has no evolution equation; it can be chosen freely. This {\em gauge} freedom affects the evolution of an ``unphysical'' part of the vector potential. Considering this along with the constraint on $\bf E$~(\ref{4:emconstraint}), one finds that the electromagnetic field has two physical degrees of freedom.

\subsection{Canonical Hamiltonian formalism}%%%jmn ok
%%%%%%%%%%%%%%%%%%%%%%%%%%%%%%%%%%%%%%%%%%%%%%%%%%%%%%%%%%%%%%%%%%%%%%
The canonical Hamiltonian formalism\cite{Nester:1991yd} is of interest because it clearly reveals the constraints, gauges, and degrees of freedom, as well as the total energy momentum---and it offers a practical way to numerically calculate solutions.

The dynamical theories of interest all have constraints. The canonical formalism for constrained Hamiltonian systems was developed mainly by Dirac\cite{Dirac58a, Dirac58,Dirac64} and Bergmann\cite{Berg49, Berg58}. For a general discussion see, e.g., Hanson, Regge, \& Teitelboim\cite{HRT}, Sundermeyer\cite{sundermeyer82,sundermeyer14}. Rosenfeld\cite{Rosenfeld} seems to have been the first to consider a Hamiltonian approach to general relativity, but this early work  was not followed up. As far as we know Pirani, Schild and Skinner\cite{PSS} were the next to address the issue. Dirac gave a rather complete treatment in 1958\cite{Dirac58a, Dirac58}. The treatment by Arnowitt, Deser \& Misner (ADM)\cite{adm} has come to be regarded as the standard.
%Certain important contributions were added by in particular by Kucha\v r\cite{Kuchar}, Teitelboim, Lichnerowicz,
%York, Choquet-Bruhat, Moncrief, Marsden, Regge \& Teitelboim, Beig \& \'O Murchadaha, Szabados, etc.
For a basic discussion see MTW\cite{MTW73} Ch 21 or Isenberg and Nester\cite{IN80}. For some critical comparison, see Kiriushcheva and Kuzmin\cite{Kiriushcheva:2008sf}. Going beyond Einstein's theory, a remarkable ``if constraint'' formalism was developed for the PG by Blagojevi\'c and Nikoli\'c\cite{BlagNic} to deal with a conditionally degenerate kinetic Hessian.
They use the  Dirac type of approach; so to construct the Hamiltonian one must first find the primary constraints---which depend on the \emph{conditional} degeneracies of the Legendre transformation.
The ``if constraint'' technique is a marvelous way to manage the technicalities involved in constructing the Hamiltonian.
In our first order approach, in contrast, one can readily formally construct the Hamiltonian and the Hamiltonian equations,
however (in line with the principle of the ``conservation of difficulties'')
a suitably adapted version of the  ``if'' constraint technique will still be needed
when one actually tries to solve the dynamical and constraint equations.
The first order approach as used in the covariant Hamiltonian formalism allows one to
investigate the general formalism and, in particular, to find covariant expressions for the ``conserved'' quantities while postponing dealing with such technical details.
%%%jmn 3c51

\subsection{The differential form of the space-time decomposition}%%%jmn ok
%%%%%%%%%%%%%%%%%%%%%%%%%%%%%%%%%%%%%%%%%%%%%%%%%%%%%%%%%%%%%%%%%%%%%%
\emph{Note: for the rest of this section, the spacetime geometry is quite general.}

\bigskip

A feature of this standard approach is the loss of manifest 4-covariance.
Now a Hamiltonian formulation essentially requires that the time derivatives to be singled out from the spatial derivatives, so in this sense it cannot be truly 4-covariant. The usual approach, however, departs far more from 4-covariance than is necessary, all the indices are 3+1 projected, leading to much extra bookkeeping.
%%%jmn322 3c52 done
In the ADM approach the spacetime metric is replaced by the spatial metric and the \emph{lapse} and \emph{shift}.
 However only the derivatives $\partial_\mu$ really need to be projected. Since interaction fields are one-form fields, this means decomposing the exterior differential $d$, decomposing $d$ will inevitably involve decomposing the differential form. One of the reasons for using differential forms is in how nicely they decompose in this fashion.

Begin from the basic first-order form Lagrangian~(\ref{3:1ordL}). Its variation~(\ref{4:the basic identity1}) identifies the first order Euler-Lagrange expressions~(\ref{4:1ordfe}). According to Hamilton's principle, the first order Euler-Lagrange expressions should vanish. This gives us our first order field equations.

We want to extract the ``time derivative'' of $p$ and $\varphi$, i.e., the change with respect to an evolution parameter (which we refer to as time) as seen by observers who move along some fixed congruence of worldlines.
This change is given by the Lie derivative in the direction of the (fixed) vector field $Z$ tangent to the congruence: i.e., $\partial_t := \pounds_Z$. The Lie derivative on the components of differential forms is given by a simple neat expression: $\pounds_Z = d \, i_Z + i_Z d $. Using this, from the differential we can extract the ``time'' derivative:
\begin{equation}
i_Z d \beta = \pounds_Z \beta - d i_Z \beta = \dot \beta - d i _Z \beta. \label{4:iNdb}
\end{equation}
(The congruence need not actually be timelike. Indeed, what we are doing here does not require a metric tensor. Even when one has a metric whether the vector field is ``timelike'' is not an important issue, our whole Hamiltonian formalism is linear in the spacetime displacement vector field $Z$, so by considering the difference between two timelike displacements one could get a spacelike displacement.
A metrically timelike displacement is important when one actually tries to find a physical solution to the equations; for evolution one wants hyperbolic equations.)
%%%jmn510 3c53 done

The description of ``time'' also includes the idea of ``instants of time''. Geometrically this is a set (foliation) of (non-intersecting) 3-dimensional hypersurfaces (we usually think of them as being spacelike). Locally, we can always choose adapted coordinates: $x^\mu = \{ t, x^k \}$, where $k = 1, 2, 3$ so that the spacelike hypersurfaces are $\Sigma_t$ with $t = \hbox{constant}$. With respect to these adapted coordinates $Z$ is the directional derivative in the time direction: $Z = \partial_t$. Note that $i_Z dt \equiv 1$.

From these considerations we are led to define the ``{\em time\/}'' and ``{\em space\/}'' projections of differential forms. We use the notations
\begin{equation}
\hat\alpha := i_Z \alpha, \qquad \underline{\alpha} := \alpha - dt \wedge \hat \alpha,
\end{equation}
to indicate the ``time'' component and the ``spatial'' part of a form. %~\cite{coham,Wallner,Hehl-Obukhov}.
These projections have simple expressions in terms of adapted
coordinates.\footnote{%
%Many of our works,
We have long used this type of decomposition beginning with \cite{IN80,Nes84}; see~\cite{MielWal88} for a similar technique.}
%%%jmn323 3c54 done

Thus a general form decomposes according to
\begin{equation}
\alpha = dt \wedge \hat \alpha + \underline{\alpha}.
\end{equation}
In our formalism $i_Z$ and $t$ are thought of as freely (except that $i_Z dt = 1$) chosen covariant fields,
then the decomposition of $\alpha$ %into $(\hat\alpha,\underline{\alpha})$
is essentially covariant.
With this notation the differential decomposes according to
%\begin{equation}
$d \alpha = dt \wedge \widehat {d \alpha} + \underline{d \alpha}$.
%\end{equation}
%
From~(\ref{4:iNdb})
\begin{equation}
\widehat{d \alpha} = \dot \alpha - d \hat \alpha;
\end{equation}
thus we can extract the part with the time derivative.

It is convenient to decompose the differential operator $d$ itself. In view of the adapted coordinate expression $d = dx^\mu \wedge \partial_\mu = dt \wedge \partial_t + dx^k \wedge \partial_k$, we define the decomposition as
\begin{equation}
d = dt \wedge \hat{d} + \underline{d}, \quad \hbox{with} \quad \hat{d} := \pounds_Z.
\end{equation}

Now we can examine the spacetime decomposition of our first order field equations~(\ref{4:1ordfe}). We first consider the time projections, which include all the time derivatives:
\begin{eqnarray}
\widehat{ \left( \frac{\delta {\cal L}^{1\text{st}}}{\delta \varphi} \right) } &=& - \varsigma ( {\underline{\dot p}} - {\underline{d}} \hat p) - \widehat{ \left( \frac{\partial \Lambda}{\partial \varphi} \right) } = 0,
\\
\widehat{ \left( \frac{\delta {\cal L}^{1\text{st}}}{\delta p} \right) } &=& (  {\underline{\dot \varphi}} - {\underline{d}} \hat { \varphi} ) - \widehat{ \left( \frac{\partial \Lambda}{\partial p} \right) } = 0.
\end{eqnarray}
%According to one of the projection identities~(\ref{4:projidents}), the spatial restriction of these equations has the same content,
they are the dynamic equations for $\underline{p}$ and $\underline{\varphi}$.
In order to use these equations to evolve $\underline p$, $\underline \varphi$, we generally need to know ${\hat p}$ and ${\hat{ \varphi}}$, which normally are provided by the initial value constraints: the spatial restriction of~(\ref{4:1ordfe}):
\begin{eqnarray}
\underline{\left( \frac{\delta {\cal L}^{1\text{st}}}{\delta \varphi} \right)} &=& - \varsigma \underline{d p} - \underline{\left( \frac{\partial \Lambda}{\partial  \varphi} \right)} = 0,
\\
\underline{\left( \frac{\delta {\cal L}^{1\text{st}}}{\delta p} \right)} &=& \underline{d \varphi} - \underline{\left(
\frac{\partial \Lambda}{\partial p} \right)} = 0.
\end{eqnarray}
If these two equations can be solved
%\footnote{If it is not possible to solve for these apparently non-dynamic components, then, we should follow
%Dirac's systematic constraint algorithm and consider the time derivatives of the constraint equations,
%which might yield the missing information. Pursuing that story would lead beyond the scope of our discussion.}
for ${\hat p}$ and ${\hat{ \varphi}}$ all is well and good (in that case the two equations are, in Bergmann's terminology, {\em second class constraints}). They then define ${\hat p}$ and $\hat{\varphi}$ for all time as functions which depend on $\underline{\varphi}$, $\underline{p}$, $\underline{d p}$ and $\underline{d \varphi}$. How to proceed for the case where these quantities cannot be found from the constraints is best understood from concrete examples. For our purposes of this work, the already-discussed Maxwell electrodynamic example is sufficient. In that case there is some undetermined gauge freedom.

\subsection{Spacetime decomposition of the variational formalism}%%%jmn ok
%%%%%%%%%%%%%%%%%%%%%%%%%%%%%%%%%%%%%%%%%%%%%%%%%%%%%%%%%%%%%%%%%%%%%%
We decomposed the equations. One could decompose the Lagrangian or its variation. Our approach easily relates these alternatives, as can be seen from the following.
%and to see that one gets the same results no matter which way one proceeds. The following chart gives the key points.
\begin{align}
{\cal L}^{1\text{st}} = d \varphi \wedge p - \Lambda \qquad & \maprt{3+1}{} & \underline{\widehat{{\cal L}^{1\text{st}}}} = (\underline{\dot{\varphi}} - \underline{d \hat{\varphi}}) \wedge \underline{p} - (\underline{\hat \Lambda} + \varsigma \underline{d \varphi} \wedge \underline{\hat p}) \label{line1}
\\
\delta \Big\downarrow \qquad\qquad & & \Big\downarrow \delta \qquad\qquad \nonumber
\\
\delta {\cal L}^{1\text{st}} = d (\delta \varphi \wedge p) \qquad & \maprt{3+1}{} & \frac{d}{dt} (\delta \underline{\varphi} \wedge \underline{p}) - \underline{d} (\delta \underline{\hat{\varphi}} \wedge \underline{p} + \varsigma \delta \underline{\varphi} \wedge \underline {\hat p})
\nonumber\\
+ \delta \varphi \wedge \frac{\delta {\cal L}^{1\text{st}}}{\delta \varphi} \qquad &  & + \delta \underline{\hat{\varphi}} \wedge \underline{\left( \frac{\delta {\cal L}^{1\text{st}}}{\delta \varphi} \right)} + \delta \underline{\varphi} \wedge \varsigma \underline{\widehat{\left( \frac{\delta {\cal L}^{1\text{st}}}{\delta \varphi} \right)}}
\nonumber\\
\quad + \frac{\delta {\cal L}^{1\text{st}}}{\delta p} \wedge \delta p \qquad & & - \varsigma \underline{\left( \frac{\delta {\cal L}^{1\text{st}}}{\delta p} \right)} \wedge \underline{\delta \hat p} + \underline{\widehat{\left( \frac{\delta {\cal L}^{1\text{st}}}{\delta p} \right) }} \wedge \underline{\delta p}
\\
{\hbox{extract}} \Big\downarrow \qquad\qquad & & \Big\downarrow {\hbox{extract}} \qquad \nonumber
\\
\frac{\delta {\cal L}^{1\text{st}}}{\delta \varphi} \qquad\quad  & \maprt{3+1}{} & \underline{\left( \frac{\delta{\cal L}^{1\text{st}}}{\delta \varphi} \right)}, \quad \underline{\widehat{\left( \frac{\delta {\cal L}^{1\text{st}}}{\delta  \varphi} \right)}} \quad
\\
\frac{\delta {\cal L}^{1\text{st}}}{\delta p} \qquad\quad  & \maprt{3+1}{} & \underline{\left( \frac{\delta {\cal L}^{1\text{st}}}{\delta p} \right)}, \quad \underline{\widehat{\left( \frac{\delta {\cal L}^{1\text{st}}}{\delta p} \right)}} \quad
\end{align}

For our objectives here we will not need to use this projection into the space and time parts of form expressions very much. Our intention here was to include enough of the details so that the reader can have some confidence that this formalism can yield a proper Hamiltonian description.
From what we have discussed, it can be seen that dynamical equations in this first order covariant form already contain both the constraint and dynamical evolution equations. It should be noted that the first line of the above set of relations~(\ref{line1}) shows how the Hamiltonian can be simply extracted from the first order Lagrangian.

%%%%%%%%%%%%%%%%%%%%%%%%%%%%%%%%%%%%%%%%%%%%%%%%%%%%%%%%%%%%%%%%%%%%%%
\section{The Hamiltonian and its boundary term}\label{S:Ham and B} %%%jmn ok
%%%%%%%%%%%%%%%%%%%%%%%%%%%%%%%%%%%%%%%%%%%%%%%%%%%%%%%%%%%%%%%%%%%%%%
%ccm 118
\emph{In this section we establish some of our main formal results concerning the covariant Hamiltonian and its boundary term. The geometry is quite general.
The energy, as well as the other conserved quantities, of a physical system can be identified with the value of the Hamiltonian. In particular for a gravitating system the associated Hamiltonian is proportional to the field equations which vanish on-shell. Therefore the corresponding conserved quantities are determined by the Hamiltonian boundary term.  The choice of Hamiltonian boundary term is associated with the specific boundary condition.\cite{Chen:1994qg,Chen:1998aw,Chang:1998wj,Chen:2000xw,Chen:2005hwa}}

\subsection{The translational Noether current}\label{S:translational Noether current}    %%%jmn ok
%%%%%%%%%%%%%%%%%%%%%%%%%%%%%%%%%%%%%%%%%%%%%%%%%%%%%%%%%%%%%%%%%%%%%%
The action should not depend on the particular way points are labeled. Thus it should be invariant under diffeomorphisms, in particular, infinitesimal diffeomorphisms---a displacement along some vector field $Z$. From a gauge theory perspective such displacements are a ``local translation''. Under a local translation quantities change according to the Lie derivative. Hence, for a diffeomorphism invariant action the key variational relation~(\ref{4:the basic identity1}) should be identically satisfied when the variation operator $\delta$ is replaced by the Lie derivative $\pounds_Z$ ($\equiv di_Z + i_Zd$):
\begin{equation}
d i_Z {\cal L}^{1\text{st}} \equiv \pounds_Z{\cal L}^{1\text{st}} \equiv  d(\pounds_Z \varphi \wedge p) + \pounds_Z \varphi \wedge \frac{\delta {\cal L}^{1\text{st}}}{\delta \varphi} + \frac{\delta {\cal L}^{1\text{st}}}{\delta p} \wedge \pounds_Z p. \label{5:noethertrans}
\end{equation}
This simply means that ${\cal L}^{1\text{st}}$ is a $4$-form which depends on position only through the fields $\varphi$, $p$.
%%%jmn323 3c56
(According to our understanding this is only possible if the set of fields in ${\cal L}^{1\text{st}}$ includes some dynamic spacetime geometric variables: gravity.)
%%%

From~(\ref{5:noethertrans}) it directly follows that the $3$-form
\begin{equation}%%%jmn 3c57
{\cal H}(Z) := \pounds_Z \varphi \wedge p - i_Z {\cal L}^{1\text{st}} \label{5:HN}
\end{equation}
satisfies the identity
\begin{equation} \label{5:IdH}
- d {\cal H}(Z) \equiv \pounds_Z \varphi \wedge \frac{\delta {\cal L}^{1\text{st}}}{\delta \varphi} + \frac{\delta {\cal L}^{1\text{st}}}{\delta p} \wedge \pounds_Z p;
\end{equation}
thus it is a conserved ``current'' {\it on shell} (i.e., when the field equations are satisfied). Substituting~(\ref{3:1ordL}) into~(\ref{5:HN}) gives the explicit expression
\begin{equation}
{\cal H}(Z) \equiv d(i_Z \varphi \wedge p) + \varsigma i_Z \varphi \wedge d p + \varsigma d \varphi \wedge i_Z p + i_Z \Lambda, \label{ham}
\end{equation}
thus this conserved {\em Noether translation current\/} can be written as a $3$-form linear in the displacement vector plus a total differential:
\begin{equation}%%%jmn 3c57
{\cal H}(Z) =: Z^\mu {\cal H}_\mu + d {\cal B}(Z). \label{5:H+dB}
\end{equation}
Compare the differential of this expression, $d{\cal H}(Z) \equiv dZ^\mu \wedge {\cal H}_\mu + Z^\mu d {\cal H}_\mu$, with~(\ref{5:IdH}); equating the $dZ^\mu$ coefficient on both sides reveals that
\begin{equation}
Z^\mu {\cal H}_\mu \equiv - i_Z \varphi \wedge \frac{\delta {\cal L}^{1\text{st}}}{\delta \varphi} + \varsigma \frac{\delta {\cal L}^{1\text{st}}}{\delta p} \wedge i_Z p.
\end{equation}
Thus, as a consequence of {\em local\/} diffeomorphism invariance, ${\cal H}_\mu$ vanishes \emph{on shell}; hence conservation of the translational Noether current~(\ref{5:IdH}) reduces to a differential identity between Euler-Lagrange expressions.
This an instance of Noether's second theorem, and, moreover, it is exactly the sort of case to which Hilbert's remark regarding the lack of a proper energy law applies.

 From the above it follows quite generally, as we remarked earlier in the special case of GR (\ref{basicHam}), the \emph{value} of the conserved quantity, $-P(Z,V)$, associated with a 3D region $V$ is determined by a 2-surface integral over the boundary, i.e., it is \emph{quasi-local}:
\begin{equation}
-P(Z, V) := \int_V {\cal H}(Z) = \oint_{\partial V} {\cal B}(Z). \label{5:EN}
\end{equation}
For {\em any\/} choice of $Z$ this expression defines a conserved quasi-local quantity. What do these values mean?
As we shall see in detail later, for a suitable timelike(spacelike) quasi-translation displacement on the boundary the expression defines a quasi-local energy(momentum), and for a suitable quasi-rotation(boost) it defines a quasi-local angular momentum(center-of-mass momentum).
However it must be noted that,  like all other conserved currents, the translational current is likewise subject to the usual ambiguity:  one can add \emph{by hand} the differential of any 2-form and still have a conserved current.  But that amounts to being able to adjust ${\cal B}$ freely, consequently one could obtain almost any quasi-local value.  The Hamiltonian perspective brings this freedom under physical control. As we shall show, the \emph{first order} translational current 3-form is something more: it is \emph{the generator of local diffeomorphisms}, i.e., \emph{the Hamiltonian}.

\subsection{The Hamiltonian formulation}\label{SS:Ham form} %%%jmn ok
%%%%%%%%%%%%%%%%%%%%%%%%%%%%%%%%%%%%%%%%%%%%%%%%%%%%%%%%%%%%%%%%%%%%%%
From the first order field equations~(\ref{4:1ordfe}), by contraction with a ``time evolution vector field'' $Z$ we get a pair of Hamiltonian-like evolution equations for the ``time derivatives'': $\pounds_Z \varphi$, $\pounds_Z p$. A key identity involving these time derivatives is revealed by comparing two relations.
Consider the projection of the Lagrangian $4$-form $i_Z {\cal L}^{1\text{st}}$, which from~(\ref{5:HN}) is just $\pounds_Z \varphi \wedge p - {\cal H}(Z)$; its variation is
\begin{eqnarray}
\delta i_Z {\cal L}^{1\text{st}} &\equiv& \delta (\pounds_Z \varphi \wedge p) - \delta{\cal H}(Z)
\nonumber\\
&\equiv& \delta (\pounds_Z\varphi) \wedge p + \pounds_Z \varphi \wedge \delta p - \delta{\cal H}(Z)
\nonumber\\
&\equiv& \pounds_Z \delta \varphi \wedge p + \pounds_Z \varphi \wedge \delta p - \delta{\cal H}(Z)
\nonumber\\
&\equiv& \pounds_Z(\delta\varphi \wedge p) - \delta \varphi \wedge \pounds_Z p + \pounds_Z \varphi \wedge \delta p - \delta{\cal H}(Z).
\end{eqnarray}
Compare this with the projection of $\delta {\cal L}^{1\text{st}}$~(\ref{4:the basic identity1}) along $Z$:
\begin{eqnarray}
i_Z \delta{\cal L}^{1\text{st}} &\equiv& i_Z d(\delta \varphi \wedge p) + i_Z \left( \delta \varphi \wedge \frac{\delta {\cal L}^{1\text{st}}}{\delta \varphi} + \frac{\delta {\cal L}^{1\text{st}}}{\delta p} \wedge \delta p \right)
\nonumber\\
&\equiv& \pounds_Z (\delta\varphi \wedge p) - di_Z(\delta \varphi \wedge p) + i_Z \left( \delta \varphi \wedge \frac{\delta {\cal L}^{1\text{st}}}{\delta \varphi} + \frac{\delta {\cal L}^{1\text{st}}}{\delta p} \wedge \delta p \right).
\end{eqnarray}
Since $Z$ is not varied the two relations are identical:
$\delta i_Z {\cal L}^{1\text{st}} \equiv i_Z \delta {\cal L}^{1\text{st}}$; consequently,
\begin{equation}
\delta {\cal H}(Z) \equiv
- \delta \varphi \wedge \pounds_Z p + \pounds_Z \varphi \wedge \delta p
+ d i_Z(\delta \varphi \wedge p) -
i_Z \left( \delta \varphi \wedge \frac{\delta {\cal L}^{1\text{st}}}{\delta \varphi}
 + \frac{\delta {\cal L}^{1\text{st}}}{\delta p} \wedge \delta p \right). \label{5:varH}
\end{equation}
The last term vanishes ``on shell''. This relation identifies the Noether translational current ${\cal H}(Z)$ as the {\em Hamiltonian $3$-form} (i.e., density), as the following considerations show. The integral of ${\cal H}(Z)$ over a 3-dimensional region,
\begin{equation}
H(Z, \Sigma) := \int_\Sigma {\cal H}(Z),
\end{equation}
is the Hamiltonian which displaces this region along $Z$, since the integral of its variation:
\begin{equation}
\delta H(Z, \Sigma) = \int_\Sigma \delta {\cal H}(Z),
\end{equation}
yields, from~(\ref{5:varH}), ``on shell'' (then the last bracketed term vanishes) the Hamilton equations:
\begin{equation}
\pounds_Z \varphi = \frac{\delta{ H}(Z,\Sigma)}{\delta p}, \qquad \pounds_Z p = -\frac{\delta{ H}(Z,\Sigma)}{\delta \varphi}, \label{5:hameqs}
\end{equation}
{\em  if\/} the boundary term in the variation of the Hamiltonian {\em vanishes}. In this case that means when $\delta \varphi$ vanishes on $\partial\Sigma$. Technically, the variational derivatives of the Hamiltonian $H(Z, \Sigma)$ displayed in~(\ref{5:hameqs}) are only defined for variations satisfying this boundary condition. In other words this Hamiltonian is ``well-defined'', i.e., {\em functionally differentiable\/}, only on the phase space of fields satisfying
the particular boundary condition $\delta\varphi|_{\partial\Sigma}=0$.

\subsection{Boundary terms: the boundary condition and reference}%%%jmn ok
%%%%%%%%%%%%%%%%%%%%%%%%%%%%%%%%%%%%%%%%%%%%%%%%%%%%%%%%%%%%%%%%%%%%%%
In some important cases the fields of physical interest do not satisfy the boundary condition naturally inherited from the Lagrangian\cite{Chen:1998aw}, this happens in particular for the spacetime metric of an asymptotically flat region. A modified formulation is needed to deal with this.

One alternative is to modify the Lagrangian 4-form itself by a total
differential. This strategy has often been adopted, beginning with Einstein~(\ref{EinsteinL}) and including many of the Hamiltonian formulations\cite{PSS,Dirac58,adm}.  But such a modification is necessarily non-covariant.  For our formalism we want to keep our Lagrangian covariant.  Furthermore
  a Lagrangian boundary term would modify the boundary condition on the whole 3-dimensional boundary of the spacetime region, thus inducing the same type of modification on the spatial boundary at large spatial distances as on the initial time hypersurface.  However we want the freedom to adjust the boundary condition on the 2-dimensional boundary of the spacelike region $\partial\Sigma$ independently of the type of initial conditions imposed within the initial time hypersurface $\Sigma_t$.
  %Moreover, we do not want to spoil our covariant variables within the dynamical region of interest
  %by considering how they differ from some reference values, we want to do that only for the values on the boundary of region.
  Thus for our objectives we turn to the Hamiltonian boundary term.\footnote{In the end it turns out that our favored Hamiltonian boundary term for GR is related to one induced by a Lagrangian boundary term.\cite{Chen:1998aw}}

Note that the Hamiltonian~(\ref{5:H+dB}) has two distinct parts; each plays a distinct role. The proper density
$Z^\mu {\cal H}_\mu$, although it has vanishing value on shell, generates the equations of motion, whereas the boundary term ${\cal B}(Z)$ determines not only the {\em quasi-local value\/}~(\ref{5:EN}) but also the {\em boundary condition\/}. Now we should make note of a very important fact:  the boundary term can be adjusted--- without changing the Hamilton equations or the conservation property~(\ref{5:IdH}).  Thus one can replace the $2$-form ${\cal B}(Z) = i_Z \varphi \wedge p$ inherited from the Lagrangian by another.

Such an adjustment is in one respect just a special case of the conserved Noether current ambiguity (i.e., for any $2$-form $\chi$, $J$ and $J' := J + d\chi$ are both conserved currents ($3$-forms) if $dJ = 0$, even though they define different conserved values).

However here, in this Hamiltonian case, any such adjustment modifies---{\em in parallel\/}---not only the value of the quasi-local quantities but also the spatial boundary conditions. Thus the boundary term ambiguity is under physical control: each distinct choice of the quasi-local expression given by the Hamiltonian boundary term is associated with a physically distinct boundary condition.

In order to accommodate suitable boundary conditions we found that, in general, one needs to introduce on the boundary for each of the dynamical fields certain reference values $\bar p$, $\bar \varphi$, which represent the ground state of the field---the ``vacuum'' (or background field) values. This is necessary in particular for fields whose natural ground state is non-vanishing; the spacetime metric is such a field.

We take our boundary terms to be linear in $\Delta\varphi := \varphi - \bar\varphi$, $\Delta p := p - \bar p$, so that they (and thus all the quasi-local quantities) vanish if the fields take on the ground state (reference) values.\footnote{Some authors use the terminology \emph{regularize}.} We presume that the reference values (like $Z$) are not varied: $\delta \bar\varphi = 0$ and $\delta \bar p = 0$, consequently $\delta \Delta \varphi = \delta \varphi$, $\delta \Delta p = \delta p$.

\subsection{Covariant-symplectic Hamiltonian boundary terms}\label{SS:co-symp Hbt}  %%%jmn ok
%%%%%%%%%%%%%%%%%%%%%%%%%%%%%%%%%%%%%%%%%%%%%%%%%%%%%%%%%%%%%%%%%%%%%%
To find an improved Hamiltonian boundary term for~(\ref{ham}) first drop the one inherited from the Lagrangian.
 Examine the boundary term generated in the variation of the $3$-form part of the Hamiltonian~(\ref{ham}); it is $-i_Z \varphi \wedge \delta p + \varsigma \delta \varphi \wedge i_Z p $.  This invites us to add a suitable complimentary boundary term.  In this way we were led to the boundary terms\cite{Chen:1994qg,Chen:1998aw,Chen:2005hwa}
\begin{equation}
{\cal B}(Z) := i_Z \left\{ \begin{array}{c} {\varphi} \\ {\bar\varphi} \end{array} \right\} \wedge \Delta p - \varsigma \Delta\varphi \wedge i_Z \left\{ \begin{array}{c} p \\ \bar p \end{array} \right\}.\label{genB}
\end{equation}
Then the associated variational Hamiltonian boundary term becomes
\begin{equation}
\delta{\cal H}(Z) \sim d\left[ \left\{ i_Z \delta\varphi \wedge \Delta p \atop - i_Z \Delta\varphi \wedge \delta p \right\} + \varsigma \left\{ -\Delta\varphi \wedge i_Z \delta p \atop \delta\varphi \wedge i_Z \Delta p \right\} \right]. \label{deltaHZbound}
\end{equation}
Here \emph{for each bracket independently} one may choose either the upper or lower term, which represent essentially a choice of Dirichlet (fixed field) or Neumann (fixed momentum) boundary conditions for the space and time parts of the fields separately.\footnote{There are more complicated possibilities, ``mixed'' choices involving some linear combination of the upper and lower expressions\cite{So:2006zz}.  We do not have any specific physical examples, but mixed boundary conditions may be of interest in certain cases.}

In each of these cases the boundary term in the Hamiltonian variation has a certain {\it symplectic} structure which pairs certain {\em control} quantities---i.e., the independent variables---with certain associated  {\em response} quantities---the dependent variables. (For discussions of this paradigm of the symplectic structure associated with variational principles which we have found to be illuminating see Refs.~[\refcite{KT79, Kijowski97}]).
%[Kijowski-Tulczyjew 1979].
The symmetry of the above expressions under an interchange of ``control'' and ``response'', formally $\delta \to \Delta$, $\Delta \to -\delta$, is noteworthy.

Thus, although it is not so well known, when the issue is examined one can readily see that there are many choices of boundary conditions and consequently really many different expressions for energy in classical field theory.  This is true especially for gravitating fields.  Actually this sort of thing is not unusual in physics; in particular one can compare the situation with that in thermodynamics (which has several physically meaningful energies: internal, enthalpy, Gibbs and Helmholtz).
%%%jmn 3c66

Nevertheless it should be noted that one of our boundary term expressions stands out: for any field which allows trivial reference values, $\bar \varphi = 0 = \bar p$, one boundary term choice vanishes (the lower choice in each bracket). Such fields,  with this choice of boundary condition, make no explicit contribution to the quasi-local boundary term. This particular boundary term has another virtue: for any field with gauge freedom it is the only gauge invariant choice. Thus there is a certain preferred boundary expression---and thus {\em a preferred boundary condition}---for this large class of fields, a class which includes all the physical fields of the standard model. There is, however, a quite important exception: \emph{gravity},  more specifically, any gravity theory formulated in terms of dynamic spacetime geometry which includes the spacetime metric as a dynamical field.  The natural reference choice for the metric \emph{is not} a vanishing metric tensor but rather the non-vanishing Minkowski metric. Consequently one must have, in general, a non-vanishing Hamiltonian boundary term.
%%%jmn323 3c67
%%%jmn510

%%%%%%%%%%%%%%%%%%%%%%%%%%%%%%%%%%%%%%%%%%%%%%%%%%%%%%%%%%%%%%%%%%%%%%
\section{Standard asymptotics}%%%jmn ok
%%%%%%%%%%%%%%%%%%%%%%%%%%%%%%%%%%%%%%%%%%%%%%%%%%%%%%%%%%%%%%%%%%%%%%
\emph{This section is concerned with suitable asymptotic conditions for our classical fields at spatial and null infinity.  It also includes a discussion of energy flux.}

\bigskip

For spatial infinity, the issue of asymptotic conditions was first investigated in GR by Regge and Teitelboim\cite{Regge:1974zd}, with later refinements by Beig and \'O Murchadha\cite{Beig:1987zz} and then Szabados\cite{Szabados:2003yn, Sza06}.  We have developed a similar idea for general fields.

\subsection{Spatial infinity}%%%jmn ok
%%%%%%%%%%%%%%%%%%%%%%%%%%%%%%%%%%%%%%%%%%%%%%%%%%%%%%%%%%%%%%%%%%%%%%
For finite regions these boundary terms in the variation of the Hamiltonian tell us exactly what needs to be held fixed (i.e., ``controlled''). For asymptotically flat regions, however, one should take into account the asymptotic fall off rates.
The various boundary terms we have constructed enable the Hamiltonian to be well defined on the phase space of fields with suitable asymptotic behavior for all typical physical fields.

For the fields it is sufficient\footnote{Sufficient, but not necessary. When examined in detail it can  be seen that one really only needs conditions on certain combinations of the components, but it is not in the spirit of our treatment to break fields up into, e.g.,  components parallel to and perpendicular to some specific boundary surface, etc.  Here we are satisfied with a formalism that includes a large class of fields.  We leave to more specific investigations finding the largest acceptable phase space with the weakest conditions.} to take the respective asymptotic fall offs for even and odd parity terms to be
\begin{equation}
\Delta \varphi \approx {\cal O}^+(1/r) + {\cal O}^-(1/r^2), \qquad \Delta p \approx {\cal O}^-(1/r^2) + {\cal O}^+(1/r^3). \label{5:asymptotics}
\end{equation}
Parity here means the parity of the components in an asymptotically Cartesian reference frame. The 2-surface area element has odd parity, so even parity $2$-forms automatically have vanishing 2-surface integral.

For asymptotically flat spaces the displacement should asymptotically be a Minkowski Killing vector, i.e.,
an infinitesimal Poincar\'e displacement.  It is sufficient to take %More precisely one can allow
\begin{equation}
Z^\mu \approx Z_0 ^\mu + \lambda_0^\mu{}_\nu x^\nu +{\cal O}^+(1/r)+{\cal O}^-(1), \label{5:asykilling}
\end{equation}
where (in terms of asymptotically Minkowski coordinates) $Z_0^\mu$ is a constant translation parameter and $\lambda_0^{\mu\nu} = \lambda_0^{[\mu\nu]}$ is a constant asymptotic infinitesimal Lorentz boost/rotation parameter.

With the asymptotics~(\ref{5:asymptotics}, \ref{5:asykilling}) it is straightforward to check that for any of the boundary term choices~(\ref{genB}) all of the quasi-local quantities have finite values; furthermore, for any of the choices all of our Hamiltonians are differentiable on the specified phase space---since the respective boundary terms in the variations of the Hamiltonians~(\ref{deltaHZbound}) vanish asymptotically. Thus our Hamiltonians are generally well-defined on a large phase space which includes physically interesting solutions.
At asymptotically-flat spatial infinity the aforementioned asymptotics are physically reasonable. Our considerations naturally straightforwardly extend to asymptotically anti-de Sitter spaces. Here the details are omitted.

\subsection{Null infinity}%%%jmn ok
%%%%%%%%%%%%%%%%%%%%%%%%%%%%%%%%%%%%%%%%%%%%%%%%%%%%%%%%%%%%%%%%%%%%%%
Let us next consider what can be expected if the boundary of our 2-surface $\partial\Sigma$ approaches future null infinity.
Long range radiation fields (e.g., electromagnetism) have slower fall offs, like $\Delta p \approx d\varphi \approx O(1/r)$. Then
the boundary terms in the variation of our various Hamiltonians will not vanish, so the Hamiltonian is no longer
functionally differentiable. This seeming calamity is actually providential---it is directing us to additional physics contained within the formalism, namely energy flux expressions.\cite{Nes84, Chen:2005hwa, Wu:2005yv}

\subsection{Energy flux}
%%%%%%%%%%%%%%%%%%%%%%%%%%%%%%%%%%%%%%%%%%%%%%%%%%%%%%%%%%%%%%%%%%%%%%
For the flux of ``energy'' there is a special way of calculating---the analog of the classical mechanics calculation (for conservative Hamiltonian systems) of
\begin{equation}
\delta H = \dot q^k \delta p_k - \dot p_k \delta q^k \qquad \Longrightarrow \qquad \dot E := \dot H \equiv 0
\end{equation}
under the replacement $\delta \to d/dt$, where the remarkable cancelation is a consequence of the particular
(symplectic) form of the Hamiltonian variation. In the present case from~(\ref{5:varH}) under the replacement $\delta \to \pounds_Z$ the same type of symplectic calculation occurs, and we are left with the respective contributions from our various boundary term choices~(\ref{deltaHZbound})
\begin{equation}
\pounds_Z {\cal H}(Z) = d\left[ \left\{ i_Z \pounds_Z \varphi \wedge \Delta p \atop - i_Z \Delta\varphi \wedge \pounds_Z p \right\} + \varsigma \left\{ -\Delta\varphi \wedge i_Z \pounds_Z p \atop \pounds_Z \varphi \wedge i_Z \Delta p \right\} \right]. \label{generalEnergyFlux}
\end{equation}
(We are presuming that $\pounds_Z \bar\varphi$, $\pounds_Z \bar p$ vanish.)

In particular, as we mentioned earlier, for all fields with vanishing reference, there is a standard choice of Hamiltonian boundary term, namely the one that vanishes. The corresponding energy flux expression is
\begin{equation}
\pounds_Z {\cal H}(Z) = d\left( - i_Z \varphi \wedge \pounds_Z p  + \varsigma \pounds_Z \varphi \wedge i_Z  p  \right). \label{preferredEnergyFlux}
\end{equation}

%%%%%%%%%%%%%%%%%%%%%%%%%%%%%%%%%%%%%%%%%%%%%%%%%%%%%%%%%%%%%%%%%%%%%%
\section{Application to Electromagnetism}
%%%%%%%%%%%%%%%%%%%%%%%%%%%%%%%%%%%%%%%%%%%%%%%%%%%%%%%%%%%%%%%%%%%%%%

\emph{To illustrate these ideas in a familiar setting, we briefly consider vacuum electromagnetism in Minkowski space (for the complete details see Ref.~[\refcite{Chen:2005hwa}]).}
\medskip

For electromagnetism in Minkowski space the formalism developed above, with the important exception of the ``on shell'' vanishing of ${\cal H}_\mu$, is still applicable. A first order Lagrangian 4-form for the (source free) Maxwell one-form (the U(1) gauge potential)  $A$ is
\begin{equation}
{\cal L}^{1\text{st}}_{\rm EM} = d A \wedge {\cal P} - \frac{1}{2}Z_0 * {\cal P} \wedge {\cal P},
\end{equation}
which yields the pair of first order equations
\begin{equation}
d {\cal P} = 0, \qquad d A - Z_0* {\cal P} = 0.
\end{equation}
%%%jmn323 3c68
These are just the vacuum Maxwell equations with $* {\cal P} = Z_0^{-1} F := Z_0^{-1} dA$; hence ${\cal P} = - Z_0^{-1} * F$, and $d * F = 0$. (Here $Z_0$ is the vacuum impedance, which has the value $\mu_0=\varepsilon_0^{-1}$ in our relativistic units in which $c=1$.
With our conventions, our conjugate momentum field ${\cal P}$ turns out to be the negative of $H$ which was introduced earlier in (\ref{maxJ}).) With $Z = \partial_t$ and the decomposition $A = (-\phi, A_k)$ we find that $i_Z F = i_Z dA = \pounds_Z A - d i_Z A$ corresponds to $F_{0k} = {\dot A}{}_k + \partial_k \phi = -E_k$. The magnetic field strength is $F_{ij} := \partial_i A_j - \partial_j A_i =: \epsilon_{ijk} B^k$. Hence ${\cal P}_{0i} = - Z_0^{-1} * F_{0i} = -\mu_0^{-1}B_i$, and ${\cal P}_{ij} = -\lambda_0 * F_{ij} = -\varepsilon_0 \epsilon_{ijk} E^k$. The natural reference choice is $\bar A = 0 = \bar {\cal P}$.

The Hamiltonian $3$-form is
\begin{equation}
{\cal H}^{\rm EM}(Z) = - i_Z A \, d {\cal P} - d A \wedge i_Z {\cal P} + i_Z \left( \frac{1}{2}Z_0 * {\cal P} \wedge {\cal P} \right) + d{\cal B}^{\rm EM}.
\end{equation}
In the usual tensor index notation the volume density part has the form
\begin{equation}
{\cal H}^{\rm EM} =    \phi \partial_k \pi^k + \frac12 (\partial_i A_j - \partial_jA_i) \epsilon^{ijk}H_k + \frac1{2 \varepsilon_0} \pi^k \pi_k -\mu_0 \frac12 H^k H_k ,
\end{equation}
where the momentum conjugate to the 3-vector potential is given the name $\pi^k$ (it works out to have in the usual terminology the value $-\varepsilon_0 E^k$, i.e., $-D^k$).
By varying $H^k$ one obtains $\mu_0H^k = \frac12\epsilon^{kij}(\partial_i A_j - \partial_j A_i)=B^k$, a 2nd class constraint that could be used to eliminate the magnetic field, then the Hamiltonian volume density would
correspond to the familiar energy density
$  \frac12 (\varepsilon_0 E^2 + \mu_0 B^2)$ plus a gauge generating term, $ \phi \partial_k  \pi^k $, which vanishes ``on shell'';
 the scalar potential in this term acts as a Lagrange multiplier to enforce the (first class) Gauss constraint $\partial_k D^k = 0$.
%%%jmn323 3c69

Let us just consider two boundary term choices, namely our preferred choice with vanishing boundary term, and the above Hamiltonian $3$-form with the boundary term
\begin{equation}
{\cal B}^{\rm EM} = i_Z A \, {\cal P} =  -\phi \pi^k dS_k. \label{can}
\end{equation}
These two are actually both well known physically, the former corresponds to the energy density from the gauge invariant energy-momentum tensor~(\ref{HilbertMax}), and the latter is the energy density of the electromagnetic canonical energy-momentum tensor~(\ref{canMax}).
Here our interest is not in the field equations but in the total differential term which, upon integration, becomes a boundary term indicating the boundary condition. Briefly: for the choice with vanishing Hamiltonian boundary term, the total derivative term in the variation of the Hamiltonian
is
\begin{equation}
-d(i_ZA \delta{\cal P}+\delta A\wedge i_Z{\cal P})\simeq \partial_k (\phi \delta \pi^k-\epsilon^{kij}\delta A_i H_j),
\end{equation}
which tells us that one should hold fixed on the boundary of the dynamical region the normal component of the electric field and the surface parallel components of the vector potential (the gauge independent part of which determines the normal component of the magnetic field).
On the other hand, for the Hamiltonian including the boundary term~(\ref{can}), one finds that the total  differential in the variation of the Hamiltonian is now
\begin{equation}
d(i_Z\delta A {\cal P}-\delta A\wedge i_Z{\cal P})\simeq -\partial_k (\delta \phi \pi^k+\epsilon^{kij}\delta A_i H_j).
\end{equation}
 This is the same boundary condition in the vector potential/magnetic sector but now for the electric sector one should instead hold fixed on the boundary of the region the scalar potential.

The physical meaning of such boundary conditions are well known. Fixing the normal component of the electric field on the boundary corresponds to fixing the surface charge density. An instructive physical example is a parallel plate capacitor. One can use a battery to charge up a capacitor with a moveable dielectric. Disconnect the battery and measure the work needed to remove/insert the dielectric (the potential varies but the charge is fixed, no current or power flow).
Alternatively leave the battery connected and measure the work needed to displace the dielectric---now the potential is fixed but the charge varies, so current and hence power flows. The respective boundary terms in the variation of the Hamiltonian are $\phi \delta \pi^k dS_k$ and $-\delta\phi \pi^k dS_k$. Both boundary condition choices are physically meaningful corresponding to real situations.
%Nevertheless there is a preferred choice.

Nevertheless for electrodynamics one expression stands out, the one with vanishing boundary term. This choice is the only one in which the value of the Hamiltonian is {\it gauge invariant}. Moreover, this is the only {\it non-negative} Hamiltonian density. Consequently the associated energy has a lower bound and the system has a natural vacuum or ground state: zero energy for vanishing fields. The value of the Hamiltonian with this boundary term can be interpreted as the {\it internal energy}, whereas the other expressions can be regarded as including some additional energy on the boundary of the system associated with maintaining the boundary condition. The associated electromagnetic energy flux expression from our formula~(\ref{preferredEnergyFlux}) reduces to just the usual Poynting energy flux:
\begin{eqnarray}
\pounds_Z {\cal H}^{\rm EM} &=&
d\left[ - i_Z A \wedge (di_Z+i_Zd) {\cal P}  -  (di_Z+i_Zd) A \wedge i_Z  {\cal P}  \right]
\nonumber\\
&\equiv& - d \left( i_Z F \wedge i_Z {\cal P} \right) =  d \left(- E_i H_j \; dx^i \wedge dx^j \right).
\end{eqnarray}
Clearly this choice, associated with fixing the normal components of the electric and magnetic fields  on the boundary, is preferred; it is the one suitable for most physical applications. It gives the usual energy density and Poynting energy flux. Similarly, for all other fields---\emph{except for dynamic spacetime geometry}---there is available a standard Hamiltonian (the one with {\emph{vanishing boundary term contribution}}) associated with a certain preferred boundary condition.

%%%jmn323 3c71
%%%%%%%%%%%%%%%%%%%%%%%%%%%%%%%%%%%%%%%%%%%%%%%%%%%%%%%%%%%%%%%%%%%%%%
\section{Geometry: covariant differential formulation}%%%jmn ok
%%%%%%%%%%%%%%%%%%%%%%%%%%%%%%%%%%%%%%%%%%%%%%%%%%%%%%%%%%%%%%%%%%%%%%
\emph{In the discussion of our covariant Hamiltonian approach, up to this point (except as was specified for a couple of specific examples), there has been no need to make any restriction on the type of geometry for our manifold.  Here in this section we discuss the specific sort of dynamic spacetime geometry that we will consider and relate it to the gauge theory paradigm.}

%ccm118
The covariant Hamiltonian formulation can apply to general theories of dynamical geometry. Standard references for differential geometry are Kobayashi \& Nomizu\cite{KobNom} and Spivak\cite{Spivak}.

%General metric affine geometry: metric, curvature, torsion and nonmetricity

%Special cases: Riemann-Cartan, teleparallel, Riemannian

\subsection{Metric and connection}%%%jmn ok
%%%%%%%%%%%%%%%%%%%%%%%%%%%%%%%%%%%%%%%%%%%%%%%%%%%%%%%%%%%%%%%%%%%%%%
For the dynamical spacetimes that we consider there are two basic  %{\em independent\/}
geometric ideas: a {\em metric tensor\/} $g = g_{\mu\nu} \vartheta^\mu \otimes \vartheta^\nu$ (which determines {\em length\/} and {\em angle\/}), and {\em parallel\/}, associated with {\em parallel transport\/}, {\em covariant derivative\/} and {\em connection\/}, i.e., we consider \emph{metric-affine} geometry.

The metric gives the {\em causal\/} structure, arc length, area and volume. Furthermore it is used to raise and lower indexes (i.e., it determines a specific isomorphism between tangent and cotangent vectors). It also provides the paths of extremal length (geodesics).  For the 4D spacetimes of interest here, the metric has the Lorentz signature.  Given such a metric there is naturally defined an associated symmetry group, the group of local Lorentz transformations: $L\in SO(1,3)\Longrightarrow g(LX,LY)=g(X,Y)$.

For the other structure, let $e_\mu$ for $\mu = 0,1,2,3$ be a basis for spacetime vector fields. The {\em covariant differential\/} $\nabla$ of each basis vector is a vector valued one-form, hence some linear combination of the $e_\beta$'s with one-form coefficients:
\begin{equation}
\nabla e_\beta = e_\alpha \, \Gamma^\alpha{}_\beta, \qquad \Gamma^\alpha{}_\beta = \Gamma^\alpha{}_{\beta i} \, dx^i,
\end{equation}
called the {\it connection one-forms}.

The covariant differential of a vector field $V = e_\mu \, V^\mu$ is
\begin{equation}
\nabla V = \nabla (e_\mu \, V^\mu) = (\nabla e_\mu) V^\mu + e_\mu \nabla V^\mu = e_\nu \Gamma^\nu{}_\mu V^\mu + e_\mu d V^\mu =: e_\mu D V^\mu.
\end{equation}
Its components are determined by the operator $D$:
\begin{equation}
D V^\mu := d V^\mu + \Gamma^\mu{}_\nu \wedge V^\nu,
\end{equation}
which extends, as indicated, to vector valued forms.
%

%{Curvature}

The notation automatically antisymmetrizes:
\begin{eqnarray}
\nabla^2 V &=& \nabla^2 (e_\mu V^\mu) = \nabla (e_\mu D V^\mu) = e_\mu D^2 V^\mu
\nonumber\\
&=& e_\mu \left[ d ( d V^\mu + \Gamma^\mu{}_\nu \wedge V^\nu) + \Gamma^\mu{}_\lambda \wedge (d V^\lambda + \Gamma^\lambda{}_\sigma \wedge V^\sigma) \right]
\nonumber\\
&=& e_\mu (d \Gamma^\mu{}_\nu + \Gamma^\mu{}_\lambda \wedge \Gamma^\lambda{}_\nu) \wedge V^\nu = e_\mu R^\mu{}_\nu \wedge V^\nu,
\end{eqnarray}
where
\begin{equation}
R^\mu{}_\nu := d \Gamma^\mu{}_\nu + \Gamma^\mu{}_\lambda \wedge \Gamma^\lambda{}_\nu=\frac12 R^\mu{}_{\nu ij} dx^i \wedge dx^j. \label{curv2}
\end{equation}
is the {\it curvature 2-form}

Exterior covariant differential form notation treats some, but not all, indices as differential forms. Rather than work with the operator $\nabla$ on geometric objects we often find it more convenient to work with $D$ on their coefficients. The covariant differential $D$ can be extended to operate on a {\em tensor valued form\/} of any type, e.g.,
\begin{eqnarray}
D P^\alpha{}_\beta &=& d P^\alpha{}_\beta + \Gamma^\alpha{}_\gamma \wedge P^\gamma{}_\beta - \Gamma^\gamma{}_\beta \wedge P^\alpha{}_\gamma,
\\
D^2 P^\alpha{}_\beta &=& R^\alpha{}_\gamma \wedge P^\gamma{}_\beta - R^\gamma{}_\beta \wedge P^\alpha{}_\gamma.
\end{eqnarray}

Generically, arranging the components of a tensor valued form $\varphi$ as a row vector we get
\begin{eqnarray}
D \varphi &=& d \varphi + \Gamma^\alpha{}_\beta \wedge \varphi\sigma_\alpha{}^\beta,
\\
D^2 \varphi &=& R^\alpha{}_\beta \wedge\varphi \sigma_\alpha{}^\beta, \label{6.13}
\end{eqnarray}
for some appropriate representation matrix $\sigma_\alpha{}^\beta$.

%{torsion and Bianchi identity}

The special case of the vector whose components are the coframe one-forms $\vartheta^\alpha = e^\alpha{}_i dx^i$ yields the {\em torsion\/} $2$-form:
\begin{equation}
T^\alpha := D \vartheta^\alpha := d \vartheta^\alpha + \Gamma^\alpha{}_\beta \wedge \vartheta^\beta = \frac12 T^\alpha{}_{ij} dx^i \wedge dx^j. \label{tor2}
\end{equation}
On the coframe $\vartheta^\alpha$, $D^2$ gives an important special case of~(\ref{6.13}):
\begin{equation}
D T^\alpha = D^2 \vartheta^\alpha = R^\alpha{}_\beta \wedge \vartheta^\beta,
\end{equation}
which is known as the {\em first Bianchi identity\/}.

Applying $D$ to the curvature 2-form $R^\alpha{}_\beta$
%(\ref{6.7})
gives
\begin{eqnarray}
D R^\alpha{}_\beta &:=& d R^\alpha{}_\beta + \Gamma^\alpha{}_\gamma \wedge R^\gamma{}_\beta - \Gamma^\gamma{}_\beta \wedge R^\alpha{}_\gamma
\nonumber\\
&\, \, \equiv& d (d \Gamma^\alpha{}_\beta + \Gamma^\alpha{}_\gamma \wedge \Gamma^\gamma{}_\beta) + \Gamma^\alpha{}_\gamma \wedge (d \Gamma^\gamma{}_\beta + \Gamma^\gamma{}_\sigma \wedge \Gamma^\sigma{}_\beta)\nonumber\\&&\qquad\qquad\qquad\qquad\quad\quad\! - \Gamma^\gamma{}_\beta \wedge (d \Gamma^\alpha{}_\gamma + \Gamma^\alpha{}_\sigma \wedge \Gamma^\sigma{}_\gamma) \equiv 0,
\end{eqnarray}
by explicit calculation. This is the {\em second Bianchi identity\/}.

%{nonmetricity}

With $g_{\mu\nu}$ the metric tensor, $D g_{\mu\nu}$ defines the {\em non-metricity\/} $1$-form.
\begin{eqnarray}
D g_{\mu\nu} &:=& d g_{\mu\nu} - \Gamma^\lambda{}_\mu g_{\lambda\nu} - \Gamma^\lambda{}_\nu g_{\mu\lambda}. \label{6.9}
\end{eqnarray}
Correspondingly, we have
\begin{eqnarray}
D^2 g_{\mu\nu} &=& - R^\lambda{}_\mu  g_{\lambda\nu} - R^\lambda{}_\nu  g_{\mu\lambda}. \label{6.12}
\end{eqnarray}

\subsection{Riemann-Cartan geometry}%%%jmn ok
%%%%%%%%%%%%%%%%%%%%%%%%%%%%%%%%%%%%%%%%%%%%%%%%%%%%%%%%%%%%%%%%%%%%%%
Here we are interested in particular in the special case where the geometry can be regarded as a local gauge theory of an appropriate spacetime symmetry group. With due consideration given to the understanding of both the geometry of and physics in Minkowski spacetime, the appropriate choice for the symmetry group is the inhomogeneous Lorentz group, generally referred to as the Poincar\'e group. This is both the symmetry group for the spacetime of special relativity and the group used to classify elementary particles in terms of mass and spin. The group is a semidirect product of the translation group and the group of rotations/Lorentz boosts. The Noether conserved quantities associated with these global symmetries are energy-momentum and angular momentum/center-of-mass momentum. The type of spacetime geometry with local Poincar\'e symmetry is known as Riemann-Cartan geometry.

In Riemann-Cartan geometry the connection is assumed to be a priori {\it metric compatible}, $D g_{\mu\nu} = 0$, via~(\ref{6.12}) this gives $R_{\alpha\beta} = R_{[\alpha\beta]}$ (i.e., a Lorentz Lie algebra valued two-form).
For our purposes it is convenient to use the orthonormal frame gauge condition, %i.e.,
then the metric components are constant and $dg_{\mu\nu}=0$, so, via~(\ref{6.12}), this gives $\Gamma_{\alpha\beta} = \Gamma_{[\alpha\beta]}$ (i.e., a \emph{Lorentz Lie algebra valued one-form}). The geometry has in general nonvanishing torsion and curvature. The metric information is encoded in the orthonormal coframe $\vartheta^\mu$, which has local Lorentz gauge freedom.

\emph{Riemannian} geometry is a special case with vanishing torsion, $T^\alpha = 0$; such a connection is called symmetric. Then the geometry is given by the curvature, which is generally non-vanishing, so the parallel transport is path dependent. On the other hand another special case is \emph{teleparallel} geometry, which has a vanishing curvature $2$-form, $R^\alpha{}_\beta = 0$. This is referred to as \emph{flat}. Parallel transport is then path independent, nevertheless it is generally nontrivial---being dependent on the torsion, which is generally non-vanishing.

\subsection{Regarding geometry and gauge}%%%jmn ok
%%%%%%%%%%%%%%%%%%%%%%%%%%%%%%%%%%%%%%%%%%%%%%%%%%%%%%%%%%%%%%%%%%%%%%
Note the respective similarities in the form of the commutator of the gauge covariant derivatives for the flat space U(1) phase case ($\nabla_\mu \phi = \partial_\mu \phi + i e A_\mu \phi$) and the Yang-Mills case ($\nabla_\mu \psi = \partial_\mu \psi + i q A_\mu^p T_p \psi$) compared with the spacetime geometric case:
\begin{eqnarray}
\left[ \nabla_\mu, \nabla_\nu \right] \phi &=& i e F_{\mu\nu} \phi, \\
\left[ \nabla_\mu, \nabla_\nu \right] \psi &=& i q F^{p}{}_{\mu\nu} T_p \psi, \\
\left[ \nabla_\mu, \nabla_\nu \right] V^\alpha &=& R^\alpha{}_{\beta\mu\nu} V^\beta - T^\gamma{}_{\mu\nu} \nabla_\gamma V^\alpha.
\end{eqnarray}
Here $\nabla_\mu$ is the \emph{covariant derivative}; the latter relation is called the \emph{Ricci identity}. On the right hand side the respective gauge field strengths appear.  One can see that for spacetime the curvature is the Lorentz field strength, and the torsion is the spacetime ``translational'' field strength, associated with the generator of infinitesimal translations, the directional derivative.

Riemann-Cartan geometry is ideally suited to admit an interpretation as a local gauge theory of the symmetry group of Minkowski space, the Poincar\'e group. (In the standard Riemannian GR formulations the torsion is \emph{a priori} assumed to vanish, then gravity does not look much like a local spacetime symmetry gauge theory. Teleparallel geometry can be regarded as a gauge theory for translations.)

We see that, when suitably formulated, gravity has both Lorentz/rotational and translational ``vector potentials'' which are similar to those of the Maxwell/Yang-Mills theories.

\subsection{On the affine connection and gauge theory}%%%jmn ok %%% jmn 2014-07-30
%%%%%%%%%%%%%%%%%%%%%%%%%%%%%%%%%%%%%%%%%%%%%%%%%%%%%%%%%%%%%%%%%%%%%%
The ``connection'' one-forms for ``translations'' and ``Lorentz'' transformations can be packaged together in a way that offers some further insight into their essential similarities and differences.

The Poincar\'e transformations on Minkowski spacetime
\begin{equation}
V^{\alpha'} = \Lambda^{\alpha'}{}_\beta V^\beta + A^{\alpha'}
\end{equation}
can be conveniently represented in matrix form as
\begin{equation}
\left( \begin{array}{c} V' \\ 1 \end{array} \right) = \left( \begin{array}{cc} \Lambda & A \\ 0 & 1 \end{array} \right) \left( \begin{array}{c} V \\ 1 \end{array} \right) = \left( \begin{array}{c} \Lambda V + A \\ 1 \end{array} \right).
\end{equation}
Then the matrix product
\begin{equation}
\left( \begin{array}{cc} \Lambda_1 & A_1 \\ 0 & 1 \end{array} \right) \left( \begin{array}{cc} \Lambda_2 & A_2 \\ 0 & 1 \\ \end{array} \right) = \left( \begin{array}{cc} \Lambda_1 \Lambda_2 \ & \ \Lambda_1 A_2 + A_1 \\ 0 & 1 \end{array} \right)
\end{equation}
reflects the semi-direct product structure. This matrix representation for infinitesimal Poincar\'e transformations has the Lie algebra
\begin{equation}
\left[ \left( \begin{array}{cc} l_1 & a_1 \\ 0 & 0 \end{array} \right), \left( \begin{array}{cc} l_2 & a_2 \\ 0 & 0 \end{array} \right) \right] = \left( \begin{array}{cc} [l_1, l_2] \ & \ l_1 a_2 - l_2 a_1 \\ 0 & 0 \end{array} \right).
\end{equation}

Now a connection can be viewed as a Lie algebra valued one-form. The spacetime ``translation'' and ``Lorentz'' connections can thus be neatly packaged in terms of the above Poincar\'e Lie algebra matrix representation:
\begin{equation}
\omega := \left( \begin{array}{cc} \Gamma \ & \ \vartheta \\ 0 \ & 0 \end{array} \right).
\end{equation}
The associated ``curvature'' Lie algebra valued $2$-form
\begin{equation}
\Omega := d\omega + \omega \wedge \omega = \left( \begin{array}{cc} d\Gamma + \Gamma \wedge \Gamma \ & \ d\vartheta + \Gamma \wedge \vartheta \\ 0 & 0 \end{array} \right) = \left( \begin{array}{cc} R \ & \ T \\ 0 \ & 0 \end{array} \right)
\end{equation}
includes the spacetime curvature and torsion $2$-forms in one package. Furthermore, the Bianchi identity for this Poincar\'e Lie algebra curvature matrix,
\begin{eqnarray}
0 &=& D\Omega := d\Omega + \omega \wedge \Omega - \Omega \wedge \omega
\\
&=& \left( \begin{array}{cc} dR + \Gamma \wedge R - R \wedge \Gamma\ & \ dT + \Gamma \wedge T - R \wedge \vartheta \\ 0 & 0 \end{array} \right) = \left( \begin{array}{cc} D R \ & \ D T - R \wedge \vartheta \\ 0 & 0 \end{array} \right), \nonumber
\end{eqnarray}
unifies the first and second spacetime Bianchi identities.

This packaging shows similarities between the connection and coframe one-forms and the curvature and torsion $2$-forms, but also some clear differences inherited from the semi-direct product structure of the Poincar\'e group.
The gauge theories of Yang-Mills\cite{YangMills} and Utiyama\cite{Utiyama, Utiyama59} also have the $\Omega = d\omega + \omega \wedge \omega$ and $D\Omega \equiv 0$ form, but the groups do not have a semi-direct product structure.

Although we find this formulation quite helpful for seeing how the coframe plays the role of the ``vector potential for translations'', we will not use it below in our treatment of the PG.  For our purposes we consider local Poincar\'e transformations to be Lorentz transformations of the coframe plus local spacetime diffeomorphisms.

%%%%%%%%%%%%%%%%%%%%%%%%%%%%%%%%%%%%%%%%%%%%%%%%%%%%%%%%%%%%%%%%%%%%%%
\section{Variational principles for dynamic spacetime geometry}%%%jmn ok 3d80
%%%%%%%%%%%%%%%%%%%%%%%%%%%%%%%%%%%%%%%%%%%%%%%%%%%%%%%%%%%%%%%%%%%%%%
%
\emph{In this section we develop the second order variational principle for gravitating material and internal gauge fields along with their associated Noether currents and differential identities. The spacetime is assumed to have Riemann-Cartan geometry, i.e., we are considering the Poincar\'e gauge theory of gravity (PG).}

We are considering geometric gravity that can be regarded as a gauge theory for the Poincar\'e group. Several authors have considered such theories, see, e.g., Refs.~[\refcite{HHKN,Hehl1980, HayashiShirafuji, Mielke, Blagojevic2002, BlagojevicHehl}].

We wish to consider the conserved Noether currents and differential identities as well as the field equations for dynamic spacetime geometry and gauge interactions. Here in this section we first work with the usual 2nd order type Lagrangian, since for that case certain expressions take a simpler form and the arguments are more transparent. Having established these results we can then present more briefly the analogous first order version which is the basis for our covariant Hamiltonian expressions.

\subsection{The Lagrangian and its variation}
%%%%%%%%%%%%%%%%%%%%%%%%%%%%%%%%%%%%%%%%%%%%%%%%%%%%%%%%%%%%%%%%%%%%%%
Rather than begin with the Lagrangian $4$-form of the type\footnote{This is sufficiently general to include all the fundamental fields of the standard model, with $\varphi$ including the Higgs and the fermions and $A^p$ the $U(1)\times SU(2)\times SU(3)$ gauge vector potential.}
\begin{equation}
{\cal L} = {\cal L}(\varphi, \vartheta^\mu, \Gamma^\alpha{}_\beta, A^p; d \varphi, d\vartheta^\mu, d \Gamma^\alpha{}_\beta, d A^p),
\end{equation}
we take the more covariant form
\begin{equation} \label{7:2}
{\cal L} = {\cal L}(\varphi, \vartheta^\mu, \Gamma^\alpha{}_\beta, A^p; D \varphi, T^\mu, R^\alpha{}_\beta, F^p),
\end{equation}
which is no less general. Here $\varphi$ is a generic $f$-form source field with total covariant differential
(the factor ordering is suitable for a matrix representation with $\varphi$ as a row ``vector'')
\begin{equation}
D \varphi = d \varphi + \Gamma^\alpha{}_\beta \wedge \varphi \, \sigma_\alpha{}^\beta + A^p \wedge \varphi \, T_p.
\end{equation}
The torsion $2$-form, curvature $2$-form and gauge field strength were given earlier~(\ref{tor2}, \ref{curv2}, \ref{F2}).
The respective variations are
\begin{eqnarray}
\delta D \varphi &=& D \delta \varphi + \delta \Gamma^\alpha{}_\beta \wedge \varphi \, \sigma_\alpha{}^\beta + \delta A^p \wedge \varphi \, T_p,
\\
\delta T^\mu &=& D \delta \vartheta^\mu + \delta \Gamma^\mu{}_\nu \wedge \vartheta^\nu,
\\
\delta R^\alpha{}_\beta &=& d \delta \Gamma^\alpha{}_\beta + \delta \Gamma^\alpha{}_\lambda \wedge \Gamma^\lambda{}_\beta + \Gamma^\alpha{}_\lambda \wedge \delta \Gamma^\lambda{}_\beta = D \delta \Gamma^\alpha{}_\beta,
\\
\delta F^p &=& d \delta A^p + C^p{}_{qr} A^q \wedge \delta A^r = D \delta A^p.
\end{eqnarray}

The variation of the total Lagrangian $4$-form~(\ref{7:2}) is: %(dropping indices and wedges):
\begin{eqnarray}
\delta {\cal L} &=& \delta D \varphi \wedge \frac{\partial {\cal L}}{\partial D \varphi}
+ \delta T^\mu \wedge \frac{\partial {\cal L}}{\partial T^\mu} + \delta R^\alpha{}_\beta \wedge \frac{\partial {\cal L}}{\partial R^\alpha{}_\beta} + \delta F^p \wedge \frac{\partial {\cal L}}{\partial F^p}
\nonumber\\
&& + \delta \varphi \wedge \frac{\partial {\cal L}}{\partial \varphi}
+ \delta \vartheta^\mu \wedge \frac{\partial {\cal L}}{\partial \vartheta^\mu} + \delta \Gamma^\alpha{}_\beta \wedge \frac{\partial {\cal L}}{\partial \Gamma^\alpha{}_\beta} + \delta A^p \wedge \frac{\partial {\cal L}}{\partial A^p}
\\
&=& (D \delta \varphi + \delta \Gamma^\alpha{}_\beta \wedge \varphi \sigma_\alpha{}^\beta + \delta A^p \wedge \varphi T_p) \wedge \frac{\partial {\cal L}}{\partial D \varphi} + \delta \varphi \wedge \frac{\partial {\cal L}}{\partial \varphi}
\nonumber\\
&& + (D \delta \vartheta^\mu + \delta \Gamma^\mu{}_\nu \wedge \vartheta^\nu) \wedge \frac{\partial {\cal L}}{\partial T^\mu} + \delta \vartheta^\mu \wedge \frac{\partial {\cal L}}{\partial \vartheta^\mu}
\nonumber\\
&& + D \delta \Gamma^\alpha{}_\beta \wedge \frac{\partial {\cal L}}{\partial R^\alpha{}_\beta} + \delta \Gamma^\alpha{}_\beta \wedge \frac{\partial {\cal L}}{\partial \Gamma^\alpha{}_\beta}
\nonumber\\
&& + D \delta A^p \wedge \frac{\partial {\cal L}}{\partial F^p} + \delta A^p \wedge \frac{\partial {\cal L}}{\partial A^p}.
\end{eqnarray}
``Integrating by parts'' and rearranging gives
\begin{eqnarray} \label{7:16}
\delta {\cal L} &=& D \left( \delta \varphi \wedge \frac{\partial {\cal L}}{\partial D \varphi}
+ \delta \vartheta^\mu \wedge \frac{\partial {\cal L}}{\partial T^\mu} + \delta \Gamma^\alpha{}_\beta \wedge \frac{\partial {\cal L}}{\partial R^\alpha{}_\beta} + \delta A^p \wedge \frac{\partial {\cal L}}{\partial F^p} \right)
\nonumber\\
&& + \delta \varphi \wedge \left( - \varsigma D \frac{\partial {\cal L}}{\partial D \varphi} + \frac{\partial {\cal L}}{\partial \varphi} \right)
+ \delta \vartheta^\mu \wedge \left( D \frac{\partial {\cal L}}{\partial T^\mu} + \frac{\partial {\cal L}}{\partial \vartheta^\mu} \right)
\nonumber\\
&& + \delta \Gamma^\alpha{}_\beta \wedge \left( D \frac{\partial {\cal L}}{\partial R^\alpha{}_\beta}
 + \frac{\partial {\cal L}}{\partial \Gamma^\alpha{}_\beta}
  + \varphi \sigma_\alpha{}^\beta \wedge \frac{\partial {\cal L}}{\partial D \varphi} %\right.
%\nonumber\\
%
+ \vartheta^\beta \wedge \frac{\partial {\cal L}}{\partial T^\alpha} \right)
\nonumber\\
&& + \delta A^p \wedge \left( D \frac{\partial {\cal L}}{\partial F^p} + \frac{\partial {\cal L}}{\partial A^p} + \varphi T_p \wedge \frac{\partial {\cal L}}{\partial D \varphi} \right),\label{varL2O}
\end{eqnarray}
which yields the conjugate momenta and Euler-Lagrange variational derivatives according to the pattern:
\begin{eqnarray} \label{7:17}
\delta {\cal L} &=& d ( \delta \varphi \wedge p %
+ \delta \vartheta^\alpha \wedge \tau_\alpha + \delta \Gamma^\alpha{}_\beta \wedge \rho_\alpha{}^\beta + \delta A^p \wedge {\cal P}_p )
\nonumber\\
&& + \delta \varphi \wedge \frac{\delta {\cal L}}{\delta \varphi}
+ \delta \vartheta^\alpha \wedge \frac{\delta {\cal L}}{\delta \vartheta^\alpha} + \delta \Gamma^\alpha{}_\beta \wedge \frac{\delta {\cal L}}{\delta \Gamma^\alpha{}_\beta} + \delta A^p \wedge \frac{\delta {\cal L}}{\delta A^p}.
\end{eqnarray}
This is the key variational relation.  According to Hamilton's principle, the 2nd order field equations are the vanishing of the Euler-Lagrange expressions named in~(\ref{7:17}) and explicitly displayed in~(\ref{varL2O}).

\subsection{Local gauge symmetries, Noether currents and differential identities}
%%%%%%%%%%%%%%%%%%%%%%%%%%%%%%%%%%%%%%%%%%%%%%%%%%%%%%%%%%%%%%%%%%%%%%
The Lagrangian $4$-form $\cal L$~(\ref{7:2}) should be ``gauge'' invariant under the local space-time gauge transformations:
\begin{eqnarray} \label{7:18}
\Delta \varphi &=& l^\alpha{}_\beta \, \varphi \sigma_\alpha{}^\beta + \alpha^p \, \varphi T_p - \pounds_Z \varphi,
\\
\Delta \vartheta^\mu &=& l^\mu{}_\nu \, \vartheta^\nu - \pounds_Z \vartheta^\mu,
\\
\Delta \Gamma^\alpha{}_\beta &=& - D l^\alpha{}_\beta - \pounds_Z \Gamma^\alpha{}_\beta,
\\
\Delta A^p &=& - D \alpha^p - \pounds_Z A^p, \label{7:22}
\end{eqnarray}
where the 6 $l^\alpha{}_\beta$ control an infinitesimal Lorentz rotation of the space-time frame $\vartheta^\alpha$, the $\alpha^p$ control an ``internal'' gauge and $Z$ is a spacetime vector field which determines a ``local translation''. The Lie derivative $\pounds_Z$ is given by $d i_Z + i_Z d$ on the components of our fields. This action on the components and on the basis one-forms correctly represents the Lie derivative on geometric objects. Under~(\ref{7:18})--(\ref{7:22}) the Lagrangian $4$-form $\cal L$~(\ref{7:2}), if it depends on position only through the indicated fields, should change according to
\begin{equation} \label{7:23}
\Delta {\cal L} = - \pounds_Z {\cal L} = - d i_Z {\cal L},
\end{equation}
which happens to be a total differential because ${\cal L}$ is a $4$-form.  With the special variations given by~(\ref{7:18})--(\ref{7:23}) eq.~(\ref{7:17}) should be satisfied identically.

One may collect all the total differential terms on the l.h.s. of the identity giving:
\begin{equation} \label{7:24}
d {\cal J}(l^\alpha{}_\beta, \alpha^p, Z) \equiv \Delta \varphi \wedge \frac{\delta {\cal L}}{\delta \varphi}
+ \Delta \vartheta^\mu \wedge \frac{\delta {\cal L}}{\delta \vartheta^\mu} + \Delta \Gamma^\alpha{}_\beta \wedge \frac{\delta {\cal L}}{\delta \Gamma^\alpha{}_\beta} + \Delta A^p \wedge \frac{\delta {\cal L}}{\delta A^p},
\end{equation}
where
\begin{equation}
{\cal J}(l^\alpha{}_\beta, \alpha^p, Z) := - i_Z {\cal L} - \Delta \varphi \wedge \frac{\partial {\cal L}}{\partial D \varphi}
- \Delta \vartheta^\mu \wedge \frac{\partial {\cal L}}{\partial T^\mu} - \Delta \Gamma^\alpha{}_\beta \wedge \frac{\partial {\cal L}}{\partial R^\alpha{}_\beta} - \Delta A^p \wedge \frac{\partial {\cal L}}{\partial F^p},
\end{equation}
is the {\it generalized total current} $3$-form.
%%%jmn Here I have changed the name of this quantity. It IS NOT the Hamiltonian (that name is ok for the 1st order version of this quantity)
Using~(\ref{7:18})--(\ref{7:22}) we have in more detail (recall the canonical stress tensor~(\ref{3:canem})
and $E := \frac{\partial L}{\partial \dot q} \dot q - L$)
\begin{eqnarray}
{\cal J}(l^\alpha{}_\beta, \alpha^p, Z) &:=& - i_Z {\cal L} + (\pounds_Z \varphi - l^\alpha{}_\beta \varphi \sigma_\alpha{}^\beta - \alpha^p \varphi T_p) \wedge \frac{\partial {\cal L}}{\partial D \varphi}
\nonumber\\
&& + (\pounds_Z \vartheta^\alpha - l^\alpha{}_\beta \vartheta^\beta) \wedge \frac{\partial {\cal L}}{\partial T^\alpha}
%\nonumber\\
%&&
+ (\pounds_Z \Gamma^\alpha{}_\beta + D l^\alpha{}_\beta) \wedge \frac{\partial {\cal L}}{\partial R^\alpha{}_\beta}
\nonumber\\
&& + (\pounds_Z A^p + D \alpha^p) \wedge \frac{\partial {\cal L}}{\partial F^p}.
\end{eqnarray}

The second Noether theorem differential identities may be obtained from~(\ref{7:24}) by comparing the coefficients of $\alpha^p, l^\alpha{}_\beta, Z^\mu, d\alpha^p, dl^\alpha{}_\beta, dZ^\mu$ on both sides. The results will be covariant, {\em but\/} the computation will not be manifestly so---because of the Lie derivative terms. However, it so happens that the Lie derivative differs from a covariant operation only by a gauge transformation, as the following short computations reveal:
\begin{eqnarray}
\pounds_Z \varphi &:=& d i_Z \varphi + i_Z d \varphi \equiv D i_Z \varphi + i_Z D \varphi - \Gamma^\alpha{}_\beta(Z) \varphi \sigma_\alpha{}^\beta - A^p(Z) \varphi T_p,
\\
\pounds_Z \vartheta^\mu &:=& d i_Z \vartheta^\mu + i_Z d \vartheta^\mu \equiv D i_Z \vartheta^\mu + i_Z D \vartheta^\mu - \Gamma^\mu{}_\nu(Z) \vartheta^\nu,
\\
\pounds_Z \Gamma^\alpha{}_\beta &:=& d i_Z \Gamma^\alpha{}_\beta + i_Z d \Gamma^\alpha{}_\beta \equiv i_Z R^\alpha{}_\beta + D ( \Gamma^\alpha{}_\beta(Z)),
\\
\pounds_Z A^p &:=& d i_Z A^p + i_Z d A^p \equiv i_Z F^p + D(A^p(Z)),
\end{eqnarray}
(in the last two relations on the r.h.s., the $D$ is formally defined by treating  $A^p(Z)$ and $\Gamma^\alpha{}_\beta(Z)$ as tensors). Thus the translation vector field $Z$ induces a modification to the gauge parameters
\begin{equation}
l'^\alpha{}_\beta := l^\alpha{}_\beta + \Gamma^\alpha{}_\beta(Z), \qquad \alpha'^p := \alpha^p + A^p(Z),
\end{equation}
which effectively replaces the non-covariant $\pounds_Z$ by the ``covariant Lie derivative'' defined by
\begin{equation}
L_Z := D i_Z + i_Z D,
\end{equation}
on the ``normal'' fields $\varphi, \vartheta^\alpha$, and by
\begin{equation}
L_Z \Gamma^\alpha{}_\beta := i_Z R^\alpha{}_\beta, \qquad L_Z A^p := i_Z F^p,
\end{equation}
on the connection one-forms.

The gauge transformations~(\ref{7:18})--(\ref{7:22}) then take the manifestly covariant form:
\begin{eqnarray} \label{7:35}
\Delta \varphi &=& l'^\alpha{}_\beta \varphi \sigma_\alpha{}^\beta \!+\! \alpha'^p \varphi T_p \!-\! L_Z \varphi = l'^\alpha{}_\beta \varphi \sigma_\alpha{}^\beta \!+\! \alpha'^p \varphi T_p \!-\! D i_Z \varphi \!-\! i_Z D \varphi,
\\
%\Delta g &=& - l' g - l' g - L_Z g,
%\\
\Delta \vartheta^\mu &=& l'^\mu{}_\nu \vartheta^\nu + 0 - L_Z \vartheta^\mu = l'^\mu{}_\nu \vartheta^\nu + 0 - D Z^\mu - i_Z D \vartheta^\mu,
\\
\Delta \Gamma^\alpha{}_\beta &=& - D l'^\alpha{}_\beta + 0 - L_Z \Gamma^\alpha{}_\beta = - D l'^\alpha{}_\beta + 0 - i_Z R^\alpha{}_\beta, \label{7:38}
\\
\Delta A^p &=& 0 - D \alpha'^p - L_Z A^p = 0 - D \alpha'^p - i_Z F^p. \label{7:39}
\end{eqnarray}
If one specializes to matter fields which \emph{are not\/} form fields---which is the only kind of matter that we know of physically---then one can see here a striking pattern which supports our identification of the geometric gauge fields. From the r.h.s. expressions one finds that most of the fields transform \emph{algebraically} under gauge infinitesimal internal gauge transformations $\alpha^p$; the only field that transforms with $D\alpha^p$ is the internal connection one-form (a.k.a. gauge vector potential) $A^p$. Similarly, most of the fields transform algebraically in $l'^{\alpha}{}_\beta$; the only field which transforms according to the differential $D l'^{\alpha}{}_\beta$ is the spacetime connection one-form $\Gamma^\alpha{}_\beta$.
Moreover (in the ``physical'' $0$-form matter case) the only field having a differential---$D Z^\mu$---rather than an algebraic transformation formula under the infinitesimal spacetime displacement $Z^\mu$ is the coframe one-form $\vartheta^\mu$. This is one more way of seeing that the coframe one-form can be identified as the ``translational gauge vector potential''.

With the above reparameterization the generalized total current $3$-form takes the form:
\begin{eqnarray}
{\cal J}(l'^\alpha{}_\beta, \alpha'^p, Z) &:=& - i_Z {\cal L} + (L_Z \varphi - l'^\alpha{}_\beta \varphi \sigma_\alpha{}^\beta - \alpha'^p \varphi T_p) \wedge \frac{\partial {\cal L}}{\partial D \varphi}
%\nonumber\\
%&& %+ (L_Z g + l' g + l' g) \frac{\partial {\cal L}}{\partial D g}
\nonumber\\
&& + (L_Z \vartheta^\alpha - l'^\alpha{}_\beta \vartheta^\beta) \wedge \frac{\partial {\cal L}}{\partial T^\alpha}
%
%&&
+ (L_Z \Gamma^\alpha{}_\beta + D l'^\alpha{}_\beta) \wedge \frac{\partial {\cal L}}{\partial R^\alpha{}_\beta}
\nonumber\\
&& + (L_Z A^p + D \alpha'^p) \wedge \frac{\partial {\cal L}}{\partial F^p}.
\end{eqnarray}

To extract the differential identities from~(\ref{7:24}) it is best to write ${\cal J}(l'^\alpha{}_\beta, \alpha'^p, Z)$ as terms algebraically linear in $l'^\alpha{}_\beta, \alpha'^p, Z$ plus a total differential:
\begin{eqnarray} \label{7:41}
{\cal J}(l'^\alpha{}_\beta, \alpha'^p, Z) &:=& - i_Z {\cal L} + d \left( i_Z \varphi \wedge \frac{\partial {\cal L}}{\partial D \varphi} + i_Z \vartheta^\mu \wedge \frac{\partial {\cal L}}{\partial T^\mu} + l'^\alpha{}_\beta \frac{\partial {\cal L}}{\partial R^\alpha{}_\beta} + \alpha'^p \frac{\partial {\cal L}}{\partial F^p} \right)
\nonumber\\
&& + i_Z D \varphi \!\wedge\! \frac{\partial {\cal L}}{\partial D \varphi}
%+
+ i_Z D \vartheta^\mu \!\wedge\! \frac{\partial {\cal L}}{\partial T^\mu} + i_Z R^\alpha{}_\beta \!\wedge\! \frac{\partial {\cal L}}{\partial R^\alpha{}_\beta} + i_Z F^p \!\wedge\! \frac{\partial {\cal L}}{\partial F^p}
\nonumber\\
&& + \varsigma i_Z \varphi \wedge D \frac{\partial {\cal L}}{\partial D \varphi} - i_Z \vartheta^\mu \wedge D \frac{\partial {\cal L}}{\partial T^\mu}
\nonumber\\
&& - l'^\alpha{}_\beta \left( D \frac{\partial {\cal L}}{\partial R^\alpha{}_\beta} + \varphi \sigma_\alpha{}^\beta \wedge \frac{\partial {\cal L}}{\partial D
\varphi} %\right.
+ \vartheta^\beta \wedge \frac{\partial {\cal L}}{\partial T^\alpha} \right)
\nonumber\\
&& - \alpha'^p \left( D \frac{\partial {\cal L}}{\partial F^p} + \varphi T_p \wedge \frac{\partial {\cal L}}{\partial D \varphi} \right).
\end{eqnarray}
The total differential will later be related to the total energy-momentum. For now merely note that it does not contribute to the l.h.s. of~(\ref{7:24}) as $d^2 \equiv 0$.

\emph{Internal gauge symmetry.} With $Z = 0 = l'^\alpha{}_\beta$ the general result~(\ref{7:41}) reduces to the expression we considered earlier for internal gauge symmetry of the Yang-Mills type~(\ref{3:N2ident}), and we again obtain
just as before (recall the argument leading to~(\ref{3:connindep}) and (\ref{3:diffeid})):
\begin{equation}\label{7:42}
\frac{\partial {\cal L}}{\partial A^p} \equiv 0, \qquad D \frac{\delta {\cal L}}{\delta A^p} + \varphi T_p \wedge \frac{\delta {\cal L}}{\delta \varphi} \equiv 0.
\end{equation}

\emph{Local Lorentz gauge symmetry.} In the same fashion, with $Z = 0 = \alpha'^p$ we obtain from~(\ref{7:24}) and~(\ref{7:35})--(\ref{7:38})
\begin{eqnarray}
d{\cal J}(l'^\alpha{}_\beta, 0, 0) &:=& - d \left\{ l'^\alpha{}_\beta \left( \frac{\delta {\cal L}}{\delta \Gamma^\alpha{}_\beta} - \frac{\partial {\cal L}}{\partial \Gamma^\alpha{}_\beta} \right) \right\}
\nonumber\\
&\equiv& l'^\alpha{}_\beta \varphi \sigma_\alpha{}^\beta \wedge \frac{\delta {\cal L}}{\delta \varphi}
%+
%\nonumber\\
%&&
+ l'^\alpha{}_\beta \vartheta^\beta \wedge \frac{\delta {\cal L}}{\delta \vartheta^\alpha} + (- D l'^\alpha{}_\beta) \wedge \frac{\delta {\cal L}}{\delta \Gamma^\alpha{}_\beta}.
\end{eqnarray}
Equating the coefficients of $l'^\alpha{}_\beta$ and $D l'^\alpha{}_\beta$ (keeping in mind that $l'^\alpha{}_\beta$ is antisymmetric) leads to the algebraic and differential identities:
\begin{eqnarray} \label{7:45}
\frac{\partial {\cal L}}{\partial \Gamma^\alpha{}_\beta} &\equiv& 0,
\\
D \frac{\delta {\cal L}}{\delta \Gamma^{[\alpha\beta]}} + \varphi \sigma_{\alpha\beta} \wedge \frac{\delta {\cal L}}{\delta \varphi}
+ \vartheta_{[\beta} \wedge \frac{\delta {\cal L}}{\delta \vartheta^{\alpha]}} &\equiv& 0. \label{7:46}
\end{eqnarray}
Formally these two relations are quite similar to those found for the internal symmetries~(\ref{7:42}).  Consequently local Lorentz symmetry is in certain respects rather like an internal gauge symmetry.

The conditions $\partial {\cal L}/\partial A^p \equiv 0 \equiv \partial {\cal L}/\partial \Gamma^\alpha{}_\beta$ mean (as we expected) that how the internal and Lorentz gauge potentials can appear is quite restricted---i.e., only via the covariant derivative or the associated field strength.
With these conditions we note that the generalized current has the neat form
\begin{equation} \label{7:47}
{\cal J}(l'^\alpha{}_\beta, \alpha'^p, Z^\mu) = Z^\mu {\cal J}_\mu - l'^\alpha{}_\beta \frac{\delta {\cal L}}{\delta \Gamma^\alpha{}_\beta} - \alpha'^p \frac{\delta {\cal L}}{\delta A^p} + d {\cal B}.
\end{equation}

\emph{Local diffeomorphism invariance.} This was already considered for general theories in first order form in section~\ref{S:translational Noether current}. Here we consider local translations in second order form while distinguishing between the source, gauge and geometric variables. This case again has some similarities to the internal and Lorentz symmetries but also some striking differences.  With $l'^\alpha{}_\beta = 0 = \alpha'^p$ in~(\ref{7:24}) we have
\begin{eqnarray} \label{7:49}
d {\cal J}(0, 0, Z^\mu) &=& d (Z^\mu {\cal J}_\mu) = D Z^\mu \wedge {\cal J}_\mu + Z^\mu D {\cal J}_\mu
\nonumber\\
&\equiv& - ( D i_Z \varphi + i_Z D \varphi) \wedge \frac{\delta {\cal L}}{\delta \varphi}
- (D i_Z \vartheta^\mu + i_Z D \vartheta^\mu) \wedge \frac{\delta {\cal L}}{\delta \vartheta^\mu}
\nonumber\\
&& - i_Z R^\alpha{}_\beta \wedge \frac{\delta {\cal L}}{\delta \Gamma^\alpha{}_\beta} - i_Z F^p \wedge \frac{\delta {\cal L}}{\delta A^p}.
\end{eqnarray}
The coefficient of $Z^\mu$ on both sides of~(\ref{7:49}) is the differential identity associated with translation invariance (related to energy-momentum) while the coefficient of $D Z^\mu$ provides a new algebraic expression for ${\cal J}_\mu$ in terms of the Euler-Lagrange variations:
\begin{equation} \label{7:50}
{\cal J}_\mu \equiv - \varphi_\mu \wedge \frac{\delta {\cal L}}{\delta \varphi} - \frac{\delta {\cal L}}{\delta \vartheta^\mu}.
\end{equation}
For ordinary (i.e., $0$-form valued) matter fields the first term vanishes, leading to the especially simple and intuitively reasonable relation:
\begin{equation}
{\cal J}_\mu \equiv - \frac{\delta {\cal L}}{\delta \vartheta^\mu}.
\end{equation}
We emphasize that such a neat formula for the translational current $3$-form is only possible is one regards the coframe as a dynamical variable.  It is also noteworthy that this remarkable relation---which does not restrict how the translation gauge potential appears in the action---was obtained from the coefficient of $DZ^\mu$, whereas the corresponding relation in both the internal and local Lorentz cases put quite severe limits, viz.~(\ref{7:42}a),~(\ref{7:45}), on how those gauge potentials could appear.

The identity~(\ref{7:50}) permits ${\cal J}(l'^\alpha{}_\beta, \alpha'^p, Z)$~(\ref{7:47}) to be written in the nice symmetrical form
\begin{eqnarray} \label{7:51}
{\cal J}(l'^\alpha{}_\beta, \alpha'^p, Z) &\equiv& - i_Z \varphi \wedge \frac{\delta {\cal L}}{\delta \varphi} - i_Z \vartheta^\mu \frac{\delta {\cal L}}{\delta \vartheta^\mu} - l'^\alpha{}_\beta \frac{\delta {\cal L}}{\delta \Gamma^\alpha{}_\beta} - \alpha'^p \frac{\delta {\cal L}}{\delta A^p}
\\
&& + d \left( i_Z \varphi \wedge \frac{\partial {\cal L}}{\partial D \varphi} + i_Z \vartheta^\mu \frac{\partial {\cal L}} {\partial T^\mu} + l'^\alpha{}_\beta \frac{\partial {\cal L}}{\partial R^\alpha{}_\beta} + \alpha'^p \frac{\partial {\cal L}}{\partial F^p} \right). \nonumber
\end{eqnarray}

Restricting to the case where the source fields are $0$-forms, we have
\begin{eqnarray} \label{7:52}
{\cal J}(l'^\alpha{}_\beta, \alpha'^p, Z) &\equiv& - Z^\mu \frac{\delta {\cal L}}{\delta \vartheta^\mu} - l'^\alpha{}_\beta \frac{\delta {\cal L}}{\delta \Gamma^\alpha{}_\beta} - \alpha'^p \frac{\delta {\cal L}}{\delta A^p}
\\
&& + d \left( Z^\mu \frac{\partial {\cal L}} {\partial T^\mu} + l'^\alpha{}_\beta \frac{\partial {\cal L}}{\partial R^\alpha{}_\beta} + \alpha'^p \frac{\partial {\cal L}}{\partial F^p} \right). \nonumber
\end{eqnarray}
This result displays the purely gauge nature of the current---it is especially noteworthy that there is no explicit appearance of the source field or its field equation.
%%%jmn This , even though the Hamiltonian does generate its evolution.

Note that if we take the variational derivatives as field equations, the numerical value of  ${\cal J}$ will be entirely from the total differential term. When the $3$-form ${\cal J}$ is integrated over a 3-dimensional region this total differential becomes, via the fundamental boundary theorem, (\ref{boundarythm}), an integral over the 2-dimensional boundary of the region.  In other words, the value is \emph{quasi-local}.

 Our results here are an application of Noether's ideas.  As expected from her 2nd theorem, with a local symmetry the conserved current becomes a differential identity.  Furthermore, we also displayed here detailed results that exactly reflect her remarks about verifying and generalizing Hilbert's assertion regarding the lack of a proper energy-momentum.  Our generalized current expressions~(\ref{7:51}) and (\ref{7:52}) are linear combinations of Euler-Lagrange expressions plus a total differential.  They have the usual conserved current ambiguity: the total differential does not contribute to the conservation law, it can be adjusted without affecting the conservation property, nevertheless it affects the value of the associated conserved quantity.  The second order Lagrangian formalism has no way to control this ambiguity.

\subsection{Interpretation of the differential identities}%%%jmn419 3d84
%%%%%%%%%%%%%%%%%%%%%%%%%%%%%%%%%%%%%%%%%%%%%%%%%%%%%%%%%%%%%%%%%%%%%%
%
Let us now specialize and consider the customary ``minimal coupled'' decomposition:
\begin{eqnarray}
{\cal L}_{\rm total} &=& {\cal L}_{\theta\Gamma}( %g,
., \vartheta^\mu,.,.; ., T^\mu, R^\alpha{}_\beta, .) + {\cal L}_A( %g,
., \vartheta^\mu, ., .; ., ., ., F^p)
\nonumber\\
&& + {\cal L}_{\varphi}( \varphi, %g,
\vartheta^\mu, ., .; D \varphi, ., ., ., .).
\end{eqnarray}
Each of these separate Lagrangian pieces is a scalar valued $4$-form. So the Noether identities we have obtained can be applied to each piece. However, it must be kept in mind that for the separate pieces our variational derivatives are, in general, no longer field equations. There is, however, one exception:
\begin{equation}
\frac{\delta {\cal L}_{\varphi}}{\delta \varphi} = \frac{\delta {\cal L}_{\rm total}}{\delta \varphi},
\end{equation}
which vanishes ``on shell''. It should also be noted that for each separate piece of ${\cal L}_{\rm total}$
 many of the terms in the various identities~(\ref{7:42}, \ref{7:46}, \ref {7:49}, \ref{7:50}) will vanish trivially.
 Consider~(\ref{7:42}): for ${\cal L}_{\vartheta\Gamma}$ it is trivial; for ${\cal L}_A$ and ${\cal L}_{\varphi}$
it reduces to the results obtained earlier~(\ref{gaugeDI}).

Let us introduce some suitable names for certain expressions.  Specifically for the material source the \emph{energy-momentum} and \emph{spin density} 3-forms are defined respectively by
\begin{equation}
\mathfrak{T}^\varphi_\mu:=\frac{\partial{\cal L}_\varphi}{\partial\vartheta^\mu},\qquad
\mathfrak{S}_\varphi^{\beta\alpha}:=2\frac{\partial{\cal L}_\varphi}{\partial\Gamma_{\alpha\beta}},
\end{equation}
with an analogous expression defining $\mathfrak{T}^A_\mu$.

Then~(\ref{7:46}) for ${\cal L}_\varphi$ and ${\cal L}_A$ becomes, respectively,
\begin{eqnarray}
D\mathfrak{S}_\varphi^{\alpha\beta}+\vartheta^\alpha\wedge\mathfrak{T}_\varphi^\beta-\vartheta^\beta\wedge\mathfrak{T}_\varphi^\alpha
&\equiv&2\varphi\sigma^{\alpha\beta}\wedge\frac{\delta{\cal L}_\varphi}{\delta\varphi} =0\ (\hbox{on shell}), \label{angmom}\\
\vartheta^\alpha\wedge\mathfrak{T}_A^\beta-\vartheta^\beta\wedge\mathfrak{T}_A^\alpha&\equiv&0.
\end{eqnarray}
The physical interpretation of these concerns angular momentum conservation (or more precisely the exchange of angular momentum with the gravitational field).  The first term in~(\ref{angmom}) is the (covariant) divergence of the source spin density, the next two terms (the anti-symmetric part of the energy-momentum density) describe the change in ``orbital'' angular momentum.  The second relation is an identity which shows that gauge fields have symmetric energy-momentum densities and trivial angular momentum conservation relations.
 For ${\cal L}_{\vartheta\Gamma}$~(\ref{7:46}) reduces to the identity
\begin{equation}
-R^\gamma{}_\alpha\wedge\frac{\partial{\cal L}_{\vartheta\Gamma}}{\partial\Gamma_{\gamma\beta}}
-R^\gamma{}_\beta\wedge\frac{\partial{\cal L}_{\vartheta\Gamma}}{\partial\Gamma_{\alpha\gamma}}
+\vartheta_{[\beta}\wedge\left(D\frac{\partial{\cal L}_{\vartheta\Gamma}}{\partial T^{\alpha]}}
+\frac{\partial{\cal L}_{\vartheta\Gamma}}{\partial \vartheta^{\alpha]}}\right)\equiv0,
\end{equation}
which is satisfied iff ${\cal L}_{\vartheta\Gamma}$ is a local Lorentz scalar.

Now let us consider the differential identity part of~(\ref{7:49}).  For a 0-form material source field it takes  the form
\begin{equation}
D\mathfrak{T}^\varphi_\mu-i_{e_\mu} T^\nu\wedge\mathfrak{T}^\varphi_\nu-\frac12 i_{e_\mu}R^{\alpha\beta}\wedge\mathfrak{S}_{\beta\alpha}\equiv D_\mu \varphi \frac{\delta{\cal L}_\varphi}{\delta\varphi}=0\ (\hbox{on shell}).
\end{equation}
When the source field equation is satisfied this relation describes the exchange of material energy-momentum with the gravitational field.
The analogous expression for the gauge field Lagrangian is a little simpler:
\begin{equation}
D\mathfrak{T}^A_\mu-i_{e_\mu} T^\nu\wedge\mathfrak{T}^A_\nu+ i_{e_\mu} F^p\wedge D\frac{\partial{\cal L}_A}{\partial F^p}\equiv0
\end{equation}
the interpretation is similar.

Finally, for the differential identity of~(\ref{7:49}) applied to ${\cal L}_{\vartheta\Gamma}$, after some straightforward calculation, we have the identity
\begin{equation}
D\frac{\partial{\cal L}_{\vartheta\Gamma}}{\partial\vartheta^\mu}- i_{e_\mu}T^\nu\wedge\left(D\frac{\partial{\cal L}_{\vartheta\Gamma}}{\partial T^\nu}+
\frac{\partial{\cal L}_{\vartheta\Gamma}}{\partial \vartheta^\nu}\right)
-i_{e_\mu}R^{\alpha\beta}\wedge D\frac{\partial{\cal L}_{\vartheta\Gamma}}{\partial R^{\alpha\beta}}
-DT^\alpha\wedge i_{e_\mu}\frac{\partial{\cal L}_{\vartheta\Gamma}}{\partial T^\alpha}\equiv0,
\end{equation}
which is satisfied if ${\cal L}_{\vartheta\Gamma}$ is a scalar valued 4-form constructed out of
the coframe, torsion and curvature.
This is the PG identity that plays a role  analogous to that of the contracted Bianchi identity in GR.

To obtain these detailed results we used the definition of the various Euler-Lagrange expressions  given in~(\ref{7:16},\ref{7:17}).

%%%%%%%%%%%%%%%%%%%%%%%%%%%%%%%%%%%%%%%%%%%%%%%%%%%%%%%%%%%%%%%%%%%%%%
%\section{First order form for Poincar\'e gauge theory}%%%jmn ok
\section{First order form and the Hamiltonian}
%%%%%%%%%%%%%%%%%%%%%%%%%%%%%%%%%%%%%%%%%%%%%%%%%%%%%%%%%%%%%%%%%%%%%%
\emph{Here for our general PG with generic matter and gauge sources we briefly present the first order form along with the associated Noether currents and differential identities and then the associated covariant Hamiltonian including our preferred boundary term which yields our quasi-local quantities.}

\subsection{First-order Lagrangian and local gauge symmetries}
%%%%%%%%%%%%%%%%%%%%%%%%%%%%%%%%%%%%%%%%%%%%%%%%%%%%%%%%%%%%%%%%%%%%%%
For certain purposes we find a first order formulation convenient. The {\em first order form\/} of our variational principle for geometry and gauge is
\begin{eqnarray} \label{7:85}
{\cal L}^{1\text{st}} &=& D \varphi \wedge p + D \vartheta^\alpha \wedge \tau_\alpha + R^{\alpha\beta} \wedge \rho_{\alpha\beta} + F^p \wedge {\cal P}_p
\nonumber\\
&& - \Lambda(  \varphi, \vartheta^\alpha, \Gamma^{\alpha\beta}, A^p; p, \tau_\alpha, \rho_{\alpha\beta}, {\cal P}_p),
\end{eqnarray}
with $\Gamma^{\alpha\beta}$, $R^{\alpha\beta}$, and $\rho_{\alpha\beta}$ being \emph{a priori\/} antisymmetric.
The variation
takes the pattern
\begin{eqnarray} \label{7:86}
\delta {\cal L}^{1\text{st}} &=:& d (\delta \varphi \wedge p + \delta \vartheta^\alpha \wedge \tau_\alpha + \delta \Gamma^{\alpha\beta} \wedge \rho_{\alpha\beta} + \delta A^p \wedge{\cal P}_p)
\nonumber\\
&& + \delta \varphi \wedge \frac{\delta {\cal L}^{1\text{st}}}{\delta \varphi} + \delta \vartheta^\alpha \wedge \frac{\delta {\cal L}^{1\text{st}}}{\delta \vartheta^\alpha} + \delta \Gamma^{\alpha\beta} \wedge \frac{\delta {\cal L}^{1\text{st}}}{\delta \Gamma^{\alpha\beta}} + \delta A^p \wedge \frac{\delta {\cal L}^{1\text{st}}}{\delta A^p}
\nonumber\\
&& + \frac{\delta {\cal L}^{1\text{st}}}{\delta p} \wedge \delta p + \frac{\delta {\cal L}^{1\text{st}}}{\delta \tau_\alpha} \wedge \delta \tau^\alpha + \frac{\delta {\cal L}^{1\text{st}}}{\delta \rho_{\alpha\beta}} \wedge \delta \rho_{\alpha\beta} + \frac{\delta {\cal L}^{1\text{st}}}{\delta {\cal P}_p} \wedge \delta {\cal P}_p,
\end{eqnarray}
where
\begin{eqnarray}
\frac{\delta {\cal L}^{1\text{st}}}{\delta \varphi} = - \varsigma D p - \frac{\partial \Lambda}{\partial \varphi}, &\qquad& \frac{\delta {\cal L}^{1\text{st}}}{\delta p} = D \varphi - \frac{\partial \Lambda}{\partial p},
\\
\frac{\delta {\cal L}^{1\text{st}}}{\delta \vartheta^\alpha} = D \tau_\alpha - \frac{\partial \Lambda}{\partial \vartheta^\alpha}, &\qquad& \frac{\delta {\cal L}^{1\text{st}}}{\delta \tau_\alpha} = D \vartheta^\alpha - \frac{\partial \Lambda}{\partial \tau_\alpha},
\\
\frac{\delta {\cal L}^{1\text{st}}}{\delta \Gamma^{\alpha\beta}} = D \rho_{\alpha\beta} - \frac{\partial \Lambda}{\partial \Gamma^{\alpha\beta}} + \varphi \sigma_{\alpha\beta} p + \vartheta_{[\beta} \wedge \tau_{\alpha]}, &\qquad& \frac{\delta {\cal L}^{1\text{st}}}{\delta \rho_{\alpha\beta}} = R^{\alpha\beta} - \frac{\partial \Lambda}{\partial \rho_{\alpha\beta}},
\\
\frac{\delta {\cal L}^{1\text{st}}}{\delta A^p} = D {\cal P}_p - \frac{\partial \Lambda}{\partial A^p} + \varphi T_p p,  &\qquad& \frac{\delta {\cal L}^{1\text{st}}}{\delta {\cal P}_p} = F^p - \frac{\partial \Lambda}{\partial {\cal P}_p}.
\end{eqnarray}
It is instructive to compare these (and subsequent) relations with the corresponding ones in the previous section.  Here we have twice as many fields and thus twice as many Euler-Lagrange expressions, but, on the other hand, the Euler-Lagrange expressions are all linear and much simpler.

The first order formulation is convenient for imposing certain {\it constraints}. In particular in order to impose one of the conditions
\begin{eqnarray}
R^\alpha{}_\beta \equiv 0 &\qquad& \hbox{teleparallel connection},
\\
T^\alpha \equiv 0 &\qquad& \hbox{symmetric connection},
\end{eqnarray}
in the first order formalism we need merely take the potential $\Lambda$ to be independent of the corresponding conjugate momenta. The momentum then functions as a Lagrange multiplier which imposes the constraint. The related ``coordinate'' equation then loses its dynamical significance and instead becomes a relation for determining the multiplier.

\subsection{Generalized Hamiltonian and differential identities}
%%%%%%%%%%%%%%%%%%%%%%%%%%%%%%%%%%%%%%%%%%%%%%%%%%%%%%%%%%%%%%%%%%%%%%
To obtain the Noether differential identities in this mode we need, in addition to~(\ref{7:35})--(\ref{7:39}), the gauge transformations of the conjugate momenta:
\begin{eqnarray}
\Delta p &=& -l'^{\alpha\beta} \sigma_\alpha{}_\beta p - \alpha'^p T_p p - L_Z p,
\\
%\\
\Delta \tau_\beta &=& -l'^\alpha{}_\beta \tau_\alpha - L_Z \tau_\beta,
\\
\Delta \rho_\alpha{}^\beta &=& l'^\beta{}_\gamma \rho_\alpha{}^\gamma - l'^\gamma{}_\alpha \rho_\gamma{}^\beta - L_Z \rho_\alpha{}^\beta,
\\
\Delta {\cal P}_q &=& - {\cal P}_p C^p{}_{qr} \, \alpha'^r - L_Z {\cal P}_q,
\end{eqnarray}
where $L_Z = D i_Z + i_Z D$. These results were deduced from relations like
\begin{equation} \label{7:71}
\Delta (p \wedge D \varphi) = \Delta p \wedge D \varphi + p \wedge \Delta (D \varphi),
\end{equation}
using the fact that $p \wedge D\varphi$ is a scalar valued $4$-form.

%:
As before $\Delta {\cal L}$ is a total differential: $- d i_Z {\cal L}$. Taking the total differential terms to the l.h.s. in~(\ref{7:86}) gives
\begin{eqnarray} \label{7:5.12}
&& d{\cal H}(l'^\alpha{}_\beta, \alpha'^p, Z)
\nonumber\\
&&\qquad\equiv \Delta \varphi \wedge {\frac{\delta {\cal L}}{\delta \varphi}}^{1\text{st}} + \Delta \vartheta^\alpha \wedge \frac{\delta {\cal L}^{1\text{st}}}{\delta \vartheta^\alpha} + \Delta \Gamma^{\alpha\beta} \wedge \frac{\delta {\cal L}^{1\text{st}}}{\delta \Gamma_{\alpha\beta}} + \Delta A^p \wedge \frac{\delta {\cal L}^{1\text{st}}}{\delta A^p}
\nonumber\\
&&\qquad\quad + {\frac{\delta {\cal L}}{\delta p}}^{1\text{st}} \wedge \Delta p + \frac{\delta {\cal L}^{1\text{st}}}{\delta \tau_\alpha} \wedge \Delta \tau_\alpha +
\frac{\delta {\cal L}^{1\text{st}}}{\delta \rho_{\alpha\beta}} \wedge \Delta \rho_{\alpha\beta} + \frac{\delta {\cal L}^{1\text{st}}}{\delta {\cal P}_p} \wedge \Delta {\cal P}_p,
\end{eqnarray}
where the first order ``generalized current'' is
\begin{eqnarray}
&& {\cal H}(l'^\alpha{}_\beta, \alpha'^p, Z)
\nonumber\\
&&\quad\quad:=  - i_Z {\cal L}^{1\text{st}} - \Delta \varphi \wedge p - \Delta \vartheta^\alpha \wedge \tau_\alpha - \Delta \Gamma^{\alpha\beta} \wedge \rho_{\alpha\beta} - \Delta A^p \wedge {\cal P}_p
\\
&&\quad\quad= i_Z \Lambda - i_Z ( D \varphi \wedge p + D \vartheta^\alpha \wedge \tau_\alpha + R^{\alpha\beta} \wedge \rho_{\alpha\beta} + F^p \wedge {\cal P}_p )
\nonumber\\
&&\quad\qquad + (L_Z \varphi - l'^\alpha{}_\beta \varphi \sigma_\alpha{}^\beta - \alpha'^p \varphi T_p) \wedge p + (L_Z \vartheta^\alpha - l'^\alpha{}_\beta \vartheta^\beta) \wedge \tau_\alpha
\nonumber\\
&&\quad\qquad + (L_Z \Gamma^{\alpha\beta} + D l'^{\alpha\beta}) \wedge \rho_{\alpha\beta} + (L_Z A^p + D \alpha'^p) \wedge {\cal P}_p
\\
&&\quad\quad= i_Z \Lambda + \varsigma D \varphi \wedge i_Z p - D \vartheta^\alpha \wedge i_Z \tau_\alpha - R^{\alpha\beta} \wedge i_Z \rho_{\alpha\beta} - F^p \wedge i_Z {\cal P}_p
\nonumber\\
&&\quad\qquad + (D i_Z \varphi - l'^\alpha{}_\beta \varphi \sigma_\alpha{}^\beta - \alpha'^p \varphi T_p) \wedge p + ( D i_Z \vartheta^\alpha - l'^\alpha{}_\beta \vartheta^\beta ) \wedge \tau_\alpha
\nonumber\\
&&\quad\qquad + D l'^{\alpha\beta} \wedge \rho_{\alpha\beta} + D \alpha'^p \wedge {\cal P}_p
\\
&&\quad\quad= i_Z \Lambda + \varsigma D \varphi \wedge i_Z p - D \vartheta^\alpha \wedge i_Z \tau_\alpha - R^{\alpha\beta} \wedge i_Z \rho_{\alpha\beta} - F^p \wedge i_Z {\cal P}_p
\nonumber\\
&&\quad\qquad + \varsigma i_Z \varphi \wedge D p - i_Z \vartheta^\alpha D \tau_\alpha
\nonumber\\
&&\quad\qquad - l'^{\alpha\beta} ( D \rho_{\alpha\beta} + \varphi \sigma_{\alpha\beta} \wedge p + \vartheta_{[\beta} \wedge \tau_{\alpha]} ) - \alpha'^p (D {\cal P}_p + \varphi T_p \wedge p)
\nonumber\\
&&\quad\qquad + D (i_Z \varphi \wedge p + i_Z \vartheta^\alpha \tau_\alpha + l'^{\alpha\beta} \rho_{\alpha\beta} + \alpha'^p {\cal P}_p). \label{7:5.16}
\end{eqnarray}
This expression is not just a Noether current, it is, as we already justified quite generally in subsection~\ref{SS:Ham form}, \emph{the (generalized) Hamiltonian 3-form}, i.e., the canonical generator for internal, local Lorentz \emph{and} local spacetime displacements (which includes the time evolution for any choice of time).  Nevertheless, like the  2nd order current, one still has the various differential identities.

\emph{Local internal gauge symmetry.} Equating coefficients of $\alpha'^p$ and $D \alpha'^p$ in~(\ref{7:5.12}) gives the algebraic and differential identities:
\begin{equation}
\frac{\partial \Lambda}{\partial A^p} \equiv 0, \qquad - D \frac{\delta {\cal L}^{1\text{st}}}{\delta A^q} \equiv \varphi T_q \wedge \frac{\delta {\cal L}^{1\text{st}}}{\delta \varphi} - \frac{\delta {\cal L}^{1\text{st}}}{\delta p} \wedge T_q p - \frac{\delta {\cal L}^{1\text{st}}}{\delta {\cal P}^p} \wedge {\cal P}_r C^r{}_{pq},
\end{equation}
which can be compared with (\ref{7:42}). The significance is the same, but now the r.h.s. includes also Euler-Lagrange variations w.r.t. momentum variables.  If these equations are imposed, the expression has the same form as~(\ref{7:42}).

\emph{Local Lorentz symmetry.} Equating coefficients of $l'^{\alpha\beta}$ and $D l'^{\alpha\beta}$ gives the algebraic and differential identities:
\begin{eqnarray} \label{7:5.18}
\frac{\partial \Lambda}{\partial \Gamma^{\alpha\beta}} &\equiv& 0,
\\
- D \frac{\delta {\cal L}^{1\text{st}}}{\delta \Gamma^{\alpha\beta}} &\equiv& \varphi \sigma_{[\alpha\beta]} \wedge \frac{\delta {\cal L}^{1\text{st}}}{\delta \varphi} + \vartheta_{[\beta} \wedge \frac{\delta {\cal L}^{1\text{st}}}{\delta \vartheta^{\alpha]}} - \frac{\delta {\cal L}^{1\text{st}}}{\delta p} \wedge \sigma_{[\alpha\beta]} p
\nonumber\\
&& + \frac{\delta {\cal L}^{1\text{st}}}{\delta \tau^{[\alpha}} \wedge \tau_{\beta]} + \frac{\delta {\cal L}^{1\text{st}}}{\delta \rho_\gamma{}^\beta} \wedge \rho_{\alpha\gamma} - \frac{\delta {\cal L}^{1\text{st}}}{\delta \rho_\gamma{}^\alpha} \wedge \rho_{\beta\gamma},
\end{eqnarray}
which should be compared with~(\ref{7:45}) and (\ref{7:46}). The significance is the same, but now the r.h.s. of the latter includes also several Euler-Lagrange variations w.r.t. momentum variables.  If these relations are imposed the expression has the same form as~(\ref{7:46}).

\emph{Local translation symmetry.} With the decomposition ${\cal H}(0, 0, Z^\mu) = Z^\mu {\cal H}_\mu + d B$,
\begin{eqnarray}
&&  D{\cal H}(0, 0, Z^\mu)
\nonumber\\
&&\quad= D Z^\mu \wedge {\cal H}_\mu + Z^\mu D {\cal H}_\mu
\nonumber\\
&&\quad\equiv - L_Z \varphi \wedge \frac{\delta {\cal L}^{1\text{st}}}{\delta \varphi} - L_Z \vartheta^\alpha \frac{\delta {\cal L}^{1\text{st}}}{\delta \vartheta^\alpha} - L_Z \Gamma^{\alpha\beta} \wedge \frac{\delta{\cal L}^{1\text{st}}}{\delta \Gamma^{\alpha\beta}} - L_Z A^p \wedge \frac{\delta {\cal L}^{1\text{st}}}{\delta A^p}
\nonumber\\
&&\qquad -\! \frac{\delta {\cal L}^{1\text{st}}}{\delta p} \wedge L_Z p \!-\! \frac{\delta {\cal L}^{1\text{st}}}{\delta \tau_\alpha} \wedge L_Z \tau_\alpha \!-\! \frac{\delta {\cal L}^{1\text{st}}}{\delta \rho_{\alpha\beta}} \wedge L_Z \rho_{\alpha\beta} \!-\! \frac{\delta {\cal L}^{1\text{st}}}{\delta {\cal P}_p} \!\wedge\! L_Z {\cal P}_p.
\end{eqnarray}
 From the coefficient of $Z^\mu$ we obtain a differential identity involving ${\cal H}_\mu$; this relation includes the conservation of energy-momentum.

We also get a new algebraic identity giving ${\cal H}_\mu$ in terms of variational derivatives: (compare~(\ref{7:50})):
\begin{equation} \label{7:5.20}
{\cal H}_\mu \equiv - \varphi_\mu \wedge \frac{\delta {\cal L}^{1\text{st}}}{\delta \varphi} - \frac{\delta {\cal L}^{1\text{st}}}{\delta \vartheta^\mu} + \varsigma \frac{\delta {\cal L}^{1\text{st}}}{\delta p} \wedge p_\mu - \frac{\delta {\cal L}^{1\text{st}}}{\delta \tau_\alpha} \wedge \tau_{\alpha\mu} - \frac{\delta {\cal L}^{1\text{st}}}{\delta \rho_{\alpha\beta}} \wedge \rho_{\alpha\beta\mu} - \frac{\delta {\cal L}^{1\text{st}}}{\delta {\cal P}_p} \wedge {\cal P}_{p\mu}.
\end{equation}
When the momentum relations are imposed, this reduces to the corresponding expression for the second order Noether translational current~(\ref{7:50}).
Inserting this new result into~(\ref{7:5.16}) gives an expression for the Hamiltonian $3$-form in terms of
 variational coefficients (which has the same form as (\ref{7:51}) if the momentum relations are imposed)
\begin{eqnarray} \label{7:5.21}
&& {\cal H}(l'^{\alpha\beta}, \alpha'^p, Z^\mu)
\nonumber\\
&&\qquad\equiv - i_Z \varphi \wedge \frac{\delta {\cal L}^{1\text{st}}}{\delta \varphi} - i_Z \vartheta^\mu \frac{\delta {\cal L}^{1\text{st}}}{\delta \vartheta^\mu} - l'^{\alpha\beta} \frac{\delta {\cal L}^{1\text{st}}}{\delta \Gamma^{\alpha\beta}} - \alpha'^p \frac{\delta {\cal L}^{1\text{st}}}{\delta A^p}
\nonumber\\
&&\qquad + \varsigma \frac{\delta {\cal L}^{1\text{st}}}{\delta p} \wedge i_Z p - \frac{\delta {\cal L}^{1\text{st}}}{\delta \tau_\alpha} \wedge i_Z \tau_\alpha - \frac{\delta {\cal L}^{1\text{st}}}{\delta \rho_{\alpha\beta}} \wedge i_Z \rho_{\alpha\beta} - \frac{\delta {\cal L}^{1\text{st}}}{\delta {\cal P}_p} \wedge i_Z {\cal P}_p
\nonumber\\
&&\qquad + D (i_Z \varphi \wedge p + i_Z \vartheta^\alpha \tau_\alpha + l'^{\alpha\beta} \rho_{\alpha\beta} + \alpha'^p {\cal P}_p).
\end{eqnarray}

This generalized current expression is again an example of applying Noether's analysis.  In accord with the second theorem, for local symmetries the current conservation expression becomes a differential identity.  And again we have a detailed expression that reflects Noether's remarks regarding a more general version of Hilbert's assertion concerning the absence of a proper energy-momentum conservation law.  Moreover there is again the Noether current ambiguity regarding the total differential term.  However, as explained in section~\ref{S:Ham and B}, the first order current is also the Hamiltonian, the canonical generator of local transformations including spacetime displacements.
The total differential (boundary) term in the Hamiltonian can and should be adjusted, as we have discussed in general terms earlier in section~\ref{SS:co-symp Hbt}.   Then, when varied, the chosen boundary term in the  variation variation of the Hamiltonian gives the associated boundary conditions.  Thereby the usual Noether current ambiguity is fixed by the chosen boundary condition.

\subsection{General geometric Hamiltonian boundary terms}%%%jmn ok
%%%%%%%%%%%%%%%%%%%%%%%%%%%%%%%%%%%%%%%%%%%%%%%%%%%%%%%%%%%%%%%%%%%%%%
Specializing our general Hamiltonian boundary term expression~(\ref{genB}) to our present variables, with the preferred choice for the material and internal gauge fields leads to boundary term expressions which
explicitly contain only the geometric variables:
\begin{eqnarray}
{\cal B}(Z) &=&
i_Z \left\{ \begin{array}{c}
 \vartheta^\alpha \\ \bar \vartheta^\alpha
 \end{array} \right\}  \Delta \tau_\alpha
+ \Delta \vartheta^\alpha \wedge i_Z \left\{ \begin{array}{c}
    \tau_\alpha \\ \bar\tau_\alpha
 \end{array} \right\}
  \nonumber \\
&+&i_Z \left\{ \begin{array}{c}
 \Gamma^\alpha{}_\beta \\ \bar\Gamma^\alpha{}_\beta
  \end{array} \right\} \Delta \rho_\alpha{}^\beta
+  \Delta \Gamma^\alpha{}_\beta \wedge i_Z \left\{ \begin{array}{c}
    \rho_\alpha{}^\beta \\\bar \rho_\alpha{}^\beta
 \end{array} \right\}\ ,
\label{B(Z)}
\end{eqnarray}
where the upper or lower line in each bracket is to be selected.  A special case of this expression, (upper, lower, upper, upper) with $\bar\tau_\alpha=0$, was first proposed in 1991\cite{Nester:1991yd}.
With the above boundary term, the total differential term in $\delta {\cal H}(Z)$ has the symplectic  form
\begin{eqnarray}
{\cal C}(Z)&=&
 \left\{ \begin{array}{c}
   i_Z\delta \vartheta^\alpha \wedge \Delta \tau_\alpha \\
   - i_Z\Delta \vartheta^\alpha \wedge \delta \tau_\alpha
   \end{array} \right\}
+
\left\{ \begin{array}{c}
   \Delta \vartheta^\alpha \wedge i_Z\delta \tau_\alpha \\
   - \delta \vartheta^\alpha \wedge i_Z \Delta \tau_\alpha
   \end{array} \right\}\nonumber\\
&&+
 \left\{\begin{array}{c}
i_Z\delta \Gamma^\alpha{}_\beta \wedge \Delta \rho_\alpha{}^\beta \\
  - i_Z\Delta \Gamma^\alpha{}_\beta \wedge \delta \rho_\alpha{}^\beta
   \end{array} \right\}
 +\left\{\begin{array}{c}
\Delta \Gamma^\alpha{}_\beta \wedge i_Z\delta \rho_\alpha{}^\beta \\
  - \delta \Gamma^\alpha{}_\beta \wedge i_Z\Delta \rho_\alpha{}^\beta
   \end{array} \right\}.
     \label{C}
\end{eqnarray}

The general geometric Hamiltonians evolve gauge dependent quantities, including the frame and connection coefficients.  Consequently they naturally include terms that are gauge dependent. These terms are in particular those with factors of $i_Z\Gamma^\alpha{}_\beta$.
However the value of the Hamiltonian boundary term will then include a contribution to the energy etc.\ due to the choice of frame gauge. From such a boundary term one could still get the correct physical value, but only if one takes on the boundary the particular frame gauge $\pounds_Z\vartheta^\alpha=0$, which means one may need a different frame for the energy-momentum and angular momentum components.  For the purposes of obtaining directly a physical value for the observable quantities, one must separate the frame gauge dependent term into a physical energy plus a gauge dependent  unphysical energy.  This issue was first considered by Hecht,\cite{Hecht93,Hecht:1995fq} and he discovered a suitable remedy. The frame gauge dependent part can be separated using the identity
\begin{equation}
\pounds_Z \vartheta^\alpha \equiv d i_Z \vartheta^\alpha + i_Z d \vartheta^\alpha \equiv D i_Z \vartheta^\alpha + i_Z D \vartheta^\alpha - i_Z \Gamma^\alpha{}_\beta \vartheta^\beta.
\end{equation}
 With the aid of this relation one can get frame gauge independent boundary terms for the quasi-local quantities. Thus, to drop the contribution from the frame gauge, one should make the replacement
\begin{equation}
i_Z\Gamma^\alpha{}_\beta\equiv D_\beta Z^\alpha+(i_Z T^\alpha)_\beta-(\pounds_Z\vartheta^\alpha)_\beta \quad \rightarrow \quad \tilde D_\beta Z^\alpha,
\end{equation}
where we have introduced the convenient \emph{transposed}
connection: %%%jmn419 3d87
\begin{equation}
\tilde D Z^\alpha := D Z^\alpha + i_Z T^\alpha,
\end{equation}
which naturally shows up whenever one expresses Lie derivative expressions in terms of a covariant derivative. In addition to our argument above, this replacement has
 been justified using (i) theoretical analysis,\cite{Hecht93,Hecht:1995fq} (ii) calculations for exact solutions,\cite{Hecht:1993np} (iii) holonomic variables (for GR),\cite{Chen:2000xw} and (iv) via a manifestly covariant Hamiltonian formulation\cite{Nester08}.
 In our presentation here we used the coframe both for convenience and for its local translational gauge role; the reference just cited provides a completely frame independent---manifestly covariant---alternative approach to the covariant Hamiltonian formalism.

\subsection{Quasi-local boundary terms}%%%jmn ok
%%%%%%%%%%%%%%%%%%%%%%%%%%%%%%%%%%%%%%%%%%%%%%%%%%%%%%%%%%%%%%%%%%%%%%
With the aforementioned replacement we get our general set of symplectic quasi-local quantity boundary terms for the PG:
\begin{eqnarray}
{\cal B}(Z) &=&
i_Z \left\{ \begin{array}{c}
 \vartheta^\alpha \\ \bar \vartheta^\alpha
 \end{array} \right\}  \Delta \tau_\alpha
+ \Delta \vartheta^\alpha \wedge i_Z \left\{ \begin{array}{c}
    \tau_\alpha \\ \bar\tau_\alpha
 \end{array} \right\}
  \nonumber \\
&+& \left\{ \begin{array}{c}
 {\tilde  D}_\beta Z^\alpha \\ {\tilde {\bar D}}_\beta Z^\alpha
  \end{array} \right\} \Delta \rho_\alpha{}^\beta
+  \Delta \Gamma^\alpha{}_\beta \wedge i_Z \left\{ \begin{array}{c}
    \rho_\alpha{}^\beta \\\bar \rho_\alpha{}^\beta
 \end{array} \right\}\ ,
\label{B(Z)q}
\end{eqnarray}
where the upper or lower line in each bracket is to be selected. As in thermodynamics, there are several kinds of ``energy'', each corresponds to the work done in a different (ideal) physical process\cite{KT79, Jezierski:1990vu}.

\subsection{A preferred choice}%%%jmn ok
%%%%%%%%%%%%%%%%%%%%%%%%%%%%%%%%%%%%%%%%%%%%%%%%%%%%%%%%%%%%%%%%%%%%%%
The cases of the PG that have been studied are mostly those where the first order potential is quadratic in the momentum fields, this leads to a linear relation between the momenta and field strengths, and corresponds to 2nd order quasi-linear equations for the geometric potentials.  The natural reference choice is Minkowsi spacetime, for which $\bar\tau_\mu$ vanishes and $\bar{\tilde D}=\bar D$.  For this class of theories, other things being equal, we would favor from among the set~(\ref{B(Z)q}) the Hamiltonian boundary term quasi-local choice (upper, lower, lower, upper), i.e.,
%%%jmn319 3d88
\begin{equation}
{\cal B}(Z) = i_Z \vartheta^\alpha \tau_\alpha + \Delta \Gamma^\alpha{}_\beta \wedge i_Z \rho_\alpha{}^\beta + {\bar {D}}_\beta Z^\alpha \Delta \rho_\alpha{}^\beta, \label{Bpref(Z)}
\end{equation}
which leads to the following boundary term in the variation of the Hamiltonian:
\begin{equation}
{\cal C}(Z) = i_Z(\delta \vartheta^\alpha \wedge \tau_\alpha - \Delta \Gamma^\alpha{}_\beta \wedge \delta \rho_\alpha{}^\beta). \label{Cpref(Z)}
\end{equation}
This corresponds to imposing boundary conditions on the coframe and the momentum conjugate to the connection.
The associated energy flux expression is
\begin{equation}
\pounds_Z {\cal H}(Z) = d i_Z (\pounds_Z \vartheta^\alpha \wedge \tau_\alpha - \Delta \Gamma^\alpha{}_\beta \wedge \pounds_Z \rho_\alpha{}^\beta).
\end{equation}

Regarding the \emph{total} energy-momentum and angular momentum/center of mass momentum, the expression~(\ref{Bpref(Z)}) matches the expression (with Minkowski reference) for the PG Hamiltonian boundary term at spatial infinity that was first proposed by Hecht in 1993.\cite{Hecht93}

\subsection{Einstein's GR}%%%jmn ok
%%%%%%%%%%%%%%%%%%%%%%%%%%%%%%%%%%%%%%%%%%%%%%%%%%%%%%%%%%%%%%%%%%%%%%
Within the PG context, the special case of GR can be reached by imposing vanishing torsion with a Lagrange multiplier. In the general the first order formulation it is sufficient to take the potential to be independent of the coframe conjugate momentum $\tau_\mu$.

A first-order Lagrangian for vacuum\footnote{For our purposes concerning the Hamiltonian boundary term, we consider here for simplicity just the vacuum case.  The results will apply to all situations where the boundary of the region of interest is in the vacuum region, outside of the domain of the matter fields.  That should include most of the cases of physical interest.}
GR is
\begin{equation}\label{GRlag}
{\cal L}_{\rm GR} = R^{\alpha\beta} \wedge \rho_{\alpha\beta}
 + D\vartheta^\mu\wedge \tau_\mu- V^{\alpha\beta} \wedge \left( \rho_{\alpha\beta} - \frac1{2\kappa} \eta_{\alpha\beta} \right),
\end{equation}
which uses a Lagrange multiplier field $V^{\alpha\beta}\equiv V^{[\alpha\beta]}$ to give the connection's conjugate momentum field a value which depends on the orthonormal frame:
\begin{equation}\label{GRmult}
\rho_{\alpha\beta} - \frac1{2\kappa} \eta_{\alpha\beta}=0.
\end{equation}
Variation of~(\ref{GRlag}) w.r.t. the coframe, connection and their respective momenta fields gives the (vacuum) equations:
\begin{eqnarray}
\delta\vartheta^\mu:\quad 0&=&D\tau_\mu +V^{\alpha\beta}\wedge \frac1{2\kappa}\eta_{\alpha\beta\mu},\label{GRdeltatheta}\\
\delta\Gamma^{\alpha\beta}:\quad 0&=& D\rho_{\alpha\beta}+\vartheta_{[\beta}\wedge\tau_{\alpha]},\label{GRdeltaGamma}\\
\delta\tau_\mu:\quad 0&=&D\vartheta^\mu,\label{GRdeltatau}\\
\delta\rho_{\alpha\beta}:\quad 0&=&R^{\alpha\beta}-V^{\alpha\beta}. \label{GRdeltarho}
\end{eqnarray}
As expected~(\ref{GRdeltatau}) gives vanishing torsion.  From the differential of~(\ref{GRmult}) one gets
\begin{equation}
D\rho_{\alpha\beta}=\frac1{2\kappa}D\vartheta^\mu\wedge\eta_{\alpha\beta\mu},
\end{equation}
which vanishes, hence~(\ref{GRdeltaGamma}) reduces to $\vartheta_{[\beta}\wedge\tau_{\alpha]}=0$, from which it follows that $\tau_\mu$ vanishes.  Then~(\ref{GRdeltatheta}) with~(\ref{GRdeltarho}) reduces to the vanishing of the Einstein 3-form:
\begin{equation}
0=\frac1{2\kappa}R^{\alpha\beta}\wedge\eta_{\alpha\beta\mu}=-\frac1{\kappa}G^\nu{}_\mu\eta_\nu.
\end{equation}

By the way, had we included a suitable source, we would have obtained
\begin{equation}
0=\frac1{2\kappa}R^\alpha{}_\beta\wedge\eta_\alpha{}^\beta{}_\mu+\mathfrak{T}_\mu,
\end{equation}
where $\mathfrak{T}_\mu$ is the Hilbert energy-momentum 3-form. Using $D\eta_\alpha{}^\beta{}_\mu=0$ this relation can be rearranged as follows:\cite{SoN09b}
\begin{eqnarray}
0&=&
\frac1{2\kappa}\left[d(\Gamma^\alpha{}_\beta\wedge\eta_\alpha{}^\beta{}_\mu)+\Gamma^\alpha{}_\beta\wedge d\eta_\alpha{}^\beta{}_\mu
+\Gamma^\alpha{}_\gamma\wedge\Gamma^\gamma{}_\beta\wedge\eta_\alpha{}^\beta{}_\mu\right]+\mathfrak{T}_\mu\nonumber\\
&\equiv&
\frac1{2\kappa}\left[d(\Gamma^\alpha{}_\beta\wedge\eta_\alpha{}^\beta{}_\mu)
-\Gamma^\alpha{}_\gamma\wedge\Gamma^\gamma{}_\beta\wedge\eta_\alpha{}^\beta{}_\mu
+\Gamma^\alpha{}_\beta \wedge\Gamma^\gamma{}_\mu\wedge\eta_\alpha{}^\beta{}_\gamma
\right]+\mathfrak{T}_\mu. \label{pseudoform}
%\\
%&=:&-\frac1{\kappa}d \mathfrak{U}_\mu+ \mathfrak{t}_\mu+\mathfrak{T}_\mu,
\end{eqnarray}
The rearrangement identifies a certain superpotential 2-form,
\begin{equation}
\mathfrak{U}_\mu=-\Gamma^\alpha{}_\beta\wedge\eta_\alpha{}^\beta{}_\mu,
\end{equation}
and gravitational energy-momentum pseudotensor 3-form,
\begin{equation}
2\kappa\mathfrak{t}_\mu=-\Gamma^\alpha{}_\gamma\wedge\Gamma^\gamma{}_\beta\wedge\eta_\alpha{}^\beta{}_\mu
+\Gamma^\alpha{}_\beta \wedge\Gamma^\gamma{}_\mu\wedge\eta_\alpha{}^\beta{}_\gamma.
\end{equation}
These manipulations and the resultant expressions are meaningful in both orthonormal and holonomic frames.
In orthonormal frames they give the expressions for the so-called \emph{tetrad-teleparallel gauge current},\cite{deAndrade:2000kr} while in holonomic frames we have obtained neat form versions of the Freud superpotential~(\ref{UF}) and the Einstein pseudotensor~(\ref{EpseudoT}).  There is a remarkable contrast between the simple short calculation given here for these quantities and the long complicated ones discussed in section~\ref{S:pseudotensors} that were done in the past using tensor calculus.
Via rearrangements of the field equations analogous to~(\ref{pseudoform}), generalized pseudotensors can be found for the PG.\cite{Nester04}

\subsection{Preferred boundary term for GR}%%%jmn ok
%%%%%%%%%%%%%%%%%%%%%%%%%%%%%%%%%%%%%%%%%%%%%%%%%%%%%%%%%%%%%%%%%%%%%%
Over twenty years ago using the covariant Hamiltonian symplectic boundary term approach we proposed a quasi-local boundary term for GR\cite{Chen:1994qg} (an equivalent quasi-local expression obtained from a Noether argument using a global background reference was proposed at about the same time by Katz, Bi{\v c}\'ak \& Lynden-Bel\cite{LBKB95, Katz:1996nr}):
\begin{equation}
{\cal B}(Z) = \frac{1}{2\kappa} (\Delta\Gamma^{\alpha}{}_{\beta} \wedge i_Z \eta_{\alpha}{}^{\beta} + \bar
D_{\beta} Z^\alpha \Delta\eta_{\alpha}{}^\beta); \qquad \eta^{\alpha\beta\dots} := * (\vartheta^\alpha \wedge \vartheta^\beta \wedge \cdots). \label{BprefGR}
\end{equation}
(This has the form of Hecht's PG expression restricted to GR and natural extended to a boundary that need not be at infinity.)
The boundary term in the variation of the Hamiltonian has the form
\begin{equation}
\delta{\cal H}(Z) \sim di_Z (\Delta\Gamma^\alpha{}_\beta \wedge \delta\eta_{\alpha}{}^\beta),
\end{equation}
Since $\eta^{\alpha\beta} = *(\vartheta^\alpha \wedge \vartheta^\beta)$,
this corresponds to fixing the orthonormal coframe $\vartheta^\mu$ (and thus the metric) on the boundary.
The energy flux expression is
\begin{equation}
\pounds_Z{\cal H}(Z) = di_Z (\Delta\Gamma^\alpha{}_\beta \wedge \pounds_Z\eta_{\alpha}{}^\beta).
\end{equation}

Like many other boundary term choices, at spatial infinity it gives the ADM\cite{adm}, MTW\cite{MTW73}, Regge-Teitelboim\cite{Regge:1974zd}, Beig-\'O Murchadha\cite{Beig:1987zz}, Szabados\cite{Szabados:2003yn} energy, momentum, angular-momentum, center-of-mass momentum.

It has some special virtues:

(i) at null infinity: the Bondi-Trautman energy \& the Bondi energy flux\cite{Chen:2005hwa},

(ii) it is ``covariant'',

(iii) it is positive---at least for spherical solutions\cite{HX14} and large spheres,

(iv) for small spheres it is a positive multiple of the Bel-Robinson tensor\cite{Nester08},

(v) first law of thermodynamics for black holes\cite{Chen:1998aw},

(vi)  for spherical solutions it has the hoop property\cite{HX14},

(vii) for a suitable choice of reference it vanishes for Minkowski space.

%%%jmn419 3d90 3d91 3d92
%%%%%%%%%%%%%%%%%%%%%%%%%%%%%%%%%%%%%%%%%%%%%%%%%%%%%%%%%%%%%%%%%%%%%%
%\section{The reference and the quasi-local quantities}%%%jmn ok
\section{A ``best matched'' reference}
%%%%%%%%%%%%%%%%%%%%%%%%%%%%%%%%%%%%%%%%%%%%%%%%%%%%%%%%%%%%%%%%%%%%%%
\emph{In this section we turn to the second ambiguity that is inherent in quasi-local energy-momentum expressions: the choice of reference.  Minkowski space is the natural choice, but one still needs to choose a specific Minkowski space.  Recently we proposed (i) 4D isometric matching on the boundary and (ii) energy optimization as criteria for selecting the ``best matched'' reference on the boundary of the quasi-local region.}
\medskip

Note: for all other fields it is appropriate to choose  vanishing reference values as the reference ground state---the vacuum. But for geometric gravity the standard ground state is the non-vanishing Minkowski metric, so a non-trivial reference is essential. One still needs to specify exactly which Minkowski space.

Reference values can be determined by choosing, in a neighborhood of the desired spacelike boundary 2-surface $S$, 4 smooth functions $y^i = y^i(x^\mu),\ i = 0,1,2,3$  with $dy^0 \wedge dy^1 \wedge dy^2 \wedge dy^3 \ne 0$ and then defining a Minkowski reference by
\begin{equation}
\bar g = -(dy^0)^2 + (dy^1)^2 + (dy^2)^2 + (dy^3)^2.
%\bar\vartheta^i:= dy^i.
\end{equation}
Geometrically this is equivalent to finding a diffeomorphism for a neighborhood of the 2-surface into Minkowski space. The associated reference connection is the pullback of the flat Minkowski connection:
\begin{equation}
\bar \Gamma^\alpha{}_\beta = x^\alpha{}_i (\bar \Gamma^i{}_j y^j{}_\beta + dy^i{}_\beta) = x^\alpha{}_i dy^i{}_\beta.
\end{equation}
Here $x^\alpha{}_i$ is the inverse of $y^i{}_\beta$, where $dy^i = y^i{}_\beta \vartheta^\beta$.

With these standard Minkowski coordinates $y^i$, a Killing field of the reference has the form $Z^k = Z^k_0 + \lambda_0^k{}_l y^l$, where $\lambda_0^{kl} = \lambda_0^{[kl]}$, with $Z^k_0$ and $\lambda_0^{kl}$ being constants.
The 2-surface integral of any one of our Hamiltonian boundary terms then has a value linear in these constant values:
\begin{equation}
 \oint_S {\mathcal B}(Z) = -Z_0^k p_k(S) + \frac12 \lambda_0^{kl} J_{kl}(S). \label{value}
\end{equation}
This implicity determines not only a quasi-local energy-momentum but also a quasi-local angular momentum/center-of-mass momentum. When specialized to GR the integrals $p_k(S),\ J_{kl}(S)$ in the spatial asymptotic limit agree with accepted expressions for these quantities: MTW\cite{MTW73} \S 20.2 and Regge-Teitelboim\cite{Regge:1974zd} with the refinements of Beig-\'O Murchadha\cite{Beig:1987zz} and Szabados\cite{Szabados:2003yn}. For the PG at spatial infinity, Hecht~\cite{Hecht:1995fq} compared in detail his expression with the other proposed expressions, e.g., Ref.~[\refcite{BlagVas}].  At spatial infinity, with the asymptotics~(\ref{5:asymptotics}), all of our PG symplectic boundary terms~(\ref{B(Z)q}) will give the same values as Hecht's expression~(\ref{Bpref(Z)}).

For energy-momentum one takes $Z$ to be a translational Killing field of the Minkowski reference. Then the second term in our preferred quasi-local boundary expressions~(\ref{Bpref(Z)},\ref{BprefGR}) vanishes.\footnote{For GR the second term in~(\ref{BprefGR}) also vanishes for 4D isometric matching on $S$, a condition we shall use below.}
With $Z^k = Z^k_0 = $ constant our preferred quasi-local expressions now take the form
\begin{eqnarray}
{\mathcal B}^{\rm PG}(Z) &=&
 Z^k_0 x^\mu{}_k[\tau_\mu+ (\Gamma^\alpha{}_\beta - x^\alpha{}_j \, dy^j{}_\beta) \wedge i_{e_\mu}\rho_{\alpha}{}^\beta]\,, \label{PGB2}\\
{\mathcal B}^{\rm GR}(Z) &=& Z^k_0 x^\mu{}_k (\Gamma^\alpha{}_\beta - x^\alpha{}_j \, dy^j{}_\beta) \wedge \eta_{\mu\alpha}{}^\beta\,. \label{B2}
\end{eqnarray}

\subsection{The choice of reference}%%%jmn ok
%%%%%%%%%%%%%%%%%%%%%%%%%%%%%%%%%%%%%%%%%%%%%%%%%%%%%%%%%%%%%%%%%%%%%%
To be completed, our Hamiltonian boundary term and the quasi-local energy-momentum/angular momentum proposal needs a prescription for choosing a reference on the boundary. There are several options; one needs a choice suited to the application.

For an extended region one may want a global background (see Refs.~[\refcite{Petrov,PetKatz02}] for this approach in GR). Consider for example solar system applications, more specifically say we want to calculate using our quasi-local energy flux formula the tidal energy flux between Jupiter and its moon Io, that is believed to be responsible for Io's volcanos. (This has already been done by several methods\cite{Purdue, BoothCr, Favata}.)
On the other hand, if one wishes to study a given a metric expressed in a specific coordinate system the analytic approach~\cite{Liu:2011jha} may be a good choice.

%***********************************
%From our unpublished paper
%***********************************

To explicitly determine the specific values of the quasi-local quantities one needs some good way to choose the reference. Minkowski spacetime is the natural choice, especially for asymptotically flat spacetimes. However, as noted above, almost any four functions will determine some Minkowski reference. With such freedom one can still get almost any value for the quasi-local quantities. This freedom is the quasi-local version of the second type of ambiguity mentioned in the introduction.

Recently we proposed a program~\cite{ae100} to fix the ``best'' choice for a \emph{quasi-local reference}, i.e., one that is determined by the dynamical fields on the boundary. It has two parts: 4D isometric matching and optimization of a certain quantity. Here we present it along with a promising alternative optimization.\cite{Chen:2013xua}
For GR we have already found that our new procedure works well for an important special case: a certain class of axisymmetric spacetimes~\cite{Sun:2013ika} which includes the Kerr metric.

For the PG, not so much has been done yet.   While PG energy-momentum and angular momentum calculations at both spatial and future null infinity were done long ago~\cite{Hecht:1993np,HN96}, and gave, in particular, the expected results for the Kerr metric, as far as we know there have not yet been any genuine \emph{quasi-local} (i.e., for a finite region) PG calculations. This is not so surprising. In general the big obstacle is the 2D isometric embedding, which we are about to discuss.  For spherical symmetry, all the calculations at least for GR can easily be done analytically.  For the aforementioned class of axisymmetric metrics the 2D embedding problem has an algebraic solution. But the boundary expressions are already sufficiently complicated that the quasi-local energy integral for GR could only be evaluated numerically.   Now that it is known how to do the case of axisymmetric GR the way is open for truly quasi-local PG energy and angular momentum calculations.  For the PG, it seems that numerical calculations will be unavoidable.

\subsection{Isometric matching of the 2-surface}
%%%%%%%%%%%%%%%%%%%%%%%%%%%%%%%%%%%%%%%%%%%%%%%%%%%%%%%%%%%%%%%%%%%%%%
We first recall an important procedure that has been used: isometric matching of the 2-surface $S$. This can be expressed in terms of quasi-spherical foliation adapted coordinates $t, r, \theta, \varphi$ as
\begin{equation}
g_{AB} \; \dot= \; \bar g_{AB} = \bar g_{ij} y^i_A y^j_B = -y^0_A y^0_B + \delta_{ij} y^i_A y^j_B\,, \label{2Diso}
\end{equation}
where $S$ is given by constant values of $t, r$, and $A, B$ range over $\theta, \varphi$.  We use $\dot=$ to indicate a relation which holds only on the 2-surface $S$. Eq.~(\ref{2Diso}) is 3 conditions on the 4 functions $y^i$.  One can regard $y^0$ as the free choice. From a classic closed 2-surface into $\mathbb R^3$ embedding theorem---as long as $S$ and $y^0(x^\mu)$ are such that on $S$
\begin{equation}
g_{AB}' := g_{AB} + y^0_A y^0_B \label{eff2Dmetric}
\end{equation}
is convex---one has a unique embedding. Wang and Yau have discussed in detail this type of
embedding of a 2-surface into Minkowski controlled by one function in their
recent quasi-local work~\cite{WY09,WYcmp09,CWY15}. %{WYcmp09,Wang:2008jy}.

Unfortunately, although there is a unique embedding, there is generally no explicit analytic formula except in special simple cases, such as spherical symmetry or axisymmetry.  The lack of an explicit formula for the solution of this 2D isometric embedding prevents exact quasi-local calculations in general.

\subsection{Complete 4D isometric matching}
%%%%%%%%%%%%%%%%%%%%%%%%%%%%%%%%%%%%%%%%%%%%%%%%%%%%%%%%%%%%%%%%%%%%%%
Our ``new'' proposal\footnote{For GR this was proposed by Szabados at a workshop in Hsinchu, Taiwan in 2000. He has since investigated it in detail\cite{Sza2000}.} is: complete 4-dimensional isometric matching on $S$:
$g\dot=\bar g$, a part of which is still the just discussed 2D isometric embedding.

In view of isometric matching, there should be a Lorentz transformation which on the boundary brings the dynamical coframe into line with the reference frame:
\begin{equation} \vartheta^i:=L^i{}_\alpha\vartheta^\alpha\dot=dy^i.\label{4Diso}
\end{equation}
The integrability condition for this equation is
\begin{equation}
d\vartheta^i|_S=0.\label{integrability}
\end{equation}
This 2-form equation restricted to the 2-surface gives 4 conditions on the 6 parameter set of Lorentz transformations $L^i{}_\alpha$.
Thus 4D isometric matching has $6-4=2$ degrees of freedom.
%\section{The best matched reference geometry}
They can be identified as the choice of time embedding function $y^0$ in~(\ref{2Diso}) plus a boost parameter $\alpha$ in the plane normal to $S$.

\subsection{Optimal energy}
%%%%%%%%%%%%%%%%%%%%%%%%%%%%%%%%%%%%%%%%%%%%%%%%%%%%%%%%%%%%%%%%%%%%%%
To fix the remaining 2 isometric matching parameters, one can regard the quasi-local value as a measure of the difference between the dynamical and the reference boundary values. This value will be a functional of $y^0,\alpha$. The critical points of this functional determine the distinguished choices for these 2 functions.

Previously we proposed~\cite{ae100} taking the optimal ``best matched'' embedding as the one which gives an extreme value to the associated invariant mass $m^2 = -p_i p_j \bar g^{ij}$. This should determine the reference up to a Poincar\'e transformation.

This is a reasonable condition, but, unfortunately, not so practical. The invariant mass is a sum of 4 terms, each quadratic in an integral over $S$. Note, however, that using the Poincar\'e freedom one can get the same $m$ value in the center-of-momentum frame from $p_0$. This leads us to our new proposal: take the preferred reference as one that gives a critical value to the quasi-local \emph{energy} given by~(\ref{value}) and~(\ref{B2}) or (\ref{PGB2}) with $Z^k = Z^k_0 = \delta^k_0$.  We expect this much simpler optimization to give the same reference geometry as that obtained from using $m^2$.

Based on some physical and practical computational arguments it seems reasonable to expect a unique solution in general. In a numerical calculation in principle one could just calculate the energy values given by~(\ref{value}) and~(\ref{PGB2}) or (\ref{B2}) with $Z^k = \delta^k_0$ for a great many choices of $y^0, \alpha$ subject to the 4D isometric matching constraint~(\ref{4Diso}) and the integrability condition~(\ref{integrability}), and then note the energy critical points.

For GR this ``best matching'' procedure already gave reasonable quasi-local energy results for spherically symmetric systems\cite{Chen:2009zd, Liu:2011jha, Wu:2011wk, Wu:2012mi}. For the Schwarzschild metric the ``best matched'' quasi-local energy has the value first found by Brown and York,\cite{Brown:1992br}
\begin{equation}
E(r)=\frac{2m}{1+\sqrt{1-2m/r}},
\end{equation}
using a closely related boundary term and a 2-surface embedding into $\mathbb{R}^3$.
Now we also have sensible results for certain axisymmetric systems including the Kerr metric.\cite{Sun:2013ika} For the surface of constant Boyer-Linquist $r$, the angular momentum is simply a constant independent of $r$, equal to its usual asymptotic value, $J$.  The quasi-local energy integral, however, is not so simple and can only be evaluated numerically.

%

%%%%%%%%%%%%%%%%%%%%%%%%%%%%%%%%%%%%%%%%%%%%%%%%%%%%%%%%%%%%%%%%%%%%%%
\section{Concluding discussion}%%%jmn118
%%%%%%%%%%%%%%%%%%%%%%%%%%%%%%%%%%%%%%%%%%%%%%%%%%%%%%%%%%%%%%%%%%%%%%
In addition to her two key theorems regarding symmetry, Emmy Noether in her 1918 paper also proved that for diffeomorphically invariant gravity there is no proper total energy-momentum density. In other words there is no covariant total energy momentum density tensor for gravitating systems. In Ch.~20 of their text\cite{MTW73} Misner, Thorne and Wheeler discuss this feature as a consequence of the equivalence principle and argue that only the total energy-momentum of gravitating systems is meaningful. But clearly the gravitational interaction is local and does allow for the \emph{local} exchange of energy-momentum. To account for this various non-covariant expressions called pseudotensors (each generated by a certain superpotential) have been proposed. There thus arose two ambiguities: which expression? in which reference frame? The modern idea is \emph{quasi-local}: energy-momentum is associated with a closed 2-surface.

One approach, which is the one we have used, is via the Hamiltonian. With the aid of differential forms and a first order variational formulation, we have developed a covariant Hamiltonian formalism. The Hamiltonian boundary term plays key roles: it determines the boundary conditions and the quasi-local values. We have shown that this approach includes all the traditional pseudotensors while clarifying the ambiguities: each superpotential is a possible Hamiltonian boundary term which is associated with a specific boundary condition, and the reference frame becomes a choice of the reference values (ground state) on the boundary.

One can regard gravity as a local gauge theory for the global symmetry group for Minkowski spacetime: the Poincar\'e group. The appropriate geometry is \emph{Riemann-Cartan}: the \emph{curvature} is the field strength for Lorentz transformations and the \emph{torsion} is the field strength for local translations (infinitesimal diffeomporhisms).   For comparison, we included in our discussion a general internal gauge gauge field. In this way we can identify the analog of the internal gauge vector potential. For the local Lorentz symmetry it is the (metric compatible) \emph{connection one-form}; for the local translation symmetry it is the orthonormal \emph{coframe}. We developed in considerable detail the associated Lagrangian and Hamiltonian Noether currents and differential identities, so one can compare the similarities and differences of the spacetime gauge symmetry expressions with those of a generic internal symmetry.  Briefly, the Lorentz rotation sector is quite similar to that of an internal gauge symmetry, but the translation symmetry has both striking similarities and differences.  This would be much less clear if we had opted for a formulation which does not include the coframe as a dynamical variable.  In the approach of Blagojevi\'c \& Hehl,\cite{BlagojevicHehl} the use of the orthonormal frame is motivated by the need to describe spin; here our basic motivation is in terms of the coframe's fundamental gauge role regarding local spacetime translations.

The geometric/gauge symmetry approach is helpful in identifying a good expression for our Hamiltonian boundary term expression for quasi-local quantities.  Our preferred expression for GR turns out to correspond to fixing the metric on the boundary---which is obviously a reasonable boundary condition choice.

A notable feature of the Hamiltonian boundary term for dynamic geometry is that it necessarily depends on the choice of some non-dynamical reference values (this is a manifestation of Noether's result regarding non-existence of a proper energy-momentum density).  One can get almost any quasi-local value if one allows free rein in the choice of reference.  One needs to fix a non-dynamical reference frame, but only on the boundary of the region.  This can be compared to choosing some flat plane to map a part of the curved surface of the Earth.  One could slice the plane through the surface of the Earth; for a spherical Earth the planar geometry would exactly match on a circle.  Similarly for spacetime.  On the 2-dimensional boundary of a spatial region one can exactly match the curved 4D spacetime metric with a flat Minkowski spacetime metric.  Detailed analysis shows that there is still two degrees of freedom.  We proposed that a good way to fix these was by considering the critical points of our quasi-local expression.  There might be other sensible options, but the main point is that  a reasonable choice is available.

Our principal results, the Hamiltonian boundary terms that are our preferred quasi-local energy-momentum expressions~(\ref{Bpref(Z)},\ref{BprefGR}) for the PG and GR,
%%%jmn419 3d93
%our Hamiltonian boundary term quasi-local energy-momentum expression~(\ref{BprefGR}),
were obtained by considering the Hamiltonian, geometry, Noether symmetry, and spacetime gauge theory.
The harmonious combination of all of these perspectives makes a strong case for the results.
Nevertheless, it should be noted that one can also be led to this result from other perspectives.  Regarding GR, essentially the same expression has been obtained (i) via a Noether approach with a global reference\cite{LBKB95, Katz:1996nr}, and (ii)
via a symplectic covariant Hamiltonian approach using the metric in a holonomic frame\cite{Chen:2000xw}.
For the PG (including GR as a special case), the same preferred expression was found
 via an entirely frame independent manifestly covariant Hamiltonian formalism\cite{Nester08}.
Although in principle there are an unlimited number of possible Hamiltonian boundary term quasi-local energy-momentum expressions---which are in a formal sense all of equal status---in practice one can---with very good reasons---discover that one expression stands out.

%%%%%%%%%%%%%%%%%%%%%%%%%%%%%%%%%%%%%%%%%%%%%%%%%%%%%%%%%%%%%%%%%%%%%%
\section*{Acknowledgements}%%%jmn117
%%%%%%%%%%%%%%%%%%%%%%%%%%%%%%%%%%%%%%%%%%%%%%%%%%%%%%%%%%%%%%%%%%%%%%

%%%jmn419 Acknowledge F.W. Hehl
We would first like to thank Prof.~F.~W. Hehl for his many recommendations
that have been very helpful in our efforts to improve the quality of this presentation.
J.M.N. would like to express his appreciation for the hospitality at the
Institute of Physics, Academia Sinica, Taipei 115, Taiwan, the
Mathematical Sciences Center, Tsinghua University, Beijing, China 100084, and
the Morningside Center of Mathematics,
Academy of Mathematics and System Science, Chinese Academy
of Sciences, Beijing
100190, China.  Part of this work was developed at those institutions during visits in 2013
and 2014.

C.M.C. was supported by the Ministry of Science and Technology of the R.O.C. under the grant MOST 102-2112-M-008-015-MY3.

%%%%%%%%%%%%%%%%%%%%%%%%%%%%%%%%%%%%%%%%%%%%%%%%%%%%%%%%%%%%%%%%%%%%%%%%%%

%%%jmn117 The references that are not cited should be commented out---but not deleted, because we might include them in a later revision

\end{document}